\documentclass[longauth]{aa} 

\usepackage{graphicx}
\usepackage{comment}
\usepackage[utf8]{inputenc}
\usepackage{enumerate}
\usepackage{amsmath} 
\usepackage{enumitem} 
\usepackage{natbib} 

\usepackage[labelfont = bf,labelsep = period]{caption}
\usepackage{subcaption}
\usepackage{caption}
\usepackage[normalem]{ulem}
\usepackage{rotating}
\usepackage{adjustbox}

\usepackage{txfonts}
\usepackage{xcolor}
\definecolor{darkgreen}{rgb}{0.0, 0.5, 0.0}

\usepackage[colorlinks=True]{hyperref}
\hypersetup{
    linkcolor=blue,  
    citecolor=blue,
    }

\usepackage{etoolbox}   
\usepackage{orcidlink}

\begin{document}

    \title{Spatially-resolved spectro-photometric SED Modeling of NGC\,253's Central Molecular Zone}
    \subtitle{I. Studying the star formation in extragalactic giant molecular clouds}

    \author{Pedro K. Humire\orcidlink{0000-0003-3537-4849} \inst{\ref{inst.USP}}
    \and Subhrata Dey\orcidlink{0000-0002-4679-0525} \inst{\ref{inst.uj}, \ref{inst.NCBJ}}
    \and Tommaso Ronconi\orcidlink{0000-0002-3515-6801} \inst{\ref{inst.SISSA},\ref{inst.IFPU},\ref{inst.OABo},\ref{inst.ICSC}}
    \and Victor H. Sasse\orcidlink{0009-0008-8763-7050} \inst{\ref{inst.UFSC}}
    \and Roberto Cid Fernandes\orcidlink{0000-0001-9672-0296} \inst{\ref{inst.UFSC}}
    \and Sergio Martín\orcidlink{0000-0001-9281-2919} \inst{\ref{inst.ESOChile},\ref{inst.JAO}}
    \and Darko Donevski\orcidlink{0000-0001-5341-2162} \inst{\ref{inst.NCBJ}, \ref{inst.SISSA}}
    \and Katarzyna Ma{\l}ek\orcidlink{0000-0003-3080-9778} \inst{\ref{inst.NCBJ}}
    \and Juan A. Fernández-Ontiveros \orcidlink{0000-0001-9490-899X}
    \inst{\ref{inst.CEFCA}}
    \and Yiqing Song \orcidlink{0000-0002-3139-3041} \inst{\ref{inst.ESOChile}, \ref{inst.JAO}}
    \and Mahmoud Hamed\orcidlink{0000-0001-9626-9642} \inst{\ref{inst.NCBJ}, \ref{inst.CAB}}
    \and Jeffrey G. Mangum\orcidlink{0000-0003-1183-9293} \inst{\ref{inst.NRAOCV}}
    \and Christian Henkel\orcidlink{0000-0002-7495-4005} \inst{\ref{inst.MPIfR},\ref{inst.Xinjiang}}
    \and Víctor M. Rivilla\orcidlink{0000-0002-2887-5859} 
    \inst{\ref{inst.CAB}}
    \and
    Laura Colzi\orcidlink{0000-0001-8064-6394} 
    \inst{\ref{inst.CAB}}
    \and N. Harada\orcidlink{0000-0002-6824-6627} \inst{\ref{inst.NAOJ},\ref{inst.ASIAA}, \ref{inst.SOKENDAI}}
    \and Ricardo Demarco\orcidlink{0000-0003-3921-2177} \inst{\ref{inst.AB_Concepcion}}
    \and Arti Goyal\orcidlink{0000-0002-2224-6664} \inst{\ref{inst.uj}}
    \and David S. Meier\orcidlink{0000-0001-9436-9471}
    \inst{\ref{inst.NMT}}
    \and Swayamtrupta Panda\orcidlink{0000-0002-5854-7426} 
    \inst{\ref{inst.gemini}, \thanks{Gemini Science Fellow}}
    \and Ângela C. Krabbe\orcidlink{0000-0003-4630-1311} \inst{\ref{inst.USP}}
    \and Yaoting Yan\orcidlink{0000-0001-5574-0549} \inst{\ref{inst.MPIfR}}
    \and Amanda R. Lopes\orcidlink{0000-0002-6164-5051} \inst{\ref{inst.IALP}}
    \and K. Sakamoto\orcidlink{0000-0001-5187-2288} \inst{\ref{inst.ASIAA}}
    \and S. Muller\orcidlink{0000-0002-9931-1313} \inst{\ref{inst.ONSALA}}
    \and K. Tanaka\orcidlink{0000-0001-8153-1986} \inst{\ref{inst.KeioUniversity}}
    \and Y. Yoshimura\orcidlink{0000-0003-0563-067X}  \inst{\ref{inst.UTokio}}
    \and K. Nakanishi\orcidlink{0000-0002-6939-0372} \inst{\ref{inst.NAOJ},\ref{inst.SOKENDAI}}
    \and Antonio Kanaan\orcidlink{0000-0002-2484-7551} \inst{\ref{inst.UFSC}}
    \and Tiago Ribeiro\orcidlink{0000-0002-0138-1365} \inst{\ref{inst.NOAO}}
    \and William Schoenell\orcidlink{0000-0002-4064-7234} \inst{\ref{inst.GMTO}}
    \and Claudia Mendes de Oliveira\orcidlink{0000-0002-5267-9065} \inst{\ref{inst.USP}}    }

    \institute{\label{inst.USP}Departamento de Astronomia, Instituto de Astronomia, Geofísica e Ciências Atmosféricas da USP, Cidade Universitária, 05508-090 São Paulo, SP, Brazil,
    \label{email}pedrokhumire@usp.br
    \and\label{inst.uj}Astronomical Observatory of the Jagiellonian University, Orla 171, 30-244 Krak\'{o}w, Poland
    \and\label{inst.NCBJ}National Center for Nuclear Research, ul. Pasteura 7, 02-093 Warsaw, Poland
    \and\label{inst.SISSA}Scuola Internazionale Superiore di Studi Avanzati (SISSA), Via Bonomea 265, IT-34136, Trieste, Italy
    \and\label{inst.IFPU}Institute for Fundamental Physics of the Universe (IFPU), Via Beirut 2, IT-34151, Trieste, Italy
    \and\label{inst.OABo}INAF - Osservatorio di Astrofisica e Scienza dello Spazio (OAS), Via Gobetti 93/3, I-40127 Bologna, Italy,
    \and\label{inst.ICSC}ICSC – Centro Nazionale di Ricerca in High Performance Computing, Big Data e Quantum Computing, Via Magnanelli 2, Bologna, Italy
    \and \label{inst.UFSC}Universidade Federal de Santa Catarina, Campus Universitário Reitor João David Ferreira Lima, 88040-900, Florianópolis, SC, Brazil
    \and \label{inst.ESOChile}European Southern Observatory, Alonso de C\'ordova, 3107, Vitacura, Santiago 763-0355, Chile  
    \and \label{inst.JAO}Joint ALMA Observatory, Alonso de C\'ordova, 3107, Vitacura, Santiago 763-0355, Chile  
    \and \label{inst.CEFCA}Centro de Estudios de F\'isica del Cosmos de Arag\'on (CEFCA), Plaza San Juan 1, E--44001 Teruel, Spain
    \and\label{inst.CAB}Centro de Astrobiolog{\'i}a (CAB), CSIC-INTA, Carretera de Ajalvir km 4, Torrej{\'o}n de Ardoz, 28850, Madrid, Spain
    \and\label{inst.NRAOCV}National Radio Astronomy Observatory, 520 Edgemont Road, Charlottesville, VA 22903-2475, USA
    \and \label{inst.MPIfR}Max-Planck-Institut f\"ur Radioastronomie, Auf-dem-H\"ugel 69, 53121 Bonn, Germany
    \and\label{inst.Xinjiang}Xinjiang Astronomical Observatory, Chinese Academy of Sciences, 830011 Urumqi, China
    \and\label{inst.NAOJ}National Astronomical Observatory of Japan, 2-21-1 Osawa, Mitaka, Tokyo 181-8588, Japan
    \and\label{inst.ASIAA}Institute of Astronomy and Astrophysics, Academia Sinica, 11F of AS/NTU Astronomy-Mathematics Building, No.1, Sec. 4, Roosevelt Rd, Taipei 106319, Taiwan
    \and\label{inst.SOKENDAI}Department of Astronomy, School of Science, The Graduate University for Advanced Studies (SOKENDAI), 2-21-1 Osawa, Mitaka, Tokyo, 181-1855 Japan
    \and\label{inst.AB_Concepcion}Institute of Astrophysics, Facultad de Ciencias Exactas, Universidad Andr\'es Bello, Sede Concepci\'on, Talcahuano, Chile
    \and \label{inst.NMT} New Mexico Institute of Mining and Technology, 801 Leroy Place, Socorro, NM, 87801, USA
    \and\label{inst.gemini}International Gemini Observatory/NSF NOIRLab, Casilla 603, La Serena, Chile
    \and \label{inst.IALP} Instituto de Astrofísica de La Plata, CONICET-UNLP, Paseo del Bosque s/n, B1900FWA, Argentina
    \and\label{inst.ONSALA}Department of Space, Earth and Environment, Chalmers University of Technology, Onsala Space Observatory, SE-43992 Onsala, Sweden
    \and\label{inst.KeioUniversity}Department of Physics, Faculty of Science and Technology, Keio University, 3-14-1 Hiyoshi, Yokohama, Kanagawa 223--8522 Japan
    \and\label{inst.UTokio}Institute of Astronomy, Graduate School of Science, The University of Tokyo, 2-21-1 Osawa, Mitaka, Tokyo 181-0015, Japan
    \and \label{inst.NOAO} NOAO, 950 North Cherry Ave. Tucson, AZ 85719, United States \and \label{inst.GMTO} GMTO Corporation, N. Halstead Street 465, Suite 250, Pasadena, CA 91107, United States
    }

   \date{Received January 27, 2025; accepted May 14, 2025}

  \abstract{Studying the interstellar medium in nearby starbursts is essential for gaining insights into the physical mechanisms driving these extreme objects, thought to be analogs of young, primeval, star-forming galaxies. This task is now feasible due to deep spectro-photometric data enabled by rapid advancements in ground- and space-based facilities. To fully leverage this wealth of information, extracting insights from the spectral line properties, and the spectral energy distribution (SED) is imperative.}
  {This study aims to produce and analyze the physical properties of the first spatially-resolved multi-wavelength SED of an extragalactic source that covers six decades in frequency (from near-ultraviolet to centimeter wavelengths) at an angular resolution of 3$^{\prime\prime}$ which corresponds to a linear scale of $\sim$51 pc at the distance of NGC\,253. We focus on the central molecular zone (CMZ) of this starburst galaxy, which contains giant molecular clouds (GMCs) responsible for half of the galaxy's star formation.}
  {We retrieve archival data from near-UV to centimeter wavelengths, covering six decades of spectral range. We compute SEDs to fit the observations using the GalaPy code and confronting results with the CIGALE code for validation. We also employ the\textsc{starlight} code to analyze the stellar optical spectra of the GMCs. }
  {Our results reveal significant differences between internal and external GMCs in terms of stellar and dust masses, star formation rates (SFRs), and bolometric luminosities, to provide a few, with internal GMCs doubling maximum values of the external ones in most of the cases. We have obtained tight relations between monochromatic stellar tracers and star-forming conditions obtained from panchromatic emission. We find that the best SFR tracers are radio continuum bands at 33~GHz, radio recombination lines (RRLs), and the total infrared (IR) luminosity range (L$_{\rm IR}$; 8--1000$\mu$m) as well as the IR emission at 60$\mu$m. The emission line diagnostics based on the BPT and WHAN diagrams suggest that the nuclear region of NGC~253 exhibits shock signatures, placing it in the composite zone typically associated with AGN/star-forming region hybrids, while the AGN fraction from panchromatic emission is negligible ($\leq$7.5\%).}
  {Our findings demonstrate the significant heterogeneity within the CMZ of NGC~253, with central GMCs exhibiting high densities, elevated SFRs, and greater dust masses compared to their external counterparts. We confirm the effectiveness of certain centimeter photometric bands as a reliable method to estimate the global SFR, in accordance with previous studies but this time on GMC scales.}


   \maketitle
%
\section{Introduction}
The panchromatic spectral energy distribution (SED) of an astrophysical system encapsulates the intricate interplay among its various components, including stars in different evolutionary phases and their remnants; molecular, atomic, and ionized gas; dust, and compact objects such as supermassive black holes. It contains the imprints of baryonic processes that drive its formation and evolution across cosmic times. A comparison of SEDs across different wavebands provides crucial constraints into the physical mechanisms governing the system’s radiation, including the relative contributions of stellar, nebular, and dust emission, as well as the processes that regulate its energy balance. Therefore, modeling the SED is one of the most effective tools, if not the most effective, for estimating the specific star formation rate (sSFR), from which the star formation rate (SFR) and the stellar mass ($M_{\rm{\star}}$) can be derived. This method relies on reconstructing the stellar emission through star formation history (SFH) models, which can be either purely theoretical or non-parametric (the latter refers to models that do not assume a specific functional form and rely on templates to infer the SFH; see, e.g., \citealt{CidFernandes2005,Cunha08,Conroy2013}). SED fitting allows for a detailed analysis of the light emitted across different wavelengths, providing insights into the age, metallicity, and dust content of the stellar population \citep[e.g.,][]{Bruzual2003}.

While SEDs have been extensively applied to high-redshift galaxies, there are inherent resolution limitations, particularly at far-infrared (FIR) wavelengths. For example, the Herschel SPIRE beam has a resolution of 36$\arcsec$ at 500~$\mu$m, corresponding to roughly 1~kpc at $z\approx0.0013$, meaning that the observed SEDs often conflate contributions from recent star formation, AGN, and/or stellar outflows, and sub-thermally excited gas, to mention a few. Considering also that a photometric aperture of 2.5 times the beam, assuming a Gaussian profile for the emission source, ensures the capture of extended emission, the problem becomes more critical. Given these constraints, to gain a deeper understanding of the relevant processes affecting galaxies, such as epochs with galaxy-galaxy interactions or times when their star formation began to decline \citep[e.g.,][]{Hopkins2006,Hopkins_mergers2010}, it is imperative to focus on local analogs. In this regard, the extensively studied nearby starburst galaxy NGC~253 serves as an ideal target due to its proximity and intense nuclear star formation, which mirrors conditions seen in distant galaxies \citep{Leroy2015}. Focusing on spatially-resolved objects, such as giant molecular clouds (GMCs), allows us to take a significant step forward in producing SED fittings with unprecedented detail, overcoming the aforementioned limitations.

NGC\,253 is a nearby starburst galaxy about 3.5~Mpc away with a steep inclination ranging from approximately 70$^\circ$ to 79$^\circ$ \citep{Rekola2005,Pence1980,Iodice2014}. Its inner kiloparsec (kpc) region is a hub of vigorous star formation, generating a rate of roughly 1.7 solar masses per year, which constitutes about half of the galaxy's total star formation activity \citep{Bendo2015,Leroy2015}. This active star-forming core encompasses a central molecular zone (CMZ), spanning around 300 parsecs in length and 100 parsec in width projection \citep{Sakamoto2011} and hosting ten giant molecular clouds (GMCs) discerned through molecular and continuum emissions at high resolution \citep{Leroy2015}. Notably, studies indicate that an active galactic nucleus (AGN) has minimal impact on the molecular gas within NGC~253's CMZ \citep{FernandezOntiveros2009,MullerSanchez2010,PerezBeaupuits2018}, making it an excellent test-bed to study pure star-forming conditions.

The CMZ of NGC\,253, along with its GMCs, has been the focus of the ALMA Comprehensive High-resolution Extragalactic Molecular Inventory (ALCHEMI) program \citep{Martin2021}. This project delves into chemical changes caused by shocks or cosmic ray flux density variations, some of them also aiming for a direct comparison with galaxy conditions, the discovery of new molecules, and the characterization of the GMCs that highlights differences between the very central part of the CMZ and its outskirts \citep{Holdship2021,Harada2021,Martin2021,Behrens2022ApJ,Holdship2022ApJ,Humire2022,Harada2022,Butterworth2024,Huang2023, Behrens2024ApJ,Bouvier2024,Tanaka2024ApJ,Harada2024}.

Previous studies of the NGC~253 spectral energy distribution (SED) have focused on specific wavelength regimes like optical/IR \citep{FernandezOntiveros2009}, mid-IR to (sub-)millimeter \citep{PerezBeaupuits2018}, far-UV to far-IR \citep{Beck2022}, the Rayleigh-Jeans tail \citep{Martin2021}, radio (1~GHz) to sub-mm (350~GHz) \citep{Belfiori2025} and low-frequency radio (MHz to 11~GHz) regimes \citep{Kapinska2017}. However, there remains significant potential in extending these investigations across a broader wavelength regime. This extension can provide valuable insights into various physical processes, including \text{attenuation} levels, the (s)SFR, and the mass-to-dust rates, to provide a few. 

In the present article, we aim to create the first extragalactic panchromatic SED (near-ultraviolet to centimeter wavelengths) performed at a linear resolution of 51~pc or 3\arcsec, at the typical physical size of a giant molecular cloud. For the specific case of this article, given the massive amount of information extracted, we will focus on the stellar processes derived per GMC, such as the SFRs and SFHs.

Along with the acquisition of our new results, we also highlight the importance of testing tracers or proxies that focus on a single wavelength range against the values provided by the full SED to understand the scope of the former in terms of luminosity, ages, object classifications, and calculated SFRs. This way, one can gauge their effectiveness on a larger spectro-photometric wavelength scale.

In the following sections, we will present the collected observations (Sect.~\ref{sec.observations}), our SED and spectroscopic modeling (Sect.~\ref{sec.SED_and_spec_modeling}), our results with small discussions inside (Sect.~\ref{sec.results}), and a general discussion on common limitations of our methods and the placement of our findings in a high-z perspective (Sect.~\ref{sec.discussion}), before providing our conclusions in Sect.~\ref{sec.conclusions}.

\section{Observations}
\label{sec.observations}

\begin{figure*}[!ht]
\centering
\includegraphics[width=\textwidth, trim={0 0cm 0 0}, clip]{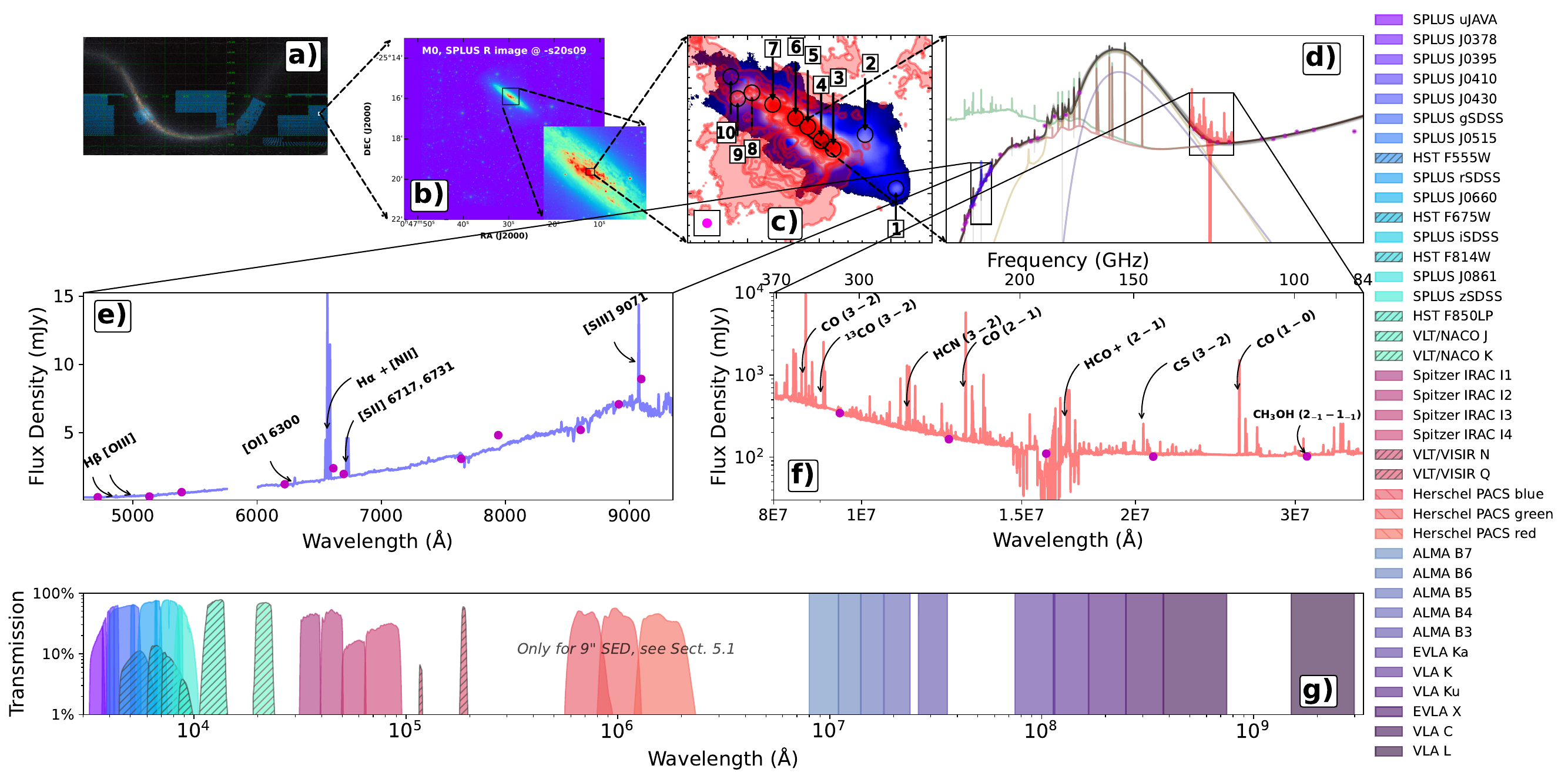}
\caption{Work overview. From left to right and top to bottom, the (labeled) panels show: a) The S-PLUS footprint scan over the southern hemisphere, where each light blue square represents a single S-PLUS observation (FoV of $1.4 \times 1.4 \, \text{deg}^2$). b) a zoom-in into the S-PLUS field where NGC~253 is located is called field $-$s20s09, where an inset provides a closer view into the nuclear regions of this galaxy. Inside the latter inset, there is another inset zooming in towards NGC~253's central molecular zone that is shown in panel c), where we indicate the H$\alpha$ (MUSE) emission as red contours and shaded areas, showing the strong stellar outflow observed at optical wavelengths. Complementing the latter, in the same panel c), we show mainly in blue and white the integrated intensity of the CN~1--0 line at $\sim$113.4~GHz taken from ALCHEMI observations at 1$\farcs$6 angular resolution. The leftmost bottom symbol of the beam in the same figure indicates this. The CN emission also presents features of molecular outflows, the most notorious one presumably coming from GMC~3, which is labeled --along with the other nine studied in this work-- in the white framed black numbers; GMC's 3$\arcsec$ apertures are indicated by black circles and are connected to the numbers by black arrows. Panel d) shows the SED of GMC~5 from GalaPy (see our Fig.~\ref{fig.SEDs} for more details), the one with the highest signal-to-noise ratio of the sample. Additionally, two zoomed-in insets are provided in the middle panels: the optical spectrum (obtained with MUSE) is shown in the middle-left panel e), and the sub-millimeter spectrum (from ALCHEMI) is presented in the middle-right panel f), while the magenta dots on top of both spectra correspond to the photometric observations listed in Table~\ref{Tab.observations}. Key emission lines are labeled in both spectra for reference. Finally, the bottom panel g), and its legend in the rightmost position of the figure, summarizes the transmission curves of all the observations used in this work to perform the SED fits at 3$\arcsec$ apertures. We also include in panel g) the Herschel PACS transmission curves used in Sect.~\ref{Sec.reliability_of_SED_fit_in_the_FIR}.}
\label{Fig.summary_plot}
\end{figure*}

To ensure consistency in the physical scales analyzed, we extract fluxes using a uniform aperture size with a 3$\arcsec$ diameter, despite the varying angular resolutions across the sample. The coordinates of all GMCs are listed in Table~\ref{tab.positions}. A summary of the dataset used in this study is provided in Table~\ref{Tab.observations}. An over-view of the full information used in this work is presented in Fig.~\ref{Fig.summary_plot}. Below, we provide a detailed description of the observations conducted with each facility.

\subsection{S-PLUS}

The Southern Photometric Local Universe Survey (S-PLUS; \citealt{MendesdeOliveira2019}) will cover $\sim$9,300~deg$^{2}$ of the southern hemisphere (currently 80\% complete), employing 7 narrow-band and 5 broad-band filters, encompassing a wavelength range from 3,533~\AA\ ($u$ filter) to 8,941~\AA\ ($z$ filter). The survey uses a dedicated robotic telescope of 0.8~m called T80-South, at the Cerro Tololo Inter-American Observatory (CTIO) in Chile. The angular resolution is subject to weather conditions, with a typical seeing range of 0\farcs8 to 2\farcs0 and an average of 1\farcs5.

For this work, we have used images in all 12 available filters from the S-PLUS fifth data release (available for internal collaboration), maintaining a consistent angular resolution of 1\farcs5 ($\sim$25.4\,pc). For further details on the S-PLUS survey, we refer the reader to \citet{MendesdeOliveira2019} and \citet{Herpich2024}.

\begin{table}[!ht]
\caption{Giant Molecular Cloud (GMC) positions.} 
\label{tab.positions}
\begin{center}
\begin{tabular}{llll}
\hline \hline
Position & $\alpha_{\rm ICRS}$ & $\delta_{\rm ICRS}$ & $v_{\rm LSR}$ \\
       & \multicolumn{1}{c}{[00$^{h}$:47$^{m}$:--$^{s}$]}  & \multicolumn{1}{c}{[-25$^{\circ}$:17':--"]}  & \multicolumn{1}{c}{[km\,s$^{-1}$]} \\
       \hline \\
GMC~1  & 31.93 & 29.00 & 304\\ 
GMC~2  & 32.36 & 18.80 & 330\\
GMC~3  & 32.81 & 21.55 & 286\\
GMC~4  & 32.97 & 19.97 & 252 \\
GMC~5  & 33.16 & 17.30 & 231\\
GMC~6  & 33.33 & 15.70 & 180\\
GMC~7  & 33.65 & 13.10 & 174\\
GMC~8  & 33.94 & 10.90 & 205 \\
GMC~9  & 34.14 & 12.00 & 201\\
GMC~10 & 34.24 & 7.84 & 144\\
\hline
\end{tabular}
\tablefoot{GMC positions taken from \citet{Harada2024} except for GMCs~3, 4, and 10, that were obtained using CPROPS \citep{Rosolowsky2006} on ALCHEMI data at 1\farcs6 angular resolution. The Local Standard of Rest (LSR) velocities were obtained by fitting a local thermodynamic equilibrium model (LTE) to the ALCHEMI spectra and averaging between the A-- and E--CH$_{3}$OH symmetry species (see \citet{Humire2022} for further information). The ALCHEMI spectral resolution of $\sim$8--9\,km\,s$^{-1}$ (see Subsect.~\ref{sec.obs.ALCHEMI}) dominates the velocity uncertainty.}
\end{center}
\end{table}

\begin{figure*}[!ht]
\centering
\includegraphics[width=\textwidth, trim={1.3cm 2cm 3cm 2.2cm}, clip]{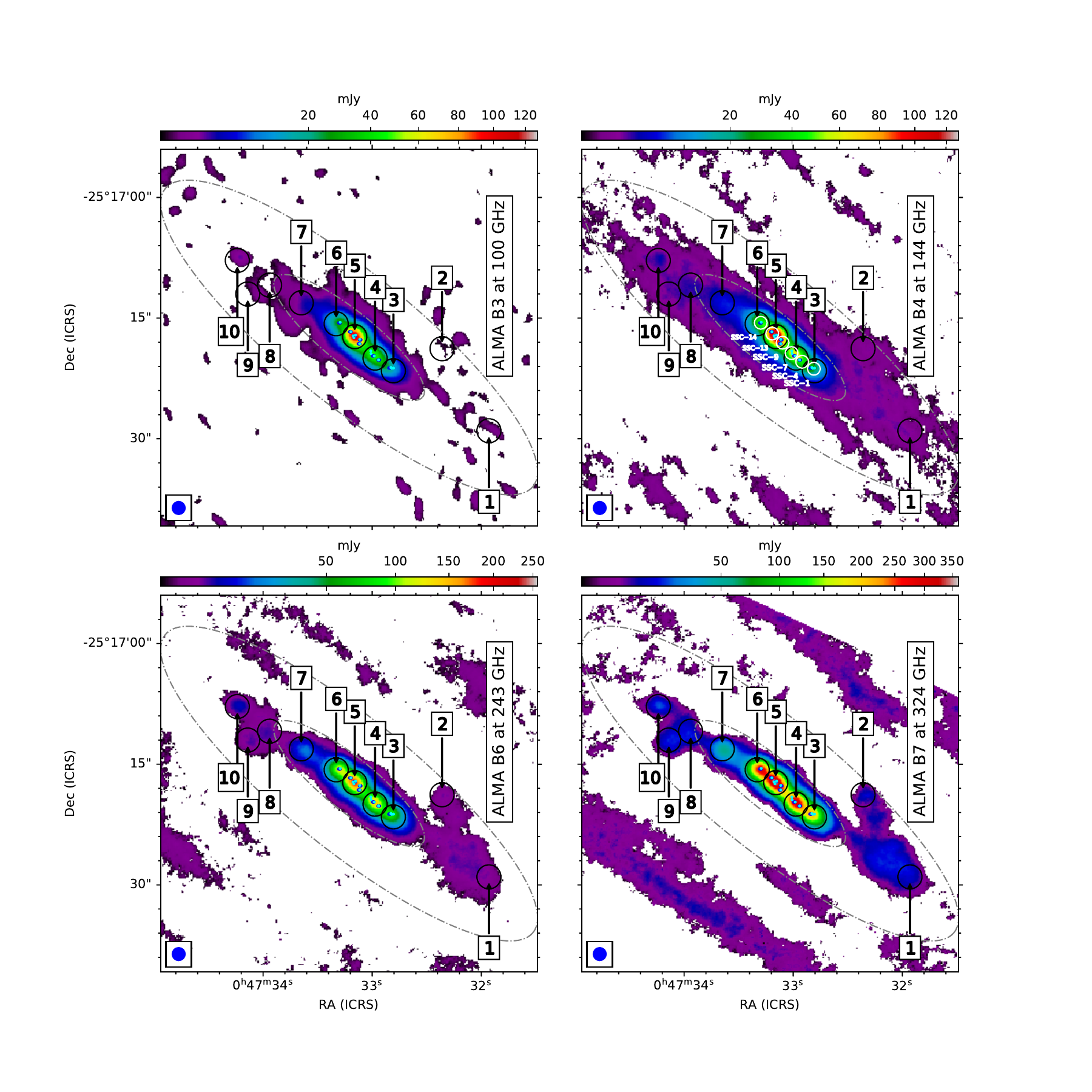}
\caption{Integrated intensity maps of NGC~253’s CMZ showing its 10 giant molecular clouds (GMCs) at the central frequencies of ALMA Bands 3, 4, 6, and 7 (i.e., at 100, 144, 243, and 324~GHz). Data from ALMA Band~5 is not shown as its central frequency of 187~GHz is strongly affected by telluric absorption from a water line at $\sim$183~GHz, as can be inferred from Fig.~\ref{Fig.summary_plot}. The ALCHEMI 1\farcs6 beam is depicted at the bottom left corner of each panel. 3$\arcsec$ apertures where the photometric points have been extracted are shown at each of the 10 GMC positions as black circles (see Table~\ref{tab.positions}). Each continuum map shows the eight dusty star-forming clouds detected by \citet{Ando2017} in blue and the 14 super star clusters identified by \citet{RicoVillas2020} in cyan. Additionally, for the ALMA~B4 continuum map, we label the six super star clusters (SSC) at 1\farcs6 resolution studied by \citet{Butterworth2024}. The color bar is in units of mJy, scaled by powers of the square-root of two, $(\sqrt{2})^n$, with $n$ starting from the 85th percentile up to the maximum emission value. Ellipses show the limits of the inner Lindblad resonances as determined by \citet{Iodice2014}. Numbers indicate the different GMCs as firstly identified and numbered by \citet{Leroy2015} but with most positions taken from \citet{Harada2024} (see Sect.~\ref{sec.GMC_positions} for details).}
\label{Fig.Cont_maps_ALCHEMI}
\end{figure*}

\subsection{VLT and HST}
\label{sec.obser_VLT}

This study incorporates the dataset presented in \citet{FernandezOntiveros2009}, including Adaptive Optics images from the Very Large Telescope/NaCo (VLT/NaCo) and diffraction-limited images from VLT/VISIR and the Hubble Space Telescope (HST). We refer the reader to \citet{FernandezOntiveros2009} for further details.

The VLT/NaCo observations were conducted on December 2 and 4, 2005, utilizing the IR wavefront sensor (dichroic N90C10) for atmospheric correction. The observation included the following filters and integration times: J (600 s; 3\% Strehl ratio\footnote{The Strehl ratio (SR) is a measure of the quality of adaptive optics correction, indicating the ratio of the peak intensity of the observed point spread function to the ideal diffraction-limited one.}), Ks (500 s; 20\% SR), L (4.375 s; 40\% SR), and NB$\_$4.05 (8.75 s, centered on Br$\alpha$; 40\% SR). The resulting images achieved different Full Width at Half Maximum (FWHM) resolutions: 0\farcs29 (J), 0\farcs24 (Ks), 0\farcs13 (L), and 0\farcs14 (NB$\_$4.05). Images from VLT/VISIR, captured on December 1, 2004, and October 9, 2005, encompassed the N (PAH2$\_$2, $\lambda$11.88$\mu$m, $\delta\lambda$0.37$\mu$m; 2826 s) and Q (Q2, $\lambda$18.72$\mu$m, $\delta\lambda$0.88$\mu$m; 6237 s) bands. The achieved FWHM resolutions were 0\farcs4 (N) and 0\farcs74 (Q). Furthermore, HST images were utilized within the central $\sim$300~pc region of NGC~253, employing the F555W (V band, FWHM = 0\farcs11), F656N (H$\alpha$, 0\farcs22), F675W (R band, 0\farcs20), F814W (I band, 0\farcs13), and F850LP (0\farcs13) filters.

\begin{table*}[!htp]
\caption{Extracted fluxes for sampled GMCs}
\label{tab.phot_points}
\scriptsize
\renewcommand{\tabcolsep}{0.16cm}
\begin{center}
\begin{tabular}{lrllllllllllll}
\hline \hline
Source & $\lambda$ [\AA]& FWHM [$\arcsec$] & $\theta_{\rm{LAS}}$ [$\arcsec$] & GMC~1         & GMC~2         & GMC~3         & GMC~4         & GMC~5         & GMC~6         & GMC~7         & GMC~8         & GMC~9         & GMC~10  \\ \hline \\
S-PLUS/uJAVA & 3,563 & $1.5$ & -- & 0.03 & 0.03 & 0.03 & 0.04 & 0.03 & 0.04 & 0.05 & 0.06 & 0.05 & 0.04 \\
S-PLUS/J0378 & 3,770 & $1.5$ & -- & 0.04 & 0.04 & 0.04 & 0.04 & 0.05 & 0.04 & 0.08 & 0.09 & 0.08 & 0.06 \\
S-PLUS/J0395 & 3,940 & $1.5$ & -- & 0.05 & 0.05 & 0.05 & 0.07 & 0.06 & 0.07 & 0.11 & 0.10 & 0.11 & 0.08 \\
S-PLUS/J0410 & 4,094 & $1.5$ & -- & 0.07 & 0.07 & 0.08 & 0.10 & 0.10 & 0.12 & 0.16 & 0.19 & 0.18 & 0.13 \\
S-PLUS/J0430 & 4,292 & $1.5$ & -- & 0.09 & 0.09 & 0.10 & 0.13 & 0.12 & 0.14 & 0.20 & 0.24 & 0.20 & 0.18 \\
S-PLUS/gSDSS & 4,751 & $1.5$ & -- & 0.13 & 0.15 & 0.24 & 0.31 & 0.26 & 0.30 & 0.40 & 0.44 & 0.38 & 0.31 \\
S-PLUS/J0515 & 5,133 & $1.5$ & -- & 0.15 & 0.17 & 0.29 & 0.37 & 0.31 & 0.35 & 0.49 & 0.54 & 0.45 & 0.37 \\
HST/F555W & 5,443 & $0.1$ & -- & -- & -- & -- & 0.74 & 0.63 & 0.70 & 0.89 & 0.79 & -- & -- \\
S-PLUS/rSDSS & 6,258 & $1.5$ & -- & 0.35 & 0.42 & 1.07 & 1.56 & 1.22 & 1.24 & 1.43 & 1.37 & 1.10 & 0.87 \\
S-PLUS/J0660 & 6,614 & $1.5$ & -- & 0.43 & 0.56 & 1.89 & 3.27 & 2.39 & 2.16 & 2.35 & 1.94 & 1.49 & 1.13 \\
HST/F675W & 6,718 & $0.1$ & -- & -- & -- & -- & 2.46 & 1.96 & 1.86 & 2.05 & 1.69 & -- & -- \\
S-PLUS/iSDSS & 7,690 & $1.5$ & -- & 0.65 & 0.84 & 2.56 & 3.94 & 3.08 & 2.90 & 2.80 & 2.53 & 1.91 & 1.53 \\
HST/F814W & 7,996 & $0.1$ & -- & -- & -- & -- & 6.28 & 4.81 & 4.15 & 3.78 & 2.98 & -- & -- \\
S-PLUS/J0861 & 8,611 & $1.5$ & -- & 0.88 & 1.24 & 3.94 & 6.46 & 5.20 & 4.71 & 4.08 & 3.40 & 2.55 & 2.01 \\
S-PLUS/zSDSS & 8,831 & $1.5$ & -- & 1.07 & 1.52 & 5.11 & 9.07 & 7.08 & 6.07 & 4.97 & 4.06 & 2.93 & 2.35 \\
HST/F850LP & 9,114 & $0.13$ & -- & -- & -- & -- & 12.53 & 8.93 & 6.95 & 5.48 & 3.87 & -- & -- \\
VLT/J & 12,650 & $0.29$ & -- & -- & -- & 22.01 & 45.75 & 31.77 & 20.08 & 9.88 & -- & -- & -- \\
VLT/K & 21,800 & $0.24$ & -- & -- & -- & 31.16 & 93.29 & 71.15 & 39.01 & -- & -- & -- & -- \\
Spitzer/IRAC 3.6$\mu$m & 36,000 & $1.66$ & -- & 2.71 & 5.72 & 28.40 & 106.38 & 109.02 & 62.01 & 23.41 & 12.60 & 6.27 & 6.30 \\
Spitzer/IRAC 4.5$\mu$m & 45,000 & $1.72$ & -- & 2.43 & 4.78 & 28.08 & 150.28 & 150.45 & 69.47 & 21.64 & 10.77 & 5.20 & 5.79 \\
Spitzer/IRAC 5.8$\mu$m & 58,000 & $1.88$ & -- & 7.23 & 17.75 & 121.03 & 357.56 & 464.85 & 250.73 & 89.61 & 45.28 & 22.27 & 22.98 \\
Spitzer/IRAC 8.0$\mu$m & 80,000 & $1.98$ & -- & 18.98 & 52.71 & 335.32 & 1016.10 & 1095.98 & 644.08 & 243.56 & 120.51 & 60.60 & 61.62 \\
VLT/N & 118,800 & $0.4$ & -- & -- & -- & 287.88 & 3207.94 & 1273.06 & 477.52 & 121.49 & -- & -- & -- \\
VLT/Q & 187,200 & $0.74$ & -- & -- & -- & 1178.25 & 18901.04 & 9984.58 & 3340.52 & 654.91 & 355.37 & -- & -- \\
ALMA/B7 & 9,252,854 & $1.6$ & 15 & 11.48 & 9.63 & 188.92 & 289.28 & 343.56 & 255.51 & 51.86 & 18.37 & 18.35 & 17.51 \\
ALMA/B6 & 12,337,138 & $1.6$ & 15 & 2.58 & 3.28 & 68.74 & 122.77 & 165.38 & 103.49 & 17.11 & 5.82 & 5.17 & 6.45 \\
ALMA/B5 & 16,031,682 & $1.6$ & 15 & 2.85 & 1.91 & 35.92 & 72.02 & 110.39 & 58.05 & 8.34 & 3.07 & 2.86 & 2.64 \\
ALMA/B4 & 20,818,921 & $1.6$ & 15 & 1.29 & 1.57 & 24.13 & 56.96 & 101.44 & 40.57 & 5.76 & 2.37 & 1.69 & 2.50 \\
ALMA/B3 & 29,979,246 & $1.6$ & 15 & -- & -- & 18.41 & 50.46 & 102.15 & 33.61 & 3.60 & -- & -- & 1.27 \\
EVLA Band Ka (DnC) & 91,330,642 & 1.76$\times$1.39 & 44 & -- & -- & 21.80 & 59.07 & 141.84 & 40.53 & 4.91 & -- & -- & 1.54 \\
VLA Band K (B) & 127,245,834 & 0.47$\times$0.26 & 7.9 & -- & -- & -- & -- & 135.60 & -- & -- & -- & -- & -- \\
VLA Band Ku (B) & 200,665,639 & 0.67$\times$0.40 & 12 & -- & -- & 30.67 & 71.92 & 181.58 & 50.21 & -- & -- & -- & -- \\
EVLA Band X (A) & 299,773,911 & 0.46$\times$0.19 & 5.3 & -- & -- & -- & 98.03 & 264.56 & 81.33 & -- & -- & -- & -- \\
VLA Band C (A)  & 616,844,217 & 0.69$\times$0.39 & 8.9 & -- & -- & 41.34 & 104.45 & 318.02 & 70.16 & 9.08 & -- & -- & -- \\
VLA Band L (A)  & 2,012,164,964 & 2.13$\times$1.25 & 36 & 6.57 & 12.38 & 89.09 & 175.32 & 274.92 & 160.83 & 36.35 & 16.68 & 15.20 & 5.69 \\
\hline
\end{tabular}
\end{center}
\tablefoot{
The fifth to last columns display the extracted fluxes [mJy] for each sampled Giant Molecular Cloud (GMC) within a 3$\arcsec$ aperture diameter. A dash -- indicates the lack of (enough) data filling the aperture. Uncertainties were assumed to be of the order of 10\% for all the sample except for ALMA data, where we assume 15\%. Additionally, observational parameters, namely, wavelength, FWHM of the beam/psf, and largest angular scales in the case of interferometric observations are indicated in columns 2, 3, and 4, respectively, while the first column specifies the source (facility or survey) and filter, plus the array configuration mode provided parenthetically for the (E)VLA observations. The largest angular scale, $\theta_{\rm{LAS}}$, is pertinent for interferometric observations only, namely: ALMA \citep{Martin2021}, EVLA\footnote{\url{https://science.nrao.edu/facilities/vla/docs/manuals/oss2012B/performance/referencemanual-all-pages}}, and VLA\footnote{\url{https://science.nrao.edu/facilities/vla/docs/manuals/oss/performance/resolution}}. }
\label{Tab.observations}
\end{table*}

\subsection{Spitzer/IRAC}

We retrieved Spitzer/IRAC mosaic images from the Spitzer Heritage Archive\footnote{\url{https://irsa.ipac.caltech.edu/data/SPITZER/Enhanced/SEIP/}}. The mean FWHM of the point spread function (PSF) of the four IRAC bands at 3.6, 4.5, 5.8, and 8.0~$\mathrm{\mu m}$ are 1\farcs66, 1\farcs72, 1\farcs88, and 1\farcs98, respectively \citep[][their Table~3]{Fazio2004}. Thanks to the large field of view (FoV) of Spitzer, which covers the entire emission from NGC~253, we were able to use data from these four IRAC filter images in all our GMCs.

\subsection{ALMA}
\label{sec.obs.ALCHEMI}
ALMA observations of NGC\,253 were conducted during Cycles 5 and 6 as part of the ALCHEMI large program.
The observations consisted of an unbiased complete spectral scan across the ALMA bands 3 to 7 (84--373~GHz, $\lambda$=3.6--0.8~mm) across the whole CMZ region (inner 500~pc; \citealt{Sakamoto2006}) of the galaxy. The observations were carried out using both 12m and 7m antenna arrays aiming to retrieve extended emission, and the data were imaged to common angular and spectral resolutions of 1\farcs6 and $\sim$8--9~km~s$^{-1}$, respectively. The flux density RMS noise ranges from 0.18 to 5.0~mJy~beam$^{-1}$ in the 8--9~km~s$^{-1}$ channels. In Fig.~\ref{Fig.Cont_maps_ALCHEMI} we provide integrated intensity maps of the continuum emission (in mJy) of the CMZ of NGC~253 at ALMA Bands 3, 4, 6, and 7 (100, 144, 243, and 324~GHz, respectively).

The absolute flux density uncertainty was reported to be of order 10 to 15\%. We conservatively assume an uncertainty of 15\% for our extracted ALCHEMI fluxes. However, the relative flux uncertainty within the individual ALCHEMI observations is $2-3\%$ after the alignment performed on the data enabled by the continuous frequency coverage. A full description of the dataset, calibration, and imaging is provided by \citet{Martin2021}.

\subsection{EVLA}

We obtained Expanded Karl G. Jansky Very Large Array (EVLA) archival data in two different ways. One was by directly searching into the VLA archive. In this way, we obtained the reduced NGC~253's CMZ image in the X Band at 3~cm (PI: N. Harada), which has been observed in A configuration mode, leading to a largest angular scale (LAS) of 5\farcs3 according to the EVLA documentation\footnote{\url{https://science.nrao.edu/facilities/vla/docs/manuals/oss2012B/performance}}. We applied a cleaning to this data using {\tt tclean} with a standard gridder and a hogbom deconvolver inside the Common Astronomy Software Applications (CASA) package\footnote{\url{https://casa.nrao.edu/index.shtml}}. Additionally, we used a 1\,cm (33\,GHz; Ka-Band) continuum image presented in \citet{Kepley2011}. These observations were done at DnC hybrid configuration, which leads to a LAS of 44$\arcsec$\footnote{according to the smaller array extension of the hybrid configuration, configuration C for this case}. We used CASA {\tt imsmooth} to convolve the X Band EVLA continuum image to a common circular beam of 1\farcs6, matching that of ALCHEMI. In the same way, for the Ka-Band (26.5--40.0~GHz) image, we choose a slightly larger (1\farcs8) beam due to an original beam of 1\farcs76$\times$1\farcs38. Since the images are in Jy/beam, we converted them into Janskys by dividing their flux by the beam area in pixels [pix/beam] available in the CASA Viewer (Statistics/BeamArea) panel. We note that the LAS, $\theta_{\rm{LAS}}$ (see Table~\ref{Tab.observations}), is smaller than the aperture size for the Ka-Band observations. However, we did not see significant deviations of these observations with respect to the rest of the dataset. In fact, the model adjusts well to the observations when they are available (GMCs~3--7), and any expected departure or uncertainty in flux is compensated for by the added uncertainties in our SED models, which is up to 30\% for the case of GalaPy \citep{Ronconi2024}.

\subsection{VLA}

We retrieve Karl G. Jansky Very Large Array (VLA) continuum maps processed with the AIPS VLA pipeline by Lorant Sjouwerman (NRAO) from the webpage\footnote{\url{https://www.vla.nrao.edu/astro/nvas/}}. Of those, VLA Band L (VLA A conf.) is the only one where all the 10 GMCs studied in this work are detectable. Details about wavelengths, fluxes per GMC, array configurations, and FWHMs, are summarized in Table~\ref{tab.phot_points}. We find very good consistency between VLA and EVLA points, delivering similar fluxes within 1 or 2$\sigma$ of the SED fitting. There is a noticeable missing flux in L-band for GMC~5 (see Fig.~\ref{fig.SEDs}), which can be associated to the free-free absorption effect \citep[e.g.,][]{Kellermann1969} observed in compact IR-bright starbursts \citep{, Condon91,Clemens10,Dey2022}.

Since we are considering flux densities, we decided not to convolve our VLA continuum maps, as we noticed that it can produce artifacts that may show up in GMC positions without clear emission, leading to false positives. Upon examining the summed fluxes within 3$\arcsec$ diameter apertures before and after convolution with CASA {\tt imsmooth}, we observed differences of less than 10\% (i.e., less than the instrumental uncertainties).

\subsection{GMC positions}
\label{sec.GMC_positions}

The selection of GMC positions to be analyzed in this study is driven by the ALMA data from the ALCHEMI project. Thus, we use the GMC positions determined by \citet{Harada2024}, which incorporate both the continuum and molecular emission peaks, except for GMC~3, GMC~4, and GMC~10. For these three clouds, and given the similarity between the dataset used in \citet{Leroy2015} and the ALCHEMI data, A.~K.~Leroy (private communication) provided modified GMC coordinates tailored to the ALCHEMI data for the collaboration, which were retained for GMCs~3, 4, and 10. These three GMC positions were derived using CPROPS \citep{Rosolowsky2006} considering intensity peaks of the CS~$J = 2{-}1$ and H$^{13}$CN~$J = 1{-}0$ transitions.

Our region coordinates are close to those used by \citet{Leroy2015}, whose reference coordinates are clarified in \citet{Behrens2022ApJ}, their Appendix~A. The differences between their coordinates and the ones we used in this study, listed in Table~\ref{tab.positions}, are up to $\sim$1$\farcs$3 in right ascension and $\sim$1\farcs4 in declination, corresponding to spatial offsets of up to 23~pc and 24~pc, respectively, at the assumed distance of NGC~253 of 3.5~Mpc. These variations fall within our photometric aperture radius of 25~pc.

The general reason for this selection is to avoid, as much as possible, overlapping between our apertures of 3$\arcsec$, while at the same time extracting most of the intensity in each aperture. We note that the variation in the GMC positions from different source extraction criteria is relatively marginal ($\sim$0\farcs2) compared to the apertures used in this paper.

Along the course of this work, we will compare our GMC characteristics with those of the GMCs in previous ALCHEMI works with the same numbering, unless it is explicitly mentioned.

\section{Analysis: SED and spectroscopic modeling}
\label{sec.SED_and_spec_modeling}

The main results from this study come from two methods used to explore in detail the spectroscopic and photometric information of the CMZ of NGC~253. We have employed GalaPy for the photometry information, from near-UV to centimeter wavelengths. To cross-validate the results obtained from GalaPy, we compared them with those from the broadly used Code Investigating GALaxy Emission (CIGALE). While CIGALE provides a refined treatment from the UV to the sub-mm regimes, it models the radio centimeter luminosities from star formation using the FIR/radio correlation coefficient (q$_{\rm IR}$) scaling relations. In contrast, GalaPy incorporates a full model based on the theoretical Simple Stellar Population (SSP) approach of \citet{Bressan02}. It self-consistently incorporates nebular and synchrotron emission within the chosen star formation history (SFH) model. This makes GalaPy a powerful tool for gaining deeper insights into the link between radio emission and stellar populations. Given that our study relies on extensive radio data, we selected GalaPy as the primary tool for SED modeling, particularly for testing the radio-SF correlation.

As we will see in Section~\ref{sec.SED_fitting}, despite differences between the SED modeling codes (e.g., in our Fig.~\ref{fig.SEDs}), we confirm the nature of GMCs and much of their main physical properties, for example, the star formation rate and the stellar mass in the studied GMCs is higher toward the inner regions of the CMZ, independently of the code used. Nevertheless, we have found certain differences among the codes’ results, which reflects the range of possibilities one can obtain when depending on the used SFH model, mass-luminosity relation, and IMF, among other assumptions.

Additionally, to complement this study, we applied \textsc{starlight} for the spectral MUSE data plus photometric (S-PLUS) information in the near-UV to near-IR range. Below, we elaborate on our procedures. In Fig.~\ref{fig.SEDs} we provide our best-fit models to the extracted flux densities summarized in Table~\ref{Tab.observations}. This Figure primarily shows GalaPy results but also presents the CIGALE best-fit models (solid blue lines). CIGALE best-fitting results are further provided in detail (i.e., with all the considered components) in Fig.~\ref{apen.fig:cigalefitting}.

\subsection{GalaPy photometric SED fitting}
\label{sec.GalaPy_model}

We perform the SED fitting of our selected GMCs with GalaPy \citep{Ronconi2024}, an open-source application programming interface developed in Python/C$++$ across a range from X-ray to radio frequencies. GalaPy assumes a Chabrier initial mass function \citep[IMF;][]{Chabrier2003} and a flat $\Lambda$CDM cosmology with approximate parameters: matter density around 0.3, baryonic density around 0.05, and a Hubble constant $H_{0}=$100 $h$ km s$^{-1}$ Mpc$^{-1}$, where $h$ is roughly 0.7 \citep{Planck2020}. GalaPy is expected to include tools for optical spectroscopy and an AGN component on top of its SED fitting algorithm, although these modules were not available at the time this work was initiated. In the subsequent subsections, we detail the relevant configurations pertinent to the objectives of this study.

\subsubsection{The In-Situ model}

GalaPy allows the user to choose among a range of SFH models, both empirical and non-parametric. For our models, we use the default In-Situ SFH model, which has proven successful in predicting the emission from both late and early type galaxies in the local Universe as well as for young highly star-forming systems up to high redshift \citep{Pantoni2019,Giulietti2023,Ronconi2024,Gentile2024}. The physically motivated In-Situ model, developed by \citet{Lapi2018}, adopts a star formation rate (\(\psi(t) \)) of:
\begin{equation}
    \psi(t) \propto e^{-x}-e^{-s\gamma x},
\end{equation}
\noindent
with $x \equiv \tau/s\tau_{\star}$, where $\tau$ is the galactic age, $\tau_{\star}$ is the characteristic star-formation timescale, \( s \approx 3 \) is related to the gas condensation, and \( \gamma \) is related to the gas dilution, recycling, and stellar feedback \citep[see][for more details]{Lapi2020}. 

In the in situ model, the evolution of the gas and dust masses and of the gas and stellar metallicities can be followed analytically as a function of the galactic age, and self-consistently with respect to the evolution of the SFR. This means that an isolated system is assumed for each molecular cloud (MC), implying an interdependence among parameters. For example, a correspondence between metallicities and ages.
This approach guarantees that the derived physical properties that directly result from the evolution of the stellar population (e.g. the components' masses and metallicities), are consistently propagated to the other models contributing to the overall emission of the object.

\subsubsection{Simple stellar populations}

Among the available Simple Stellar Population (SSP) libraries in GalaPy, we have selected the ``refined'' version of $parsec22.NTL$ \citep{Bressan2012,Yan2019,Ronconi2024} as our preferred choice. This library ensures a dense wavelength grid with a minimum of 128 values per order of magnitude within its 2189-point grid. It incorporates emission from dusty Asymptotic Giant Branch (AGB) stars and accounts for nebular emission and free-free continuum in addition to the stellar continuum and non-thermal synchrotron radiation from core-collapse supernovae.

For the purposes of our study, the $parsec22.NTL$ SSP library was modified to correctly process the photometric points obtained from the narrow bands of the VLT/VISIR instrument (see Sect.~\ref{sec.obser_VLT}). This adjustment will be available in the next release of GalaPy.

\subsubsection{Dust model}

GalaPy implements an age-dependent two component dust model which is constructed in order not to assume an attenuation curve but to derive it instead from structural parameters (e.g. density and extension) which can be obtained by tuning the model to best represent an observed dataset.

The two different components comprise the contribution of (1) a hot molecular cloud phase (normally referred in the literature as a ``birth-molecular cloud''; see, e.g., \citealt{CharlotFall2000}) of new stellar populations and (2) a diffuse and extended dusty medium which further attenuates the stellar emission. Both components also contribute to the IR emission which is then the combination of two separate modified grey-bodies.

Among the several possible free parameters on which this dust model depends, we chose to fix some known values and ranges, in order to both ensure physically plausible results and speed up the computing process. This consists in setting a range of reliable physical terms, such as the total number of molecular clouds to \(N_\text{MC} = 1\), as in this work we are assuming isolated GMCs, and their dusty and molecular radii, which were set to be within a 10--100~pc range, although they will likely be closer to the lower limit considering the 30~pc diameter inferred from their sub-mm emission \citep{Leroy2015}. A list of all common parameter ranges and individual values used to model the SEDs in the ten GMCs is provided in Table~\ref{tab.gmcs_GalaPy_constant_parameters}. There, the redshift was taken from the SIMBAD Astronomical Database \citep{Wenger2000}. These ranges are fixed for all GMCs. On the other hand, the parameters that were fine-tuned for specific GMCs along with their corresponding adjusted ranges are presented in Table~\ref{tab.gmcs_variable_parameters}. The parameters listed in this second table vary across different GMCs, with their respective ranges for each GMC. For example, the range of the molecular cloud radius, R$_{\rm{CM}}$ (pc), was fine-tuned and is set between 0.0 and 1.7 for GMCs 2 and 8, while for the other GMCs, this radius remains between 0.0 and 2.0 (in log10 scale).

We use different initial conditions for unknown parameter ranges such as the GMC ages (age), the time of maximum star-formation (sfh.tau\_star), the time that stars need to escape from its parental molecular cloud (ism.tau\_esc), the maximum SFR (sfh.psi\_max), and the average radius of the molecular cloud (ism.R\_MC). We mainly modified the R\_MC, ism.R\_MC, and sfh.psi\_max in cases where the results were bimodal, in order to achieve a single result in the parameter space distribution. For example, the R\_MC was fixed to not exceed 50~pc in GMCs 2 and 8, motivated by Table~3 in \citet{Leroy2015}, where their molecular radii are expected to be around 52 and 44~pc, respectively. Interestingly, these two GMCs are the largest in the sample, according to the mentioned study.

We recommend the reader consult Table~B.3 in \cite{Ronconi2024} for a complete description of all GalaPy parameters. This table is also described on the project's $Read the Docs$ web page\footnote{\url{https://galapy.readthedocs.io/en/latest/general/free_parameters.html}}. The information there goes beyond what is covered in the present work, which is limited to the In-Situ model.

\begin{table}[htbp]
\small
\setlength{\tabcolsep}{0.12cm}
\caption{GalaPy common inputs.}
\begin{tabular}{lll}
\hline \hline
Parameter & Value/Range & Brief description \\
\hline
redshift & 0.000864                 & \\
sfh.tau\_quench & 1e+20             & Star formation quenching (yrs.)\\
ism.f\_MC & ([0.0, 1.0], lin)         & MCs' fraction into the ISM\\
ism.norm\_MC & 100.0                & MCs' normalization factor\\
ism.N\_MC & 1.0                     & number of MCs\\
ism.dMClow & 1.3                    & Extinction index $<$100$\mu$\\
ism.dMCupp & 1.6                    & Extinction index $\gtrsim$100$\mu$\\
ism.norm\_DD & 1.0                  & Diffuse dust norm. factor\\
ism.Rdust & ([0.0, 2.0], log)         & Radius of the diffuse dust (DD, pc)\\
ism.f\_PAH & ([0.0, 1.0], lin)        & DD fraction radiated by PAH\\
ism.dDDlow & 0.7                    & DD extinction index $<$100$\mu$\\
ism.dDDupp & 2.0                    & DD extinction index $\gtrsim$100$\mu$\\
syn.alpha\_syn & 0.75               & Spectral index\\
syn.nu\_self\_syn & 0.2             & Self-absorption frequency (GHz)\\
f\_cal & ([-5.0, 0.0], log)           & Calibration uncertainty\\
\hline
\end{tabular}
\tablefoot{Parameter values or ranges common to the ten GMCs studied in this work provided as inputs to GalaPy. The terms ``log'' and ``lin'' next to the ranges indicate logarithmic (log10) and linear values, respectively. For a detailed explanation, see Sect.~\ref{sec.GalaPy_model}.}
\label{tab.gmcs_GalaPy_constant_parameters}
\end{table}

\begin{table*}[htbp]
\caption{GalaPy varying inputs.}
\begin{adjustbox}{max width=\textwidth}
\begin{tabular}{llll}
\hline \hline
Parameter & GMC(s) with varied parameters & Common parameter ranges & Brief description \\
\hline
age & GMC2: ([6.0, 12.0]) & ([6.0, 11.0]) & Age of the MC \\
sfh.tau\_star  & GMC2: ([6.0, 12.0]) & ([6.0, 11.0]) & Characteristic timescale\\
ism.R\_MC & GMC2: ([0.0, 1.7]), GMC8: ([0.0, 1.7]) & ([0.0, 2.0]) & Average radius of a MC\\
ism.tau\_esc & GMC2: ([6.0, 12.0]) & ([6, 11.0]) & Stars' escape time from MC\\
sfh.psi\_max & GMC3: ([-2.0, 1.5]), GMC5: ([-2.0, 1.5]) & ([-3.0, 1.5]) & Maximum SFR, at sfh.tau\_star\\
\hline
\end{tabular}
\end{adjustbox}
\tablefoot{Parameter ranges set to GalaPy that vary depending on the modeled GMC. All values are given in log10 scale. For a detailed explanation, see Sect.~\ref{sec.GalaPy_model}.}
\label{tab.gmcs_variable_parameters}
\end{table*}

\begin{figure*}[htpb]
\centering
\includegraphics[width=0.825\textwidth, trim={0 0 0 0}, clip]{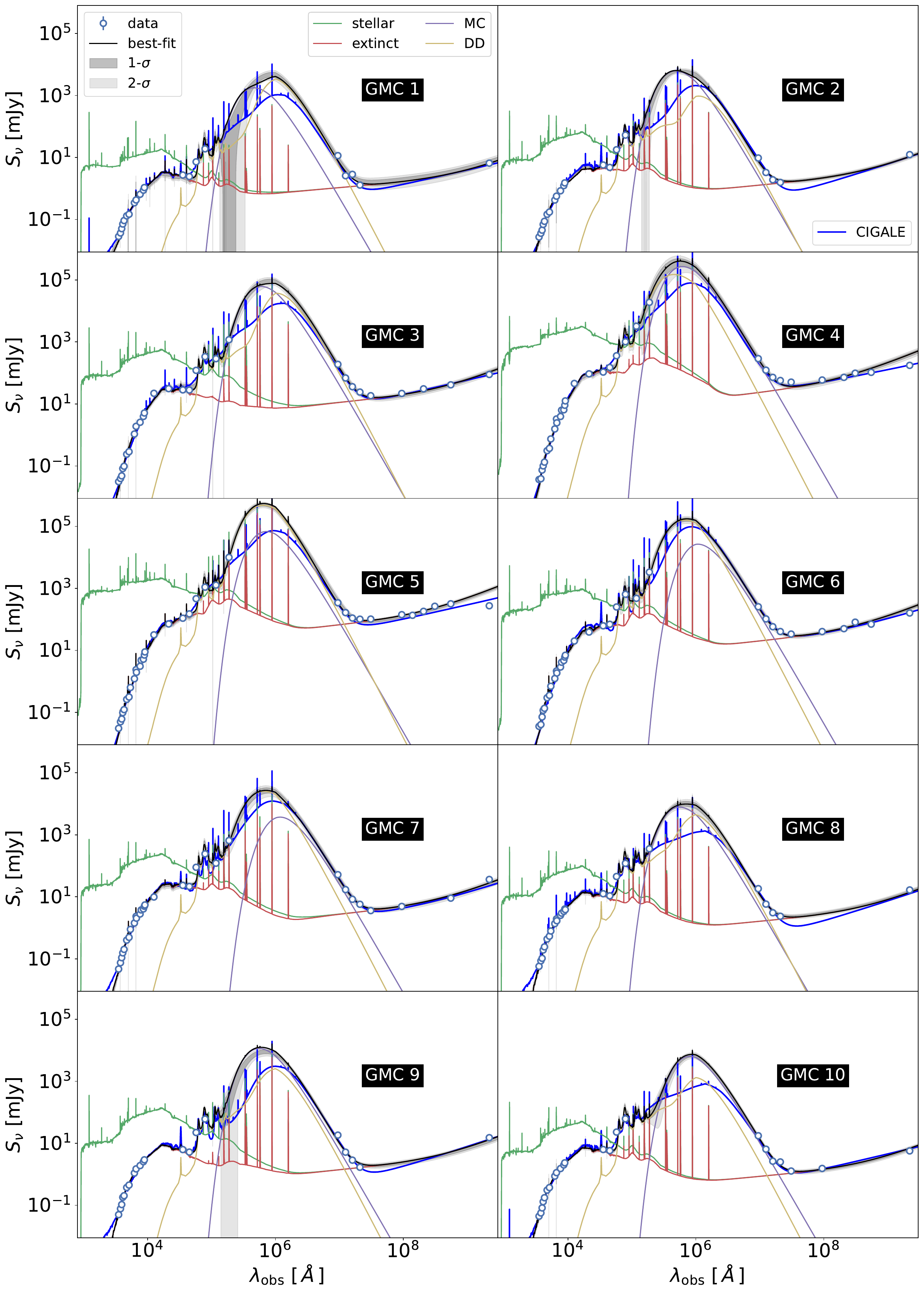}
\caption{SED models obtained with the GalaPy software for the ten GMCs studied in this work, each of them covering a diameter of 3$\arcsec$ ($\sim$50~pc; see Sect.~\ref{sec.GalaPy_model}). Points correspond to the photometric measurements obtained from the sample (Sects~\ref{sec.observations} and \ref{sec.SED_fitting}), while solid lines correspond to the unattenuated stellar emission (green), molecular cloud component (MC, purple), stellar emission considering extinction (extinct, red), and diffuse dust (DD, yellow). Best-fitting modeling results from CIGALE, devoid of an AGN component, are overlaid as solid blue lines (see also Fig.~\ref{apen.fig:cigalefitting} for a detailed view). Detailed information about the CIGALE modeling is given in Appendix~\ref{apen.CIGALE_results}.}
\label{fig.SEDs}
\end{figure*}

\subsection{CIGALE photometric SED fitting}\label{cigale_sed}

In addition to the main results obtained from GalaPy, for comparison we also provide simultaneous UV-to-radio SED modeling with the CIGALE 2022.1 version \citep[][]{Yang22}. This software operates on the principle of energy balance, where energy attenuation in the UV-to-optical range is re-emitted in the infrared in a self-consistent manner. Its modular and parallelized design, along with its user-friendly interface, has made it a popular choice in the literature for estimating the astrophysical characteristics of various targets. Below, we provide a brief overview of the models employed to simulate the galaxy emission.

To construct the SEDs with CIGALE, we utilized the full photometric coverage available for our GMCs. The modeling was based on the widely-used \citet{Bruzual2003} SSPs, adopting the IMF of \citet{Chabrier2003} and assuming a solar-like metallicity.
The stellar populations evolved using a delayed SFH, a model in which the star formation rate initially rises, peaks after a delay, and then declines exponentially, with an additional burst component \citep{ciesla2015}. This SFH approach is more realistic compared to the simple delayed model and has been demonstrated to minimize biases in the estimation of SFR and stellar mass \citep{Schreiber2018}. Additionally, the inclusion of the burst is consistent with the ALMA detections \citep{Hamed2021, hamed2023a}. For the SFH parameters, we maintained flexibility in the age and the e-folding time of the main stellar population to avoid introducing artificial biases in the efficiency of the initial burst.

Dust attenuation was modeled following the \citet{Calzetti2000} attenuation law, allowing the color excess of the young stellar population to vary as a free parameter. Dust emission was a critical component of our SED modeling, given the extensive IR coverage, particularly from ALMA. We used the models provided by \citet{Draine2014}. To ensure a good fit, we explored the parameter space for PAH fractions and the radiation field, U$_{\rm{min}}$, aiming to minimize $\chi^2$ while keeping the models physically realistic. Considering the rich radio data available for the GMCs, we extended our UV-IR modeling to the radio regime taking into account the power-law nature of the synchrotron spectrum and the ratio of the FIR/radio (q$_{\rm IR}$) correlation in the CIGALE fitting. 

CIGALE minimizes $\chi^2$ by measuring the difference between observed data and the model predictions, weighted by the uncertainties in the observations. The goal of minimization is to find the best-fitting model parameters that reduce this difference \citep{Boquien19}. Additional details, including the addition of AGN-related parameters, are provided in Appendix \ref{apen.CIGALE_results}.

\subsection{Stellar population fitting with {\textsc{starlight}}}

GalaPy and CIGALE are used in the full SED fits (all wavelengths from UV to Radio), while \textsc{starlight} will be used for the optical data only to fit stellar population models. \textsc{starlight} has been successfully applied to fit the stellar component of a diverse range of galaxy types, such as low-luminosity AGN, Seyfert nuclei, and normal galaxies \citep{CidFernandes2005,CidFernandes2013,Krabbe2017}.

While traditionally used for optical spectra, we applied the latest version of the code, described by \citet[][]{Werle2019}, to fit simultaneously the combined MUSE and S-PLUS data, thus covering the 3,500--10,000 \AA\ range. The combination of these datasets provides a comprehensive view of the galaxy's stellar populations, spanning a wide range of wavelengths. In particular, the S-PLUS bands densely cover the highly informative region around the 4,000 \AA\ break which falls outside the range of MUSE spectra that start at 4,600\AA .

\textsc{starlight} uses a chi-squared minimization technique to find the best-fit non-parametric combination of stellar populations of different ages and metallicities from a pre-defined library. This analysis allows us to estimate properties, such as mean age, metallicity, and extinction, as well as the velocity dispersion of the stellar component. The base is built from SSP models from the Charlot \& Bruzual 2019 library models \citep{paredes2023} using PARSEC (\textit{PA}dova T \textit{R}ieste \textit{S}tellar \textit{E}volution \textit{C}ode) evolutionary tracks \citep{Bressan2012,ChenBressan2015} and a Chabrier IMF. A base of composite stellar population is built from the SSP models assuming constant SFRs in 16 logarithmically spaced age bins with 5 different metallicities. Extinction is modeled with two components: one applied to all populations ($A_{\rm{V}}$) and an extra one applied exclusively to $\leq 10$ Myr old stars ($\delta A_{V}$).

The list of parameters to fit includes the flux (or mass) fractions  of the different populations, global and selective extinction, the velocity shift and velocity dispersion.

We have adjusted the weight scheme for the code used to extract residuals for H$\alpha$, [N II], [O III], and H$\beta$ by augmenting them around those lines. This modification allows a detailed examination on the contribution of stellar absorption features to these emission lines, compared to a general fit to the MUSE+S-PLUS spectra/photometry. These lines will subsequently serve as input for diagnostic diagrams and to estimate the extinction affecting the emission line regions.

\section{Results}
\label{sec.results}

In the following, we supplement our SED analysis with existing results from the literature, including those obtained by the ALCHEMI collaboration. Since the GMC selection was based on ALMA observations, our primary sources of information are the ALCHEMI results \citep{Martin2021}. After detailing the characteristics of the ten GMCs by dividing them in two main groups, we will compare our panchromatic results -- derived from SED fitting -- on star formation rates (SFRs) and histories (SFHs) with those obtained from pure-optical and pure-sub-mm/cm observations, that is, from monochromatic tracers such as the H$\alpha$ and H40$\alpha$ emission lines and different infrared and radio continuum regimes or bands. This approach will help us understand the key limitations to consider when working with mono-wavelength datasets. For a more detailed description of our results in light of previous literature information and for each of the ten GMCs, we refer the reader to our Appendix~\ref{apen.individual_GMCs}.

\subsection{SED fitting}
\label{sec.SED_fitting}

The wealth of archival data described in Sect.~\ref{sec.observations} does not always cover the ten GMCs. Only S-PLUS, Spitzer/IRAC, ALCHEMI, VLA L (1.5~GHz), and Ku (15~GHz) Bands fully cover the NGC~253's CMZ (see Table~\ref{tab.phot_points}). EVLA Band X does not detect emission outside GMCs~3 to 6, similar to the case in VLA K (23.6~GHz) band observation, which also detects emission only in GMCs~3 to 6. Also, the VLA C (4.9~GHz) band does not detect emission in GMCs~1 and 2, and our aperture of 3\arcsec\ is not fully covered by emission in GMCs~8--10, hence we discarded the latter contribution. From the above, we can say that the best-studied GMCs are GMCs 3 to 6, corresponding to the core of NGC~253 and encapsulating most of the star formation activity across the central starburst region \citep{Bendo2015}. It is worth saying that, in addition to the possible lack of continuum emission, the not imaged/detected GMCs may fall below the detection limit of our ALMA Band~3 observations and the different centimeter bands mentioned above. The detection limit at 3~$\sigma$ of EVLA Band X, and VLA Bands K and C are 2.4$\times$10$^{-2}$, 5.4$\times$10$^{-3}$, and 6.6$\times$10$^{-4}$~Jy, respectively.

\begin{table*}[htbp]
\caption{GalaPy results for star-formation-related properties.}
\centering
\begin{adjustbox}{max width=\textheight}
\begin{tabular}{llllllllllllllr}
\hline \hline
     GMC                       & SFR                          & $M_{\bigstar}$          & $\tau_{\bigstar}$ & $Z_{\bigstar}$       & $L_{\rm{bol}}$             & Age                    \\

  & [\(\times 10^{-4} M_{\odot}\) yr$^{-1}$]            & [log$_{10}$ ($M_{\odot}$)]            & [log$_{10}$ (yrs)]  &    [$\times$10$^{-3}$]                               & [log$_{10}$ ($L_{\odot}$)]   & [log$_{10}$ (yrs)]        \\  
\hline
 1  & 50$_{-10}^{+10}$ & 6.535$_{-0.097}^{+0.159}$ & 10.88$_{-0.21}^{+0.08}$ & 0.35$_{-0.08}^{+0.18}$ & 8.065$_{-0.073}^{+0.052}$  & 9.30$_{-0.13}^{+0.15}$  \\
 2  & 80$_{-10}^{+10}$ & 7.534$_{-0.086}^{+0.106}$ & 9.21$_{-0.14}^{+0.14}$ & 3.88$_{-0.15}^{+0.19}$ & 8.391$_{-0.062}^{+-0.002}$ & 9.66$_{-0.12}^{+0.13}$  \\
 3  & 870$_{-90}^{+110}$ & 8.573$_{-0.100}^{+0.108}$ & 9.07$_{-0.46}^{+0.28}$ & 7.88$_{-0.94}^{+2.27}$ & 9.354$_{-0.042}^{+0.071}$  & 9.62$_{-0.23}^{+0.15}$  \\
 4  & 2740$_{-280}^{+390}$ & 8.850$_{-0.136}^{+0.251}$ & 7.48$_{-0.10}^{+0.13}$ & 20.84$_{-1.47}^{+2.31}$ & 10.132$_{-0.114}^{+0.050}$ & 8.54$_{-0.08}^{+0.16}$  \\
 5  & 6450$_{-590}^{+600}$ & 8.620$_{-0.105}^{+0.124}$ & 7.57$_{-0.11}^{+0.12}$ & 17.59$_{-1.14}^{+1.17}$ & 10.184$_{-0.065}^{+0.016}$ & 8.40$_{-0.03}^{+0.13}$  \\    
 6  & 1870$_{-170}^{+180}$ & 8.605$_{-0.095}^{+0.072}$ & 8.17$_{-0.12}^{+0.10}$ & 12.87$_{-0.63}^{+0.82}$ & 9.602$_{-0.045}^{+0.061}$  & 8.56$_{-0.04}^{+0.06}$  \\
7 & 220$_{-20}^{+30}$ & 8.156$_{-0.087}^{+0.096}$ & 8.50$_{-0.18}^{+0.31}$ & 8.12$_{-1.18}^{+0.87}$ & 8.903$_{-0.080}^{+0.041}$ & 9.39$_{-0.14}^{+0.23}$  \\
8 & 110$_{-10}^{+10}$ & 7.837$_{-0.127}^{+0.098}$ & 9.24$_{-0.24}^{+0.24}$ & 4.45$_{-0.40}^{+0.71}$ & 8.533$_{-0.048}^{+0.045}$ & 9.76$_{-0.16}^{+0.15}$  \\
9 & 90$_{-10}^{+10}$ & 7.752$_{-0.099}^{+0.106}$ & 9.12$_{-0.18}^{+0.24}$ & 4.53$_{-0.53}^{+0.62}$ & 8.605$_{-0.212}^{+0.102}$ & 9.73$_{-0.10}^{+0.12}$  \\
10 & 50$_{-10}^{+10}$ & 7.581$_{-0.150}^{+0.095}$ & 9.55$_{-0.25}^{+0.21}$ & 3.31$_{-0.32}^{+0.37}$ & 8.507$_{-0.090}^{+0.075}$ & 9.64$_{-0.12}^{+0.12}$  \\ 
 \hline 
\end{tabular}
\end{adjustbox}
\tablefoot{From left to right, the columns correspond to the GMC number, the instantaneous star formation rate (SFR), the stellar mass ($M_{\bigstar}$), the characteristic star formation timescale ($\tau_{\bigstar}$), the stellar metallicity ($Z_{\bigstar}$), the bolometric luminosity ($L_{\rm{bol}}$), and the stellar age; all including 1$\sigma$ uncertainties. A complete list of results is provided in Table~\ref{Tab.SED_outputs_GalaPy}.}
\label{Tab.star_formation_results_GalaPy}
\end{table*}

\begin{table*}[htbp]
\caption{CIGALE results for star-formation-related properties.}
\centering
\begin{adjustbox}{max width=\textwidth}
\setlength{\tabcolsep}{0.1cm}
\begin{tabular}{llllllllllllllll}
\hline \hline
GMC & SFR$_{\rm{INST}}$ & SFR$_{\rm{100Myr}}$ & SFR$_{\rm{10Myr}}$ & $M_{\bigstar}$ & $\tau_{\rm{burst}}$ & $M_{\bigstar, \text{old}}$ & $M_{\bigstar, \text{young}}$ & Age \\
& [\(\times 10^{-4} \, M_{\odot} \, \text{yr}^{-1}\)] & [\(\times 10^{-4} \, M_{\odot} \, \text{yr}^{-1}\)] & [\(\times 10^{-4} \, M_{\odot} \, \text{yr}^{-1}\)] & [log$_{10}$(\(M_{\odot}\))] & [log$_{10}$(yrs)] & [log$_{10}$(\(M_{\odot}\))] & [log$_{10}$(\(M_{\odot}\))] & [log$_{10}$(yrs)] \\ \hline
1 & 20.76$\pm$3.81 & 23.27$\pm$4.43 & 20.95$\pm$3.82 & 6.70$\pm$0.07 & 8.15$\pm$0.33 & 6.70$\pm$0.04 & 7.87$\pm$0.09 & 9.41$\pm$0.57 \\
2 & 31.02$\pm$4.81 & 26.37$\pm$7.53 & 31.80$\pm$4.82 & 7.17$\pm$0.06 & 8.32$\pm$0.31 & 7.17$\pm$0.04 & 8.10$\pm$0.09 & 9.52$\pm$0.93 \\
3 & 262.01$\pm$73.46 & 292.05$\pm$60.68 & 264.88$\pm$75.60 & 7.82$\pm$0.07 & 8.10$\pm$0.33 & 7.82$\pm$0.04 & 7.74$\pm$0.09 & 9.36$\pm$0.55 \\
4 & 642.50$\pm$71.84 & 446.83$\pm$66.94 & 655.24$\pm$73.26 & 8.50$\pm$0.02 & 8.39$\pm$0.33 & 8.50$\pm$0.01 & 8.16$\pm$0.09 & 9.54$\pm$1.03 \\
5 & 1956.59$\pm$262.37 & 204.02$\pm$24.33 & 2024.70$\pm$239.81 & 8.17$\pm$0.02 & 8.26$\pm$0.33 & 8.17$\pm$0.01 & 7.92$\pm$0.09 & 9.44$\pm$0.83 \\
6 & 476.20$\pm$68.99 & 509.41$\pm$79.90 & 481.12$\pm$70.22 & 7.94$\pm$0.06 & 8.23$\pm$0.33 & 7.94$\pm$0.04 & 7.93$\pm$0.09 & 9.44$\pm$1.04 \\
7 & 191.88$\pm$38.04 & 214.98$\pm$40.54 & 193.54$\pm$37.91 & 7.68$\pm$0.06 & 8.03$\pm$0.33 & 7.68$\pm$0.04 & 7.68$\pm$0.09 & 9.33$\pm$0.52 \\
8 & 47.72$\pm$12.44 & 46.49$\pm$10.59 & 49.32$\pm$12.65 & 7.51$\pm$0.07 & 8.02$\pm$0.33 & 7.51$\pm$0.04 & 7.75$\pm$0.09 & 9.36$\pm$0.64 \\
9 & 35.29$\pm$6.93 & 41.35$\pm$8.98 & 35.72$\pm$6.92 & 7.18$\pm$0.06 & 8.15$\pm$0.31 & 7.18$\pm$0.04 & 7.71$\pm$0.09 & 9.35$\pm$0.64 \\
10 & 32.41$\pm$5.98 & 38.11$\pm$9.19 & 32.98$\pm$6.07 & 7.17$\pm$0.07 & 8.22$\pm$0.33 & 7.17$\pm$0.04 & 7.82$\pm$0.09 & 9.40$\pm$0.58 \\
\hline
\end{tabular}
\end{adjustbox}
\tablefoot{From left to right, the columns correspond to the GMC number, the instantaneous star formation rate (SFR), the integrated SFR over the last 100~Myr, the SFR over the last 10~Myr, the stellar mass ($M_{\bigstar}$), the age of the last stellar burst ($\tau_{\rm{burst}}$), the old stellar population mass ($M_{\bigstar, \text{old}}$), the young stellar population mass ($M_{\bigstar, \text{young}}$), and the stellar age; all including 1$\sigma$ uncertainties. Additional results are listed in Tables~\ref{Apen.Tab.Extinction_Factors_CIGALE}, \ref{Apen.Tab.Modified_Parameters_CIGALE}, and \ref{Apen.Tab.CIGALE_AGN}.}
\label{Tab.CIGALE_main_results}
\end{table*}

In general, the GMC's SED shapes, which can be seen in Fig.~\ref{fig.SEDs}, are similar to what is found in local and high-redshift starbursts \citep[e.g.,][]{Swinbank2010}. These SEDs exhibit significant absorption in the near-UV and optical wavelength regimes ($\sim 3\times10^{3}$--$10^{4}$~\text{\AA}), caused by an exceptionally large amount of dust. This leads to a substantial difference between the predicted and observed emissions (i.e., stellar emission with and without extinction), illustrated by the red and green lines in Fig.~\ref{fig.SEDs}. At the mentioned wavelengths, the flux difference can reach up to four orders of magnitude (e.g., from $\sim10^{3}$~mJy down to $\sim10^{-1}$~mJy in GMCs~4--6 at 10$^{4}$~$\AA$). The total extinction ($A_{\rm{v}}$) derived from the stellar component exceeds 5.0 magnitudes in GMCs~3--6, based on analyses from GalaPy, CIGALE, and the Balmer lines (H$\alpha$ and H$\beta$; see Sect.~\ref{subsec.attenuations}).

Tables~\ref{Tab.star_formation_results_GalaPy} and \ref{Tab.CIGALE_main_results} respectively contain GalaPy and CIGALE results related to star-formation processes, such as the SFR, the stellar mass ($M_{\bigstar}$), the characteristic timescale ($\tau_{\bigstar}$), or the stellar age (Age). In addition, results on the gas and dust characteristics, the attenuation, the radiation field (U$_{\rm min}$), the time for stars to escape from their parental MC (ism.tau\_esc), and uncertainties of the model, to mention a few, are given in Tables~\ref{Tab.SED_outputs_GalaPy} and \ref{Apen.Tab.Modified_Parameters_CIGALE} for GalaPy and CIGALE, respectively.

We summarize below the main characteristics of the molecular clouds studied here by dividing them into two groups: the central GMCs, numbered from 3 to 6, and the external ones, GMCs 1, 2, and 7–10. We find strong differences among them in terms of stellar and dust masses, and star formation rates. Appendix~\ref{apen.individual_GMCs} provides detailed information of individual GMCs.

\subsubsection{GMC characteristics}
\label{subsubsect.GMC_characteristics}

The Central Molecular Zone (CMZ) of NGC~253 can mainly be divided into two primary groups based on the physical and chemical properties found in their giant molecular clouds (GMCs): internal and external GMCs. The internal GMCs, corresponding to GMCs 3, 4, 5, and 6, are located in the very nuclear region of the CMZ, about 120~pc extension (see, e.g., the scale bar in Fig.~\ref{Fig.SFR_H40a}), and are characterized by a large occurrence of molecular tracers indicating high densities ($n_{\rm{gas}}>$10$^{7}$~cm$^{-3}$), elevated temperatures ($T_{\rm{kin}}>$100~K), and strong/fast shocks ($v_{\rm{shock}}\gtrsim$15-20~km~s$^{-1}$) likely triggered by a tremendous star formation activity \citep{Meier2015,Bouvier2024,Bao2024}. 

\begin{figure}[!ht]
\centering
\includegraphics[width=0.5\textwidth, trim={0 0 0 0}, clip]{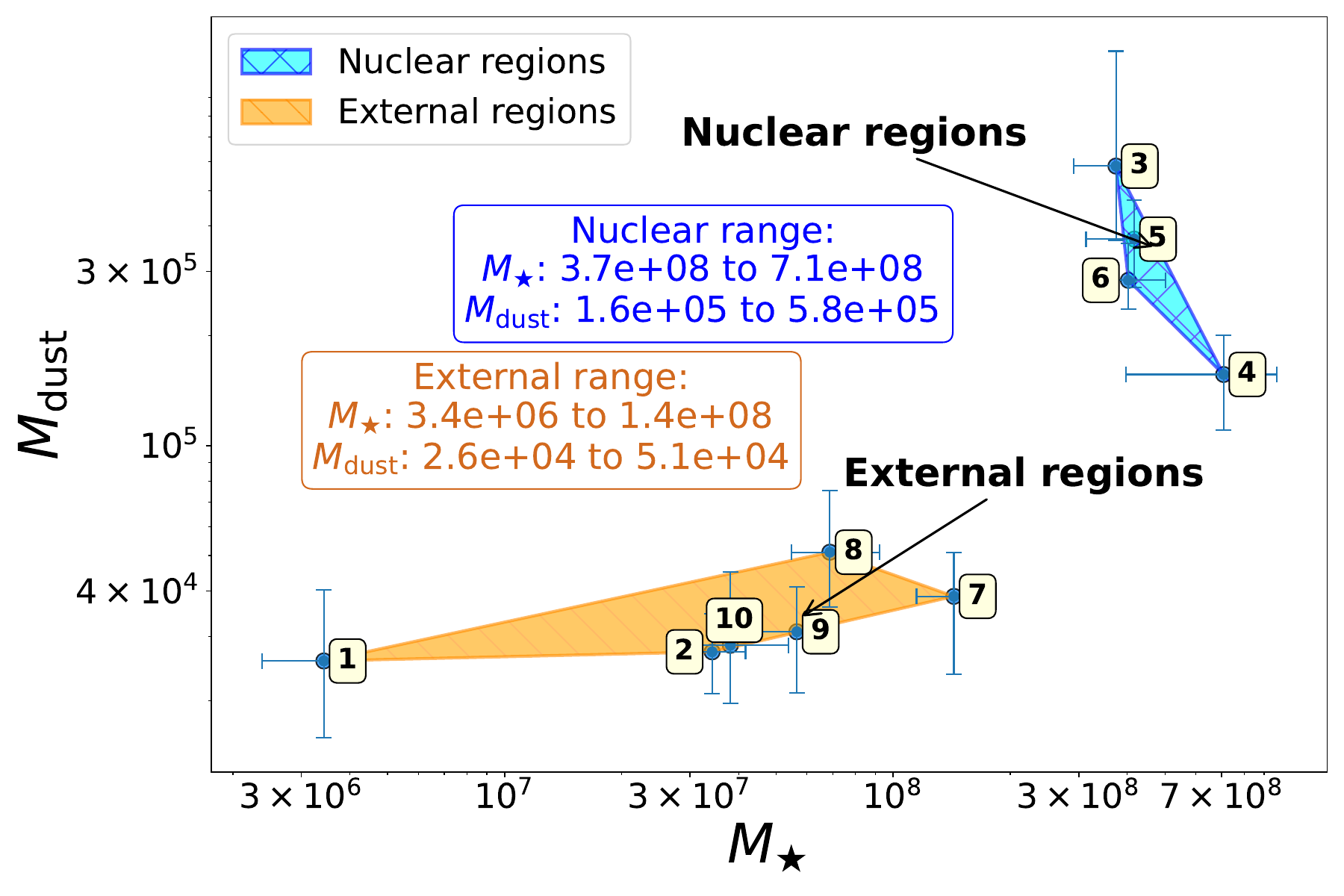}
\caption{Stellar and dust masses of the analyzed GMCs derived by GalaPy (see Subsect.~\ref{subsubsect.GMC_characteristics}). Their different ranges are indicated in the central labels.}
\label{Fig.Mstar_vs_Mdust}
\end{figure}

Conversely, the external GMCs encompass GMCs 1, 2, 7, 8, 9, and 10, located near (GMC~7) or fully immersed in the periphery of the internal bar. We note that, unlike some ALCHEMI studies \citep[e.g.,][]{Huang2023} but in line with others \citep[e.g.,][]{Harada2024}, based on the findings of the current study --namely, the star formation rate (SFR), dust ($M_{\rm{dust}}$), and stellar ($M_{\bigstar}$) mass--  GMC~7 is better defined as an external region (see the text below and also Fig.~\ref{Fig.Mstar_vs_Mdust}). Due to their locations, these external GMCs may experience the effect of orbital intersections such as inner Lindblad resonances \citep{Iodice2014}, bar/spiral arm interactions --also known as $x_{1}/x_{2}$ interactions \citep[e.g.,][]{Kim2012}--, and cloud-cloud collisions \citep[][see their Sect. 4.3]{Ellingsen2017} take place (see, e.g., ellipses in Fig.~\ref{Fig.Cont_maps_ALCHEMI}). These GMCs are dominated by a lower-density molecular gas, as compared to the internal GMCs, and exhibit signatures of slow shocks (up to $\sim$15~km~s$^{-1}$) rather than intense stellar feedback. This division reflects significant dynamical and chemical variations across the CMZ, previously noted by ALCHEMI studies \citep[e.g.,][]{Tanaka2024ApJ, Behrens2024ApJ, Harada2024}.

The study by \citet{Harada2024} utilized ALMA observations to conduct a detailed survey of molecular lines, applying principal component analysis (PCA) to distinguish patterns of chemical excitation and physical conditions across the CMZ. Their results indicate that the internal GMCs (GMCs 3–6) exhibit strong emissions from dense gas tracers such as HCN and HC$_{3}$N, along with radio recombination lines (RRLs), which are associated with active star formation and high-excitation regions \citep[e.g.,][]{Bendo2015}. In contrast, the external GMCs (GMCs 1, 2, and 7 to 9 in the work of \citet{Harada2024}, which did not account for GMC~10) are characterized by tracers of slow shocks, such as CH$_3$OH and an increment of HNCO over SiO \citep{Meier2015,Humire2022}, reflecting less energetic dynamical processes. The PCA revealed that molecular emission correlates strongly with dynamical features, with high-excitation species dominating the central regions, while shock tracers are more prevalent in the outskirts. 

In the current analysis, the differentiation between internal and external GMCs is further supported by variations in star formation rate (SFR), stellar mass, and dust mass. The internal GMCs exhibit SFRs above 0.087~$M_{\odot}$~yr$^{-1}$, as found by GalaPy, which provides an instantaneous SFR listed in the second column of Table~\ref{Tab.star_formation_results_GalaPy}. They also have greater stellar masses (third column; Table~\ref{Tab.star_formation_results_GalaPy}) and dust masses ($M_{\rm{dust}}$; Table~\ref{Tab.SED_outputs_GalaPy}) compared to the external GMCs. This is consistent with the high-density, high-excitation environments described by \citet{Harada2024}.

For the sake of clarity, we provide a plot showing the dust mass, $M_{\rm{dust}}$ versus stellar mass $M_{\rm{star}}$ in Fig.~\ref{Fig.Mstar_vs_Mdust}, from GalaPy. In the latter figure, we label the ranges for these quantities for nuclear and external regions or GMCs. The internal or nuclear GMCs (GMCs 3-6) display stellar masses in the 3.7$^{+1.1}_{-0.7}$--7.1$^{+5.5}_{-1.4}\times$10$^{8}~M_{\odot}$ range, and dust masses in the 1.6$^{+0.7}_{-0.3}$--5.8$^{+3.5}_{-3.0}\times$10$^{5}\,M_{\odot}$ range, in strong contrast to the external GMCs, where the stellar and dust masses range between 0.034$^{+0.01.5}_{-0.00.7}$--1.4$^{+0.4}_{-0.3}\times$10$^{8}~M_{\odot}$ and 2.6$^{+1.6}_{-0.9}$--4.8$^{+1.7}_{-1.2}\times$10$^{4}\,M_{\odot}$, respectively. 

The above indicates that the dust mass is the clearest distinction between the two primary groups, while the stellar mass, although enhanced in the internal GMCs, is only a factor of two larger in GMC~6 (internal) with respect to GMC~7 (external).

The external GMCs are also characterized by a lower SFR than the inner ones, with values in the  0.005--0.022~$M_{\odot}$~yr$^{-1}$ range. These findings align with the chemical and dynamical distinctions previously observed in multiple ALCHEMI studies \citep[e.g.,][]{Tanaka2024ApJ,Behrens2024ApJ,Harada2024}, reinforcing the conclusion that the physical properties of the CMZ are tightly coupled with its molecular and dynamical characteristics.

\subsection{Optical spectral fitting}
\label{sec.MUSE_starlight}

We fit archival MUSE data (ID:0102.B-0078(A), PI:~Laura Zschaechner) in the 4600--9350\AA \ range with S-PLUS photometric data using \textsc{starlight} \citep{CidFernandes2005} in its latest version which is capable to fit spectra and photometry simultaneously \citep{LopesFernandez2016, Werle2019}. 

To correct for the mismatch between spectroscopic and photometric fluxes, we scale MUSE spectra to match the iSDSS band flux of S-PLUS. The S-PLUS bands used for the fits are uJAVA, J0395, J0410 and J0430, iSDSS. The filter F0378 was not used to avoid contamination by [O II] 3727 emission. Of these, the blue ones are the most important, because they extend the MUSE coverage to the age sensitive 4,000 \AA\ region.

\begin{figure*}[!ht]
\centering
\includegraphics[width=0.9\textwidth, trim={0 0.2cm 0 0}, clip]{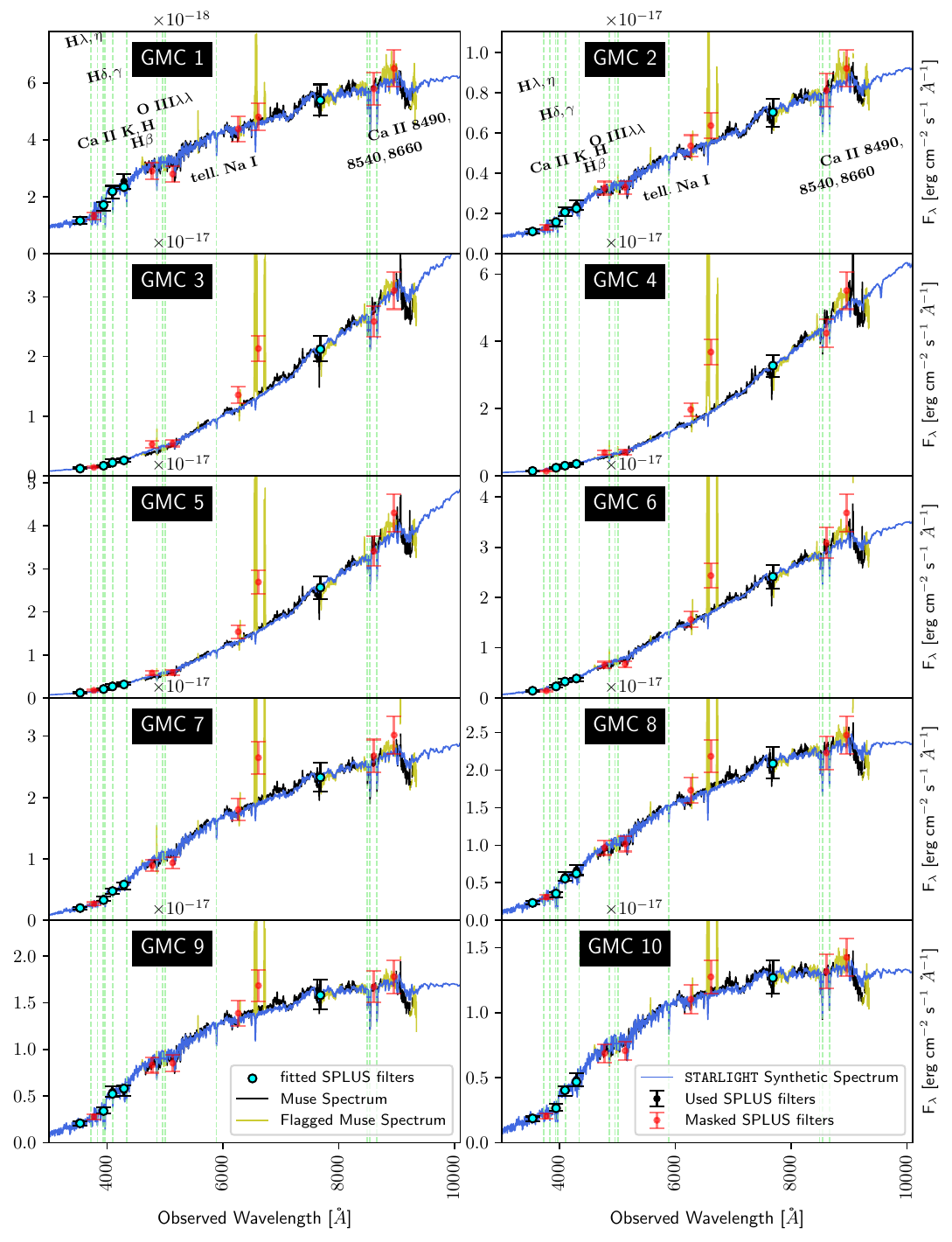}
\caption{The \textsc{starlight} fits (shown by blue lines) are applied to the MUSE spectra (represented in black, with yellow lines indicating masked data) along with S-PLUS photometry fits (depicted with black and red circles, depending on whether they were used in the fit, with red points marking the masked data, and cyan points for the S-PLUS photometric data fitted by \textsc{starlight}). Vertical dashed lines indicate the position of the stellar features labeled in the top panels.}
\label{Fig.STARLIGHT_fit}
\end{figure*}

The \textsc{starlight} fits to the MUSE spectra and the fitted S-PLUS photometric points for the ten GMCs are shown by the blue lines and the cyan points, respectively, in Fig.~\ref{Fig.STARLIGHT_fit}. Some important stellar absorption features are labeled in the top panels and indicated by vertical dashed green lines for all the GMCs. The outputs of the fitting are summarized in Table~\ref{Tab.STARLGIGHT_results}.

\begin{figure*}[!ht]
\centering
\includegraphics[width=0.9\textwidth, trim={0 0.2cm 0 0}, clip]{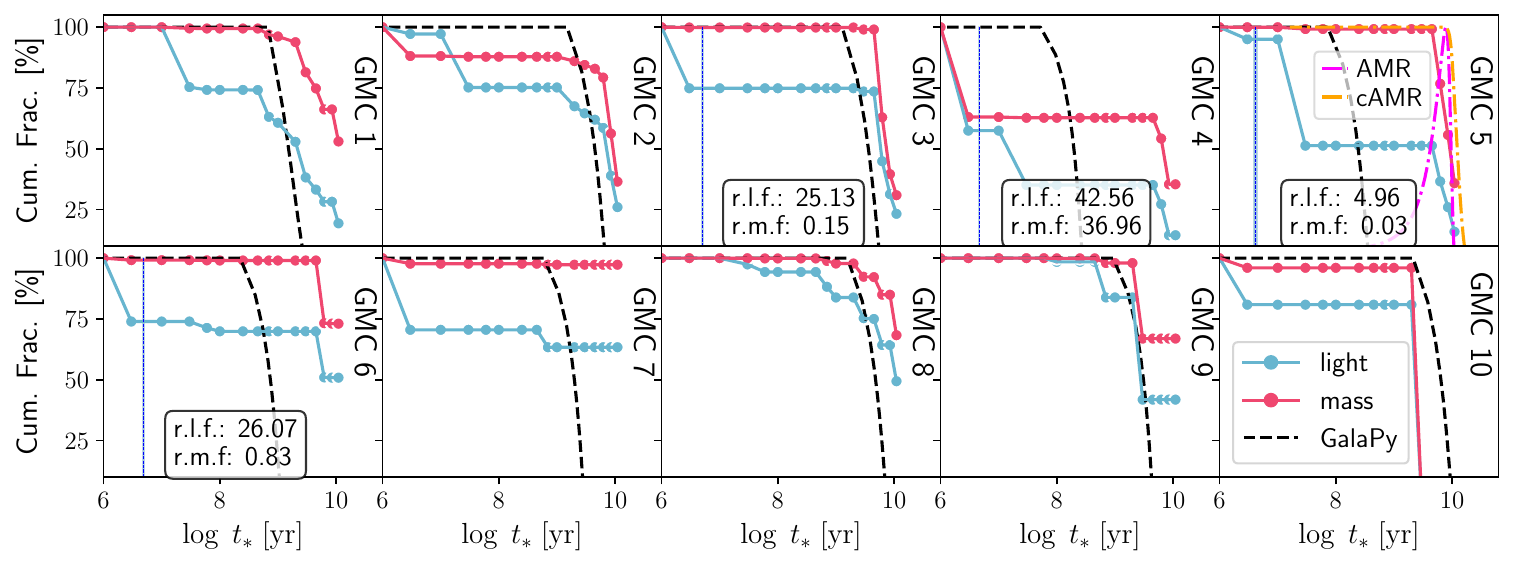}
\caption{\textsc{starlight} and GalaPy star-formation histories for the GMCs studied in this work. \textsc{starlight} discriminates between light and mass contributions to the total emission, contribution mainly coming from recent (blue) and old (red) stellar populations, respectively. GalaPy (black) generally provides ages between \textsc{starlight}'s light- and mass-weighted ages. For GMC~5, we have over-plotted the cumulative (in yellow) and normalized (in magenta) skewed Gaussian derived from the age-metallicity relation (see Appendix~\ref{sec.AMR}). GMCs~3 to 6 also show the estimated ages from super star clusters derived by \citet{Butterworth2024} in vertical dashed blue lines surrounded by 1~$\sigma$ uncertainties as shaded areas; the remaining cumulative fraction (Cum. Frac.) to reach 100\% of the emission at those ages, that is to say, the percentage of stellar formation that these radio observations account for, are denoted as r.l.f. and r.m.f. for remaining light and mass fractions, respectively.}
\label{Fig.SFHs}
\end{figure*}

As is commonly known \citep[see, e.g.,][]{Conroy2013}, low-mass stars dominate both the total mass and the number of stars in a galaxy, but they contribute only a small fraction to the overall light output, or bolometric luminosity, since young stars outshine older stars. This can be seen in the star formation histories per GMC, which are plotted in Fig.~\ref{Fig.SFHs}. Independent of the molecular cloud, the light fraction is always shifted towards young stars, while the mass fraction is concentrated in old stars. For example, in GMC~5, nearly 100\% of the stellar mass was already formed 10~Gyr ago while $\sim$ 50\% of the light comes from stars younger than 30~Myr.

Also in Fig.~\ref{Fig.SFHs}, for GMC~5, we have over-plotted (in magenta) the normalized skewed Gaussian derived from the age-metallicity relation (AMR; see Appendix~\ref{sec.AMR}). This same distribution was adapted to be cumulative (cAMR) and plotted (in yellow). GMCs~3 to 6 also show the estimated ages from super star clusters derived by \citet{Butterworth2024} in vertical dashed blue lines.

\begin{table*}[htbp]
\caption{Best-fitting results derived for each studied GMC by \textsc{starlight}.} 
\centering
\begin{adjustbox}{max width=\textwidth}
\begin{tabular}{rrrrrrrrrrrrrrr}
\hline \hline
   GMC &   adev &   Lum$_{\rm{tot}}$ &   $\log M_\bigstar$ &     $v_0$ &   $\sigma_\bigstar$ &   $A_{\rm{V}}$ &   $\delta A_{\rm{V}}$ &   $x(\delta A_{\rm{V}}>0)$ &   $\tilde{A_{\rm{V}}}$ &   $\langle \log t_{*} \rangle_L$ &   $\langle \log t_{*} \rangle_M$ &   $\langle \log Z/Z_{\odot} \rangle_L$ &   $\langle \log Z/Z_{\odot} \rangle_M$ \\

    &  [\%] & [$L_{\odot}$] & [$M_{\odot}$] & [km~s$^{-1}$] & [km~s$^{-1}$]& & & [\%] & & [yr] & [yr] &  \\
   
\hline
     1 &   1.85 &               9.82 &         4.94 &  -2.41 &       107.86 &    2.12 &      3.85 &          0.00    &            2.12 &                             8.95 &                             9.79 &                                   0.06 &                                   0.15 \\
     2 &   2    &             234.68 &         5.27 & -11.02 &       113.8  &    2.22 &      7.24 &          2.76 &            2.42 &                             9.15 &                             8.77 &                                  -0.07 &                                  -0.07 \\
     3 &   4.3  &             160.71 &         6.28 &  11.63 &       154.93 &    4.44 &      0    &         23.78 &            4.44 &                             8.72 &                             9.86 &                                   0.02 &                                   0.13 \\
     4 &   3.56 &            6241.17 &         6.23 &   1.37 &        83.82 &    3.35 &      5.77 &         40.23 &            5.67 &                             7.33 &                             7.22 &                                   0.22 &                                   0.17 \\
     5 &   3.46 &             229.04 &         6.35 & -18.47 &       126.71 &    4.66 &      0    &          4.75 &            4.66 &                             8.51 &                             9.89 &                                  -0.09 &                                   0.26 \\
     6 &   3.03 &             251.04 &         6.05 & -39.4  &       125.21 &    3.59 &      2.04 &         24.95 &            4.1  &                             8.66 &                             9.88 &                                  -0.11 &                                  -0.25 \\
     7 &   2.73 &             310.66 &         5.8  & -62.53 &       133.2  &    1.85 &      3.94 &         28.56 &            2.97 &                             8.57 &                             9.82 &                                   0.46 &                                   0.54 \\
     8 &   2.77 &              40.41 &         5.63 & -59.2  &       106.07 &    2.28 &      4.46 &          0.00    &            2.28 &                             9.69 &                             9.95 &                                  -0.4  &                                  -0.51 \\
     9 &   2.65 &              22.74 &         5.35 & -56.9  &       116.32 &    1.84 &      4.22 &          0.00    &            1.84 &                             9.57 &                             9.78 &                                  -0.12 &                                  -0.18 \\
    10 &   2.14 &              74.93 &         5.01 & -69.57 &       124.54 &    1.43 &      3.61 &         18.89 &            2.12 &                             8.6  &                             9.08 &                                   0.55 &                                   0.55 \\
\hline           
\end{tabular}
\end{adjustbox}
\tablefoot{
From left to right, the columns represent region number (GMC), mean deviation from the model to the observed data (adev), total luminosity ($\rm{Lum_{tot}}$), stellar mass ($M_\star$), velocity offset ($v_0$), stellar velocity dispersion ($\sigma_\bigstar$), attenuation at V band ($A_{\rm{V}}$), extra extinction applied to young stellar populations ($\delta A_{\rm{V}}$), percentage of stellar populations affected by the extra extinction ($ x(\delta A_{\rm{V}}>0)$), the effective extinction ($ \tilde {A_{\rm{V}}} $), which is equivalent to $A_{\rm{V}}+\delta A_{\rm{V}} \times x(\delta A_{\rm{V}}>0)$ and corresponds to the total extinction; the mean light-pondered ages ($ \langle \log t_\star \rangle_{L}$), mean logarithmic mass-pondered ages ($\langle logt_\star \rangle _M$), mean logarithmic light-pondered metallicity ($ \langle \log Z/Z_\odot \rangle _L $), and mean logarithmic mass-pondered metallicity ($ \langle \log Z/Z_\odot \rangle _M $). Further information is provided in Sect.~\ref{sec.MUSE_starlight}.}

\label{Tab.STARLGIGHT_results}
\end{table*}

\begin{figure*}[!ht]
    \centering
    \begin{minipage}{0.49\textwidth}
        \centering
        \includegraphics[width=\textwidth, trim={0 0 0 0}, clip]{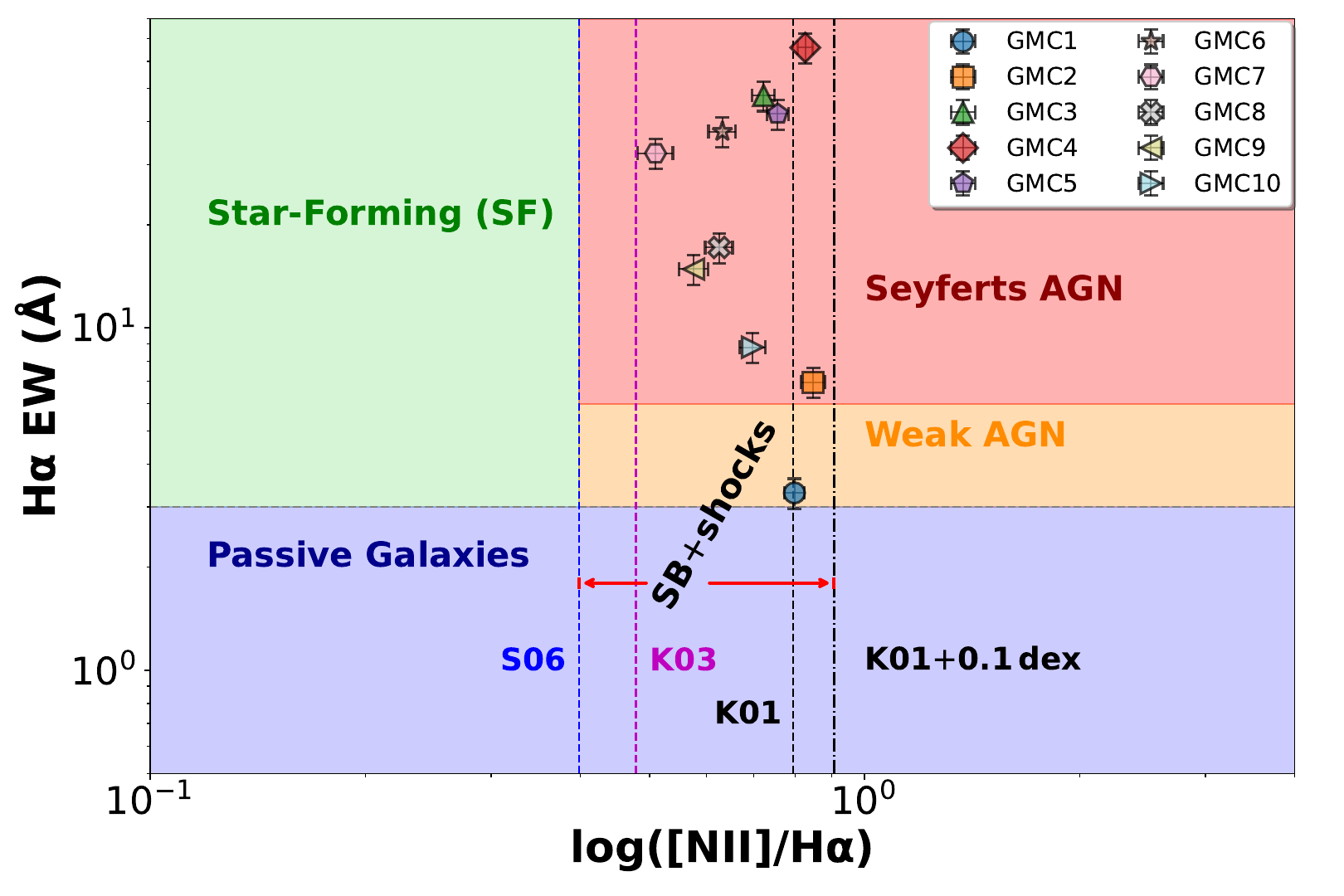}
    \end{minipage}%
    \begin{minipage}{0.49\textwidth}
        \centering
        \includegraphics[width=\textwidth, trim={0 0 0 0}, clip]{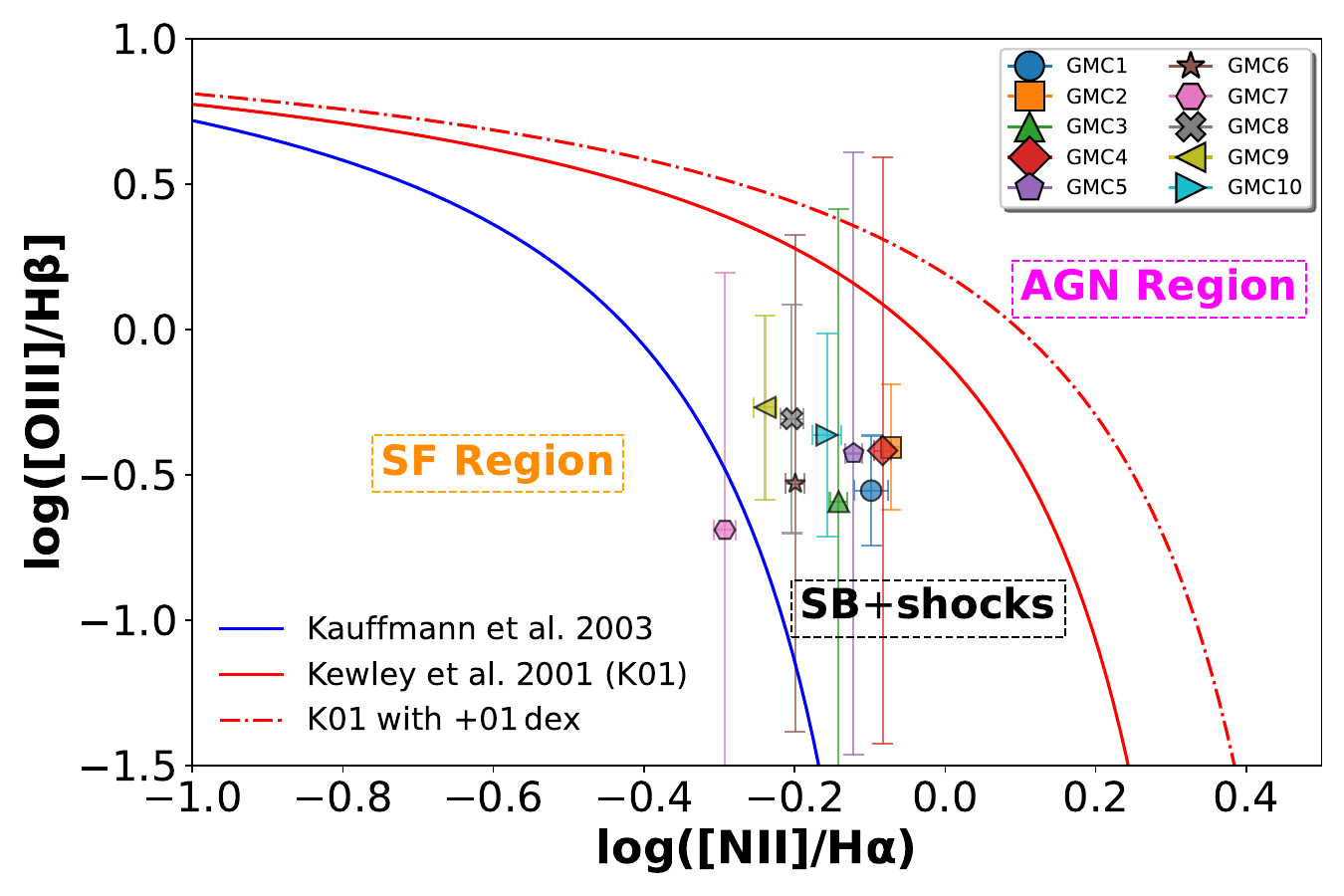}
    \end{minipage}
    \caption{WHAN (left) and BPT (right) diagnostic diagrams (see Sect.~\ref{subsec.WHAN_BPT}) for the ten GMC studied using emission line ratios of the optical spectra from MUSE. We prefer the term ``starburst+shocks'' (SB+Shocks) over the commonly used ``Composite'' given the findings highlighted in Sect.~\ref{subsec.WHAN_BPT}. 
    }
    \label{fig.WHAN_BPT}
\end{figure*}

In most cases (GMCs~1--9), GalaPy yields stellar age estimates that are consistent with those from \textsc{starlight}, falling between the young and old stellar populations that best reproduce the optical stellar contribution modeled by \textsc{starlight}. This is more clearly seen in Fig.~\ref{fig.stellar_ages_GalaPy_STARLIGHT_CIGALE}, where we plot, for all ten GMCs studied in this work, the average old and young stellar population ages from \textsc{starlight}, and the global stellar ages from GalaPy. Both codes agree in that GMC~4 is the youngest in the sample, with younger stellar populations in GMCs~3 to 7 and older ones located in GMCs~1, 2, and 8 to 10. In addition, in the same plot we added results from CIGALE, whose output delivers youngest ages in general with the youngest stellar burst in GMC~5, very close to the youngest burst in GMC~4 derived by GalaPy.

\subsubsection{Emission line diagnostic diagrams}
\label{subsec.WHAN_BPT}

The fluxes of emission lines from the MUSE spectra were measured in the pure nebular spectra, which were obtained by subtracting the stellar population continuum from the observed spectra. The contribution of the stellar population was determined using the stellar population synthesis code \textsc{starlight} \citep{CidFernandes2005}.

We show our results in Fig.~\ref{fig.WHAN_BPT}, where we extract the emission lines in our common 3$\arcsec$ diameter aperture using archival MUSE observations (PI:~Laura Zschaechner). Fig.~\ref{fig.WHAN_BPT} displays two diagnostic diagrams to analyze the dominant ionization mechanism of the emitting gas in the galaxy:

\begin{enumerate}[label=(\alph*)]
    \item The [O III] $\lambda5007$/H$\beta$ versus [N II] $\lambda6584$/H$\alpha$ diagnostic diagram proposed by \citet{Baldwin1981}, commonly known as the BPT diagram. This diagram includes the theoretical separation between H\textsc{ii}-like and AGN-like objects proposed by \citet{Kewley2001} [K01], and the empirical star-forming limit proposed by \citet{Kauffmann2003} [K03]. The region between the theoretical and empirical limits is commonly referred to as the ``composite region,'' although there are claims against this designation (see below).
    \item The WHAN diagram, namely, $\log(\mathrm{EW~H}\alpha)$ versus $\log([\mathrm{N~II}]\lambda6584)$ \citep{CidFernandes2010}, which is used to differentiate the nature of the ionization sources, classifying objects as star-forming, strong AGNs, weak AGNs, or retired galaxies. The line at $\log(\mathrm{EW~H}\alpha) = 6\AA\,$ represents the limit between weak and strong AGN emission. We note that in this diagram, the chosen log10([N II]$\lambda6584$/H$\alpha$; x-axis) limit to disentangle between SF and AGN gaseous ionization origin is normally taken from \citet{Stasinska2006} [S06] in the literature, although for NGC~253's CMZ we can rely on K03 as there are no GMCs between S06 and K03 limits.
\end{enumerate}

Although molecular gas and star-forming regions are known to spatially decorrelate at sub-100~pc scales \citep[e.g.,][]{Kruijssen2014,Chevance2020b}, in the CMZ of NGC~253, we do observe gas and dust in regions where star formation is taking place, especially in the inner regions \citep[e.g.,][]{Bendo2015,Ando2017,RicoVillas2020} (see also the super star clusters in Fig.~\ref{Fig.Cont_maps_ALCHEMI}). Nevertheless, our GMCs do not occupy the regions in these diagnostic diagrams that are typically associated with star-forming regions. Instead, they are located in what is often referred to as the ``composite'' zone, lying between star-forming and AGN-dominated areas. According to some studies, this behavior can occur in starbursts where stellar shocks are strong \citep[e.g.,][]{Kewley2001a}, and we consider this as the most likely scenario for the CMZ of NGC~253, which is also consistent with recent ALCHEMI studies on shocks \citep[e.g.,][]{Bao2024, Bouvier2024}. However, we cannot completely discard a decorrelation between star-forming regions and molecular gas in our line of sight given the large inclination (70$^\circ$--79$^\circ$; e.g., \citealt{Pence1980}) of NGC~253.

However, given the lack of evidence for an actual AGN in NGC~253, and the fact that these line-ratios are observed well away from the nucleus, our GMCs are most likely not starburst$+$AGN composites. Instead, we prefer to describe them as ``starburst$+$shocks'' composites, where  
the observed line ratios are a mixture of star-formation and shocks from stellar winds and supernovae. Shocks are indeed known to be present in NGC~253  \citep{Meier2015,Holdship2022ApJ,Humire2022,Behrens2022ApJ,Harada2022,Huang2023,Behrens2024ApJ,Bouvier2024,Bao2024}, whereas there is extensive literature arguing against the presence of an AGN \citep[e.g.,][]{FernandezOntiveros2009,Brunthaler2009,MullerSanchez2010}.

From the SED perspective, a clear argument against the existence of an AGN is provided in Appendix~\ref{apen.CIGALE_AGN}, where the AGN fraction, $f_{\rm{AGN}}$ in Table~\ref{Apen.Tab.CIGALE_AGN}, is negligible ($<$7.5\%) and also does not peak at the centrally located GMC~5 as one might expect in the presence of an AGN. Although this behavior could potentially suggest the presence of a low-luminosity AGN (LLAGN), it is more plausibly interpreted as an artifact of CIGALE's attempt to accurately fit the AGN-free SED. Further arguments specifically against the presence of an LLAGN can be found, for instance, in \citet[][their Sect. 7.1]{Mangum2019}.

\subsubsection{Attenuation estimations}
\label{subsec.attenuations}

The attenuation from dust in H\textsc{ii} regions can be estimated using the Balmer lines H$\alpha$ and H$\beta$ by contrasting their observed ratios with their expected ratios without attenuation. The intrinsic H$\alpha$/H$\beta$ flux ratio is 2.86 under typical conditions of electron density ($n_{\rm{e}} \sim 100$~cm$^{-3}$) and temperature ($T_{\rm{e}} \sim 10,000$~K), although variations in these conditions can change the ratio by up to 4\% \citep{Osterbrock2006}. The so called Balmer extinction is given by

\begin{equation}
A_{\rm{V}}^{\rm Balmer}  = \frac{2.5}{ \frac{A_{\rm{H}\beta}}{A_{\rm{V}}} - \frac{A_{\rm{H}\alpha}}{A_{\rm{V}}} } \log_{10}\left(\frac{(F_{\rm{H}\alpha}/F_{\rm{H}\beta})_\text{obs}}{2.86}\right),
\label{eq.attenuation.1}
\end{equation}

\noindent
where \( A_{\mathrm{H}\beta}/A_{\rm{V}} = 1.14 \) and \( A_{\mathrm{H}\alpha}/A_{\rm{V}} = 0.82 \) for a \citet{Calzetti2000} reddening law. In the following, we will apply this method to the GMCs studied in this work by measuring \( (F_{\mathrm{H}\alpha}/F_{\mathrm{H}\beta})_{\text{obs}} \) for each region and estimating their respective attenuations.

For comparison, we also considered the attenuation model proposed by \citet{CharlotFall2000}, a widely used approach for estimating dust extinction in stellar populations. This method distinguishes between the contributions of young and old stellar populations, assigning different power-law attenuation slopes to each. The attenuation for young populations, attributed to birth clouds, is represented by \(\alpha_{\rm{BCs}}\), while that for older populations, associated with the ISM, is denoted as \(\alpha_{\rm{ISM}}\), as described in Eq.~\ref{eq.CF2000}. According to this framework, stars are initially formed within interstellar birth clouds (BCs) and migrate to the ISM after approximately \(10^7\) years. The transition between these two attenuation regimes occurs at a wavelength of 5,500~\(\text{\AA}\). In CIGALE, this model can be adopted using the {\tt dustatt\_2powerlaws module}. The functional form of the law is commonly expressed in the literature as:

\begin{equation}\label{eq.CF2000_orientative}
A(\lambda) = 
\begin{cases} 
A_{\lambda}(\text{BC}) + A_{\lambda}(\text{ISM}), & \text{for young stars, age} < 10^7 \, \text{years}, \\
A_{\lambda}(\text{ISM}), & \text{for old stars, age} > 10^7 \, \text{years}.
\end{cases}
\end{equation}

\noindent
where \( A_{\lambda} \) is the wavelength-dependent attenuation expressed as \( A_{\lambda} = A_{\rm{V}} \left( \frac{\lambda}{\lambda_{\rm{V}}} \right)^\delta \), with \( \lambda_{\rm{V}} = 5,500 \, \text{Å} \), and \( A_{\rm{V}} \) represents the attenuation in the V band. In practice, this translates to the following expression:

\begin{equation}\label{eq.CF2000}
A(\lambda) = A_\mathrm{V} \left[ \left( \frac{\lambda}{\lambda_0} \right)^{\alpha_{\rm{BCs}}}  (\lambda < 5500 \, \text{\AA}) + \left( \frac{\lambda}{\lambda_0} \right)^{\alpha_{\rm{ISM}}} (\lambda \geq 5500 \, \text{\AA}) \right],
\end{equation}

\noindent
where \(A_{\rm{V}}\) is the overall attenuation at 5,500 \(\text{\AA}\) that we assume to be unity for simplicity. The exponents \(\alpha_{\rm{BCs}}\) and \(\alpha_{\rm{ISM}}\) describe the slopes for the birth clouds and the ISM, respectively. In the implementation, the values for \(\alpha_{\rm{BCs}}\) range from --0.69 to $-$0.9, while \(\alpha_{\rm{ISM}}\) ranges from --0.75 to --0.3. We find these values starting from those commonly used in \citet{CharlotFall2000} and also being influenced by other works that assume a -0.7 (an original assumption from C\&F 2000) value for both stellar populations. Then, looking at our GalaPy-derived curves from the SED, we set values that better reproduce the observations.

\begin{figure*}[!ht]
\centering
\includegraphics[width=0.515\textwidth, trim={0 0 0 0}, clip]{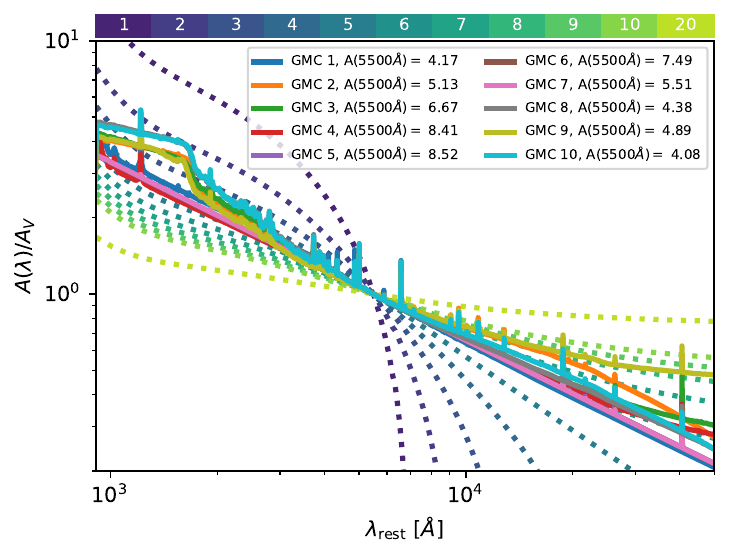}
\hfill
\includegraphics[width=0.468\textwidth, trim={1.5cm 0.05cm 0 -0.09}, clip]{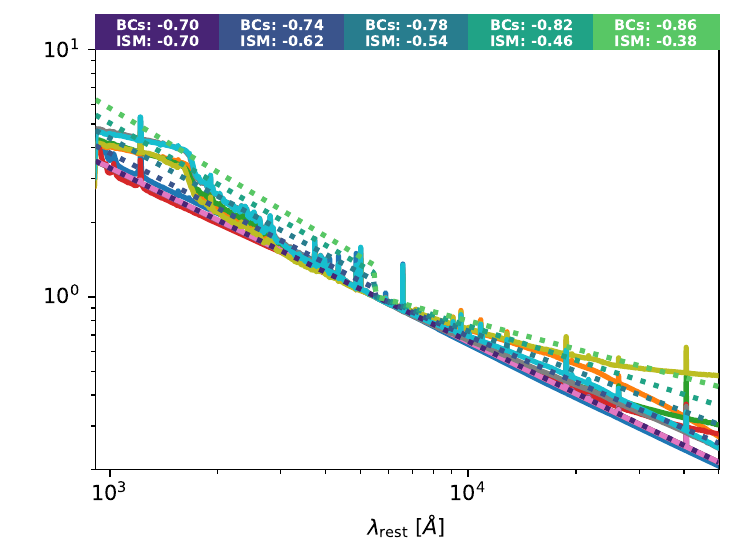}
\caption{
Attenuation curves normalized at $A(\lambda)/A_{V}=$1 when $\lambda=$5500\AA\ over wavelength in the rest frame, as derived by GalaPy SED fittings. The $A(\lambda)$ value at 5500\AA\ is labeled in the legend of the left panel in linear scale and per GMC. This legend also indicates the color corresponding to each GMC in the solid lines. Curves based on $R_{\rm{V}}$ values according to the methodology of \citet{Calzetti2000} are included for comparison (as dotted curves) in the left panel, following the colorbar. On the other hand, the dotted curves on the right panel show the attenuation curves from the models of \citet{CharlotFall2000}, which differentiate between young and old stellar populations, providing different exponential factors to the attenuation coming from birth clouds (BCs) and the ISM (ISM), related to these populations (see Sect.~\ref{subsec.attenuations}), which are labeled in the colorbar.}\label{fig.attenuation_curves}
\end{figure*}

We compute the attenuation curves over a wide wavelength range, from the near-UV to the near-infrared, using these parameter ranges. The curves are normalized at 5,500 \(\text{\AA}\) to allow a consistent comparison. As shown in Fig.~\ref{fig.attenuation_curves}, the model's attenuation curves are sensitive to the chosen values of $R_{\rm{V}}$ for the \citet{Calzetti2000} attenuation law, which assumes a single attenuation slope for all stellar populations, and the values of \(\alpha_{\rm{BCs}}\) and \(\alpha_{\rm{ISM}}\) in the case of the \citet{CharlotFall2000} models. The better agreement of our model-inferred attenuation laws with the \citet{CharlotFall2000} prescription compared to \citet{Calzetti2000} stems from fundamental differences in their treatment of the star-dust geometry, which are particularly critical in the dense, young stellar environments of NGC 253’s giant molecular clouds. While \citet{Calzetti2000} provides a robust empirical description for homogeneous starburst systems by assuming a single attenuation component with a gray slope ($R_{\rm{V}}$ $\sim$ 4.05) and suppressed 2175\(\text{\AA}\) UV bump \citep{Conroy2013, Samir2020}, it fails to capture the spatially segregated attenuation effects expected in galaxies with distinct birth clouds and diffuse ISM components. Our results align with theoretical studies demonstrating that the two-component \citet{CharlotFall2000} models – which separate the attenuation between optically thick birth clouds and a diffuse ISM with shallower attenuation – better reproduces the observed SEDs of systems with ongoing star formation in dense environments. This is consistent with NGC 253’s molecular clouds, where young stars remain embedded in their natal dusty clouds, creating a non-uniform attenuation geometry that the Calzetti law’s simplified single-screen approximation cannot adequately model. The widespread adoption of the Calzetti law in literature reflects its utility for global galaxy-scale attenuation in starburst-dominated systems, but our findings emphasize that its limitations become pronounced in resolved studies of individual star-forming regions.

Table~\ref{tab:attenuation_comparison} lists the measured Balmer decrement, the corresponding \( A_{\rm{V}}^{\mathrm{Balmer}} \), as well as the \( A_{\rm{V}} \) values obtained with CIGALE and \textsc{starlight}. 
As expected from the extremely red shape of the optical spectra (see Fig.\ \ref{Fig.STARLIGHT_fit}), we obtain large values of
\( A_{\rm{V}}^{\mathrm{Balmer}} \): from 1 to 6 mag. The largest values are found for GMCs 3, 4, 5, and 6, which are closer to the galactic bar and in dense star-forming areas. We note, however, the very substantial uncertainties in \( (F_{\mathrm{H}\alpha}/F_{\mathrm{H}\beta})_{\text{obs}} \) (ranging from 10 to nearly 90\%). These happen because of the weak \( \mathrm{H}\beta \) emission, which is barely visible in most cases in Fig.\ \ref{Fig.STARLIGHT_fit}.

Table~\ref{tab:attenuation_comparison} further lists the \textsc{starlight} results for the extinction. 
In the case of \textsc{starlight} the table gives $\tilde{A}_{\rm{V}}$, which we define as the average of the extinctions affecting all populations ($A_{\rm{V}}$) and that affecting only those younger than 10 Myr ($A_{\rm{V}} + \delta A_{\rm{V}}$), weighting them by the fluxes of the respective components as derived from the fits. Again, the extinction values are large (1.8--5.7 mag), as expected from the 
redness of the optical continuum. 
The derived values are close to, but in general somewhat smaller than those derived from the Balmer decrement. CIGALE values of $A_{\rm{V}}$ are, in median, 16\% lower than those from the Balmer decrement, $A_{\rm{V}}^{\rm Balmer}$. On the other hand, \textsc{starlight}-based values also differ from those of $A_{\rm{V}}^{\rm Balmer}$ by a median value of 27\%.

\begin{table}[ht]
    \caption{GMC attenuations.}
    \label{tab:attenuation_comparison}
    \centering
    \setlength{\tabcolsep}{0.12cm}
    \begin{tabular}{lccccc}
    \hline\hline
    GMC & $(F_{\rm{H}\alpha}/F_{\rm{H}\beta})_\text{obs}$ & $A_{\rm{V}}^\text{Balmer}$ & $A_{\rm{V}}^\text{CIGALE}$ & $\tilde{A_{\rm{V}}}^\textsc{starlight}$ \\
        &                                       & [mag]                         & [mag]               & [mag]                            \\ \hline
    1   & $3.81 \pm 0.32$                       & $1.08 \pm 0.11$               & $3.57 \pm 0.02$     & $2.12$                           \\
    2   & $8.32 \pm 1.70$                       & $2.89 \pm 0.30$               & $3.58 \pm 0.02$     & $2.42$                           \\
    3   & $29.79 \pm 16.90$                     & $5.06 \pm 0.56$               & $6.04 \pm 0.11$     & $4.44$                           \\
    4   & $45.69 \pm 37.94$                     & $5.87 \pm 0.67$               & $6.06 \pm 0.00$     & $5.67$                           \\
    5   & $43.43 \pm 37.37$                     & $5.81 \pm 0.65$               & $6.24 \pm 0.02$     & $4.66$                           \\
    6   & $29.14 \pm 16.45$                     & $5.04 \pm 0.54$               & $6.08 \pm 0.01$     & $4.10$                           \\
    7   & $21.21 \pm 10.11$                     & $4.38 \pm 0.47$               & $5.07 \pm 0.02$     & $2.97$                           \\
    8   & $19.18 \pm 10.04$                     & $4.18 \pm 0.45$               & $3.53 \pm 0.01$     & $2.28$                           \\
    9   & $17.01 \pm 7.84$                      & $3.96 \pm 0.42$               & $3.54 \pm 0.01$     & $1.84$                           \\
    10  & $11.13 \pm 5.05$                      & $3.36 \pm 0.37$               & $3.54 \pm 0.12$     & $2.11$                           \\
    \hline
    \end{tabular}
    \tablefoot{Attenuation estimates for the studied giant molecular clouds (GMCs) derived from the Balmer decrement ($A_{\rm{V}}^\text{Balmer}$), CIGALE ($A_{\rm{V}}^\text{CIGALE}$), and effective $A_{\rm{V}}$ ($\tilde{A_{\rm{V}}}^\textsc{starlight}$) from \textsc{starlight}. The first two columns correspond to the GMC number and the H$\alpha$ over H$\beta$ observed flux ratio, from left to right, respectively.}
\end{table}

The only GMC where $\tilde{A}_V$ exceeds $A_{\rm{V}}^{\rm Balmer}$ is GMC~1. This discrepancy may suggest additional sources of attenuation, potentially linked to warm or dense dust. One possible explanation is the influence of strong shocks, which could sputter dust grains and enhance local dust densities. Supporting evidence for the presence of shocks in GMC~1 comes from its elevated SiO\,$J=$5--4/HNCO\,(10$_{0,10}$--9$_{0,9}$) transition ratios observed at about 220~GHz (ALMA Band~6). This region shows the highest ratio values, averaging around 1.3, as noted in the right panel of Fig.~11 in \citet{Humire2022}. In contrast, the mean ratio in other regions is 0.5, excluding GMC~10, where this ratio cannot be measured. The above values have been derived using an aperture of 3\farcs0 and the coordinates from this work, listed in Table~\ref{tab.positions}. These elevated ratios are well-known tracers of strongly shocked gas, indicating a prevalence of strong (traced by SiO) over weak/mild shocks (traced by HNCO). Additionally, GMC~1 hosts methanol masers with the strongest deviations from LTE expectations among the 10 GMCs, as illustrated in the left panel of Fig.~13 in \citet{Humire2022}. While these observations strongly point to the presence of shocks, they are not direct indicators of warm or dense dust. However, shocks may indirectly affect the dust environment by altering its spatial distribution or properties. Future studies are needed to assess whether gas and dust are tightly coupled in this region and to determine the extent to which shocks influence the observed attenuation.

\begin{figure}[!ht]
\centering
\includegraphics[width=0.49\textwidth, trim={0 0 0 0}, clip]{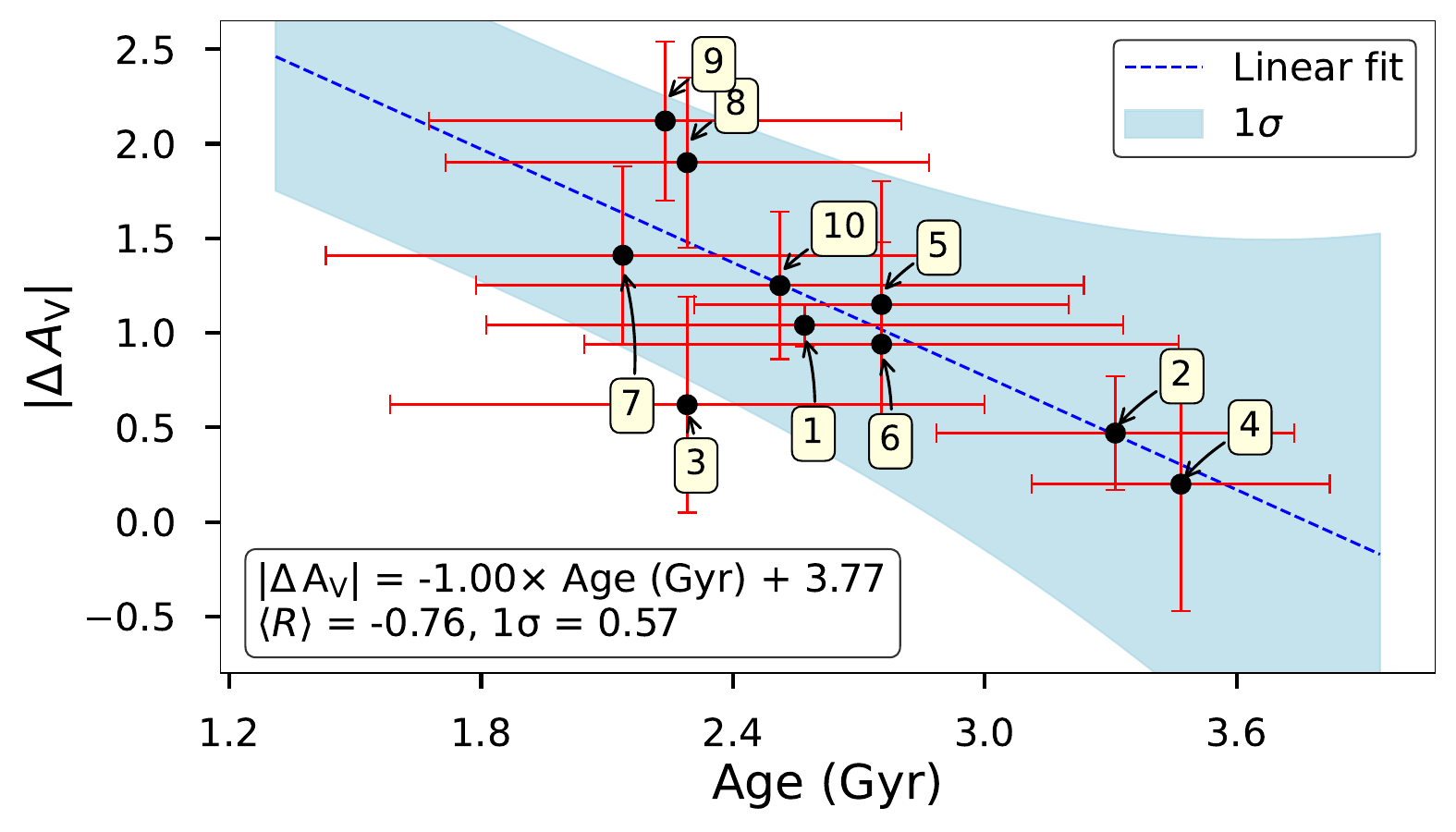}
\caption{Relation between absolute attenuation differences, $|\Delta\, A_{\rm{V}}|$, obtained from the Balmer decrement, $A_{\rm{V}}^{\rm{Balmer}}$, representing the gas attenuation, and the stellar attenuation obtained by considering the continuum emission from the near UV to the near IR by \textsc{starlight}. These attenuation differences are plotted in the y-axis against the stellar ages derived by CIGALE (x-axis).}
\label{fig.attenation_diff_vs_CIGALE_age}
\end{figure}

We compared the stellar continuum attenuation in the V-band, $A_{\rm{V}}$, with the corresponding stellar colour excess, E(B-V)s (represented by the \texttt{attenuation.E\_BVs} parameter), both derived from the CIGALE spectral energy distribution fitting results (Tables~\ref{tab:attenuation_comparison} and \ref{Apen.Tab.Extinction_Factors_CIGALE}). We find that the ratio $A_{\rm{V}}$/E(B-V)s ranges from $\sim$4.0 to 6.3 across our sample. Notably, this ratio systematically differs from the value $R_{\rm V} = 3.1$ that was set as an input parameter in our CIGALE configuration.

This apparent discrepancy arises from the specific implementation within the \texttt{dustatt\_modified\_starburst} module used in our analysis. This module employs distinct attenuation laws for the stellar continuum and the nebular emission lines. The input parameter $R_{\rm V}=3.1$, characteristic of the diffuse interstellar medium in the Milky Way \citep[e.g.,][]{Cardelli1989}, governs the shape of the Cardelli et al. extinction curve applied specifically to the nebular emission lines (as we selected \texttt{Ext\_law\_emission\_lines = 1}). However, the attenuation affecting the stellar continuum (from which $A_{\rm{V}}$ and E(B-V)s are primarily determined) is modeled using the empirical starburst attenuation law from \citet{Calzetti2000}, which can be further modified by a power-law slope ($\delta$) and a UV bump. The baseline Calzetti law itself possesses an effective ratio $A_{\rm{V}}$/E(B-V) $\approx 4.05$, inherently different from the Milky Way value. The flexibility introduced by the allowed modifications ($\delta$ and the UV bump) within the module accounts for the observed range of $A_{\rm{V}}$/E(B-V)s extending above this baseline value, consistent with the methodology described in \citet[][Sect. 3.4.2]{Boquien19}. Therefore, the $A_{\rm{V}}$ and E(B-V)s values derived for the stellar component are not expected to follow the relation $A_{\rm{V}} = 3.1 \cdot$E(B-V)s, but rather reflect the properties of the modified Calzetti law applied to the integrated starlight.

GalaPy derived attenuation curves are shown in Fig.~\ref{fig.attenuation_curves}, where curves indicate the attenuation in the visible band and legends show its value at 5500\AA, on a linear scale, to be compared with \textsc{starlight} outputs (Table~\ref{Tab.STARLGIGHT_results}). The results from this code are normalize at that wavelength value. $R_{V}$ values following \citet{Calzetti2000} are also included for comparison. These values describe the relationship between the total amount of dust extinction and the amount of reddening it causes in the context of the Milky Way’s dust extinction curve and corresponds to the ratio between the total extinction $A_{V}$ and the color excess E(B-V). Considering the ten GMCs studied, most $R_{\rm{V}}$ values range between 6 and 9, far beyond the Galactic value of 3.1. Surprisingly, the extreme values are found in clouds located at the boundaries of the CMZ, with the highest $R_{\rm{V}}$ values in GMCs~2 and 9. These clouds are closer to the modeled curves (dashed lines in the left panel of Fig.~\ref{fig.attenuation_curves}) corresponding to $R_{\rm{V}} \sim 8$--9. In contrast, the lowest $R_{\rm{V}}$ value is found in GMC~1, which is best matched by a modeled curve with $R_{\rm{V}} \sim 4$.

As a final remark, when comparing the different attenuation estimates, it is important to recognize that the gas attenuation (from the Balmer decrement) and stellar attenuation (from GalaPy, CIGALE, and STARLIGHT in our case) have distinct origins. Recent studies suggest that these two types of attenuation should become increasingly similar as the observed region ages \citep{Li2024}.

However, we do not see a clear trend for attenuation differences between the ones derived from the Balmer decrement and those obtained from CIGALE, namely $ \Delta A_{\rm{V}} = A_{\rm{V}}^{\rm{Balmer}} - A_{\rm{V}}^{\rm{CIGALE}}$, taken from Table~\ref{tab:attenuation_comparison}, versus derived stellar ages from either GalaPy or CIGALE (last columns of Tables~\ref{Tab.star_formation_results_GalaPy} and \ref{Tab.CIGALE_main_results}, respectively), we do see a trend when comparing attenuation differences between Balmer lines and the gas attenuation derived from the stellar continuum by \textsc{starlight}, namely $ \Delta A_{\rm{V}} = A_{\rm{V}}^{\rm{Balmer}} - A_{\rm{V}}^{\textsc{\rm{starlight}}}$, versus stellar ages derived by CIGALE. For an easy visualization of that finding, in Fig.~\ref{fig.attenation_diff_vs_CIGALE_age} we present such a relation.

\subsubsection{Linear regression}
\label{subsubsec.lin_regrees}

The procedure for all linear regressions in this work, including Fig.~\ref{fig.attenation_diff_vs_CIGALE_age}, followed two main steps. First, we used the {\tt linregress} function from the {\tt scipy} Python library to obtain initial priors. For power-law relationships, we performed the regression in log-space when required (using $\log_{10}$ values as inputs), with results either maintained in logarithmic form ($\log y = m\log x + c$) or converted back to linear space depending on the analysis.

The second step applied Bayesian inference through Markov Chain Monte Carlo (MCMC) sampling using the {\tt emcee} Python package \citep{Foreman-Mackey2013}. Unlike {\tt linregress}, MCMC incorporates uncertainties in both axes (x and y), which were rigorously included in our modeling. We ran 1,000 iterations per fit with 50 walkers, discarding the first 100 as burn-in to ensure convergence. The resulting posterior distributions provided robust estimates of the slope and intercept, with their $1\sigma$ uncertainties represented as shaded regions in all plots.

\subsection{Star formation rates}

\subsubsection{Kennicutt (1998): H$\alpha$–SFR calibration}
\label{subsec.KS_rel}

\begin{figure}[!ht]
\centering
\includegraphics[width=0.5\textwidth, trim={0.4cm 0.6cm 0.2cm 0.3cm}, clip]{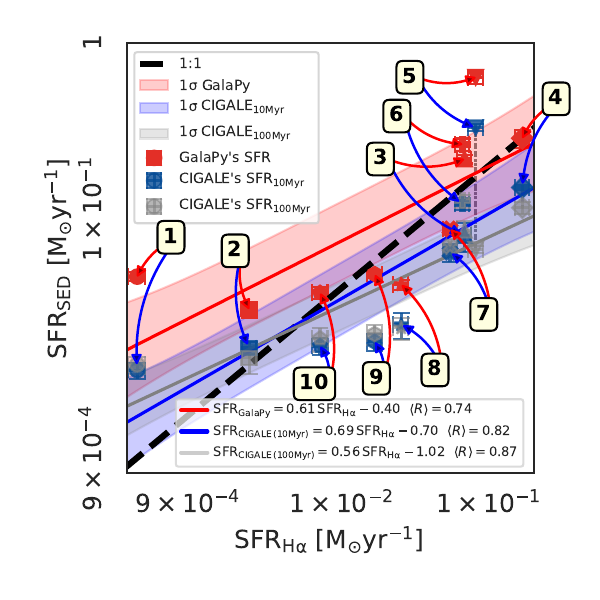}
\caption{Star formation rate (SFR) obtained from H$\alpha$, SFR$_{\rm{H\alpha}}$, following Eq.~\ref{eq.KS_rel} versus the ones obtained from our SED fittings using GalaPy (red points; SFR in Table~\ref{Tab.star_formation_results_GalaPy}) and CIGALE (blue points; SFR$_{\rm{10Myr}}$ in Table~\ref{Tab.CIGALE_main_results}), following the legend. The numbers correspond to the identification of the GMCs. Shaded areas indicate the 1$\sigma$ dispersion for GalaPy (red) and CIGALE (SFRs averaged over 10~Myr in blue and over 100~Myr in grey) datapoints. The dashed black line shows the 1:1 expected correlation \citep{Kennicutt1998}. The SFR obtained from the H$\alpha$ emission line has been corrected by Balmer inferred attenuations $A_{\rm{V}}^{\rm{Balmer}}$ listed in Table~\ref{tab:attenuation_comparison}, which already incorporates a correction of $\times$0.82 (see Subsect.~\ref{subsec.attenuations}). Grey symbols are CIGALE's SFRs averaged over 100~Myrs, they are connected by dots to the CIGALE's SFR$_{\rm{10~Myr}}$ (only noticeable for the GMC~5 case). Best fits and correlation coefficients are shown in the bottom legend.}
\label{Fig.SFR_SED_MUSE_Ha}
\end{figure}

There are many ways to obtain the SFR from monochromatic emission. One of the most commonly used is the relation found by \citet{Kennicutt1998} regarding H$\alpha$, which we will hereafter refer to as the ``Kennicutt relation'':\\
\begin{equation}
\rm{SFR}~ [\mathrm{M_{\odot}~year^{-1}}] = 7.9 \times 10^{-42} L(\mathrm{H\alpha})~[\mathrm{erg~s^{-1}}].
\label{eq.KS_rel}
\end{equation}

\noindent where $L(\mathrm{H\alpha})$ is corrected for extinction using $A_{\rm{H}\alpha} = 0.82 \times A_{\rm{V}}^{\rm Balmer}$.

Fig.~\ref{Fig.SFR_SED_MUSE_Ha} shows a comparison between the SFRs obtained from the above considerations and the one obtained from the SED fitting using GalaPy and CIGALE, while the dashed-black line indicates a one-to-one proportion. To create this plot we considered the attenuation of the H$\alpha$ emission line derived SFR considering the $A_{\rm{V}}^{\rm{Balmer}}$, since it is expected to be the most accurate method: \textsc{starlight} and SED derived attenuation rely on the continuum level. 

We notice that SFRs derived by GalaPy (red symbols in Fig.~\ref{Fig.SFR_SED_MUSE_Ha}) tend to be larger than expected by the Kennicutt relation. This result is expected if the instantaneous SFR is greater than the average SFR over 10~Myr, as the SFR derived from H$\alpha$ corresponds to the stellar formation averaged over that period of time \citep[see, e.g.,][]{Stroe2015}. When considering the 10~Myr and 100~Myr averaged SFRs from CIGALE (blue and gray symbols in Fig.~\ref{Fig.SFR_SED_MUSE_Ha}, respectively), the Kennicutt relation is followed more closely at 100~Myr than at 10~Myr (see Table~\ref{Tab.CIGALE_main_results}). However, the slope which is closer to 1 is obtained from CIGALE's SFR$_{\rm{10~Myr}}$, in line with what is expected from deriving the SFR from H$\alpha$ emission.

\begin{figure}[!ht]
\centering
\includegraphics[width=0.49\textwidth, trim={0 0 0 0}, clip]{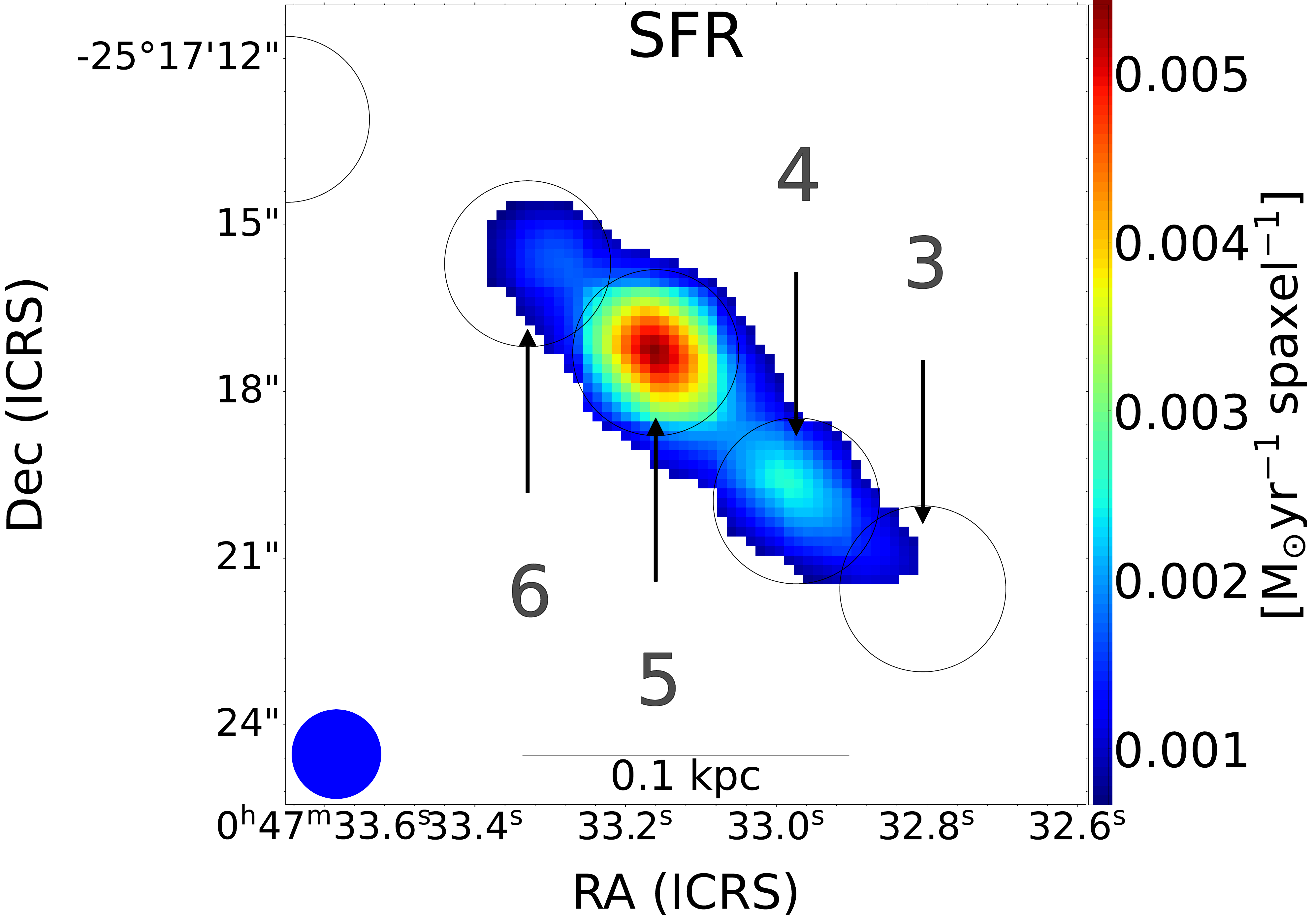}
\caption{Star formation rate (SFR) of the inner regions of NGC\,253's CMZ as accounted from the H$40\alpha$ RRL. The original ALCHEMI beam of 1\farcs6 is denoted in the bottom-left corner. Circles correspond to the photometric aperture (3\arcsec diameter) used for the flux extraction and posterior SED fitting.}
\label{Fig.SFR_H40a}
\end{figure}

\begin{figure}[!ht]
\centering
\includegraphics[width=0.49\textwidth, trim={0 0 0 0}, clip]{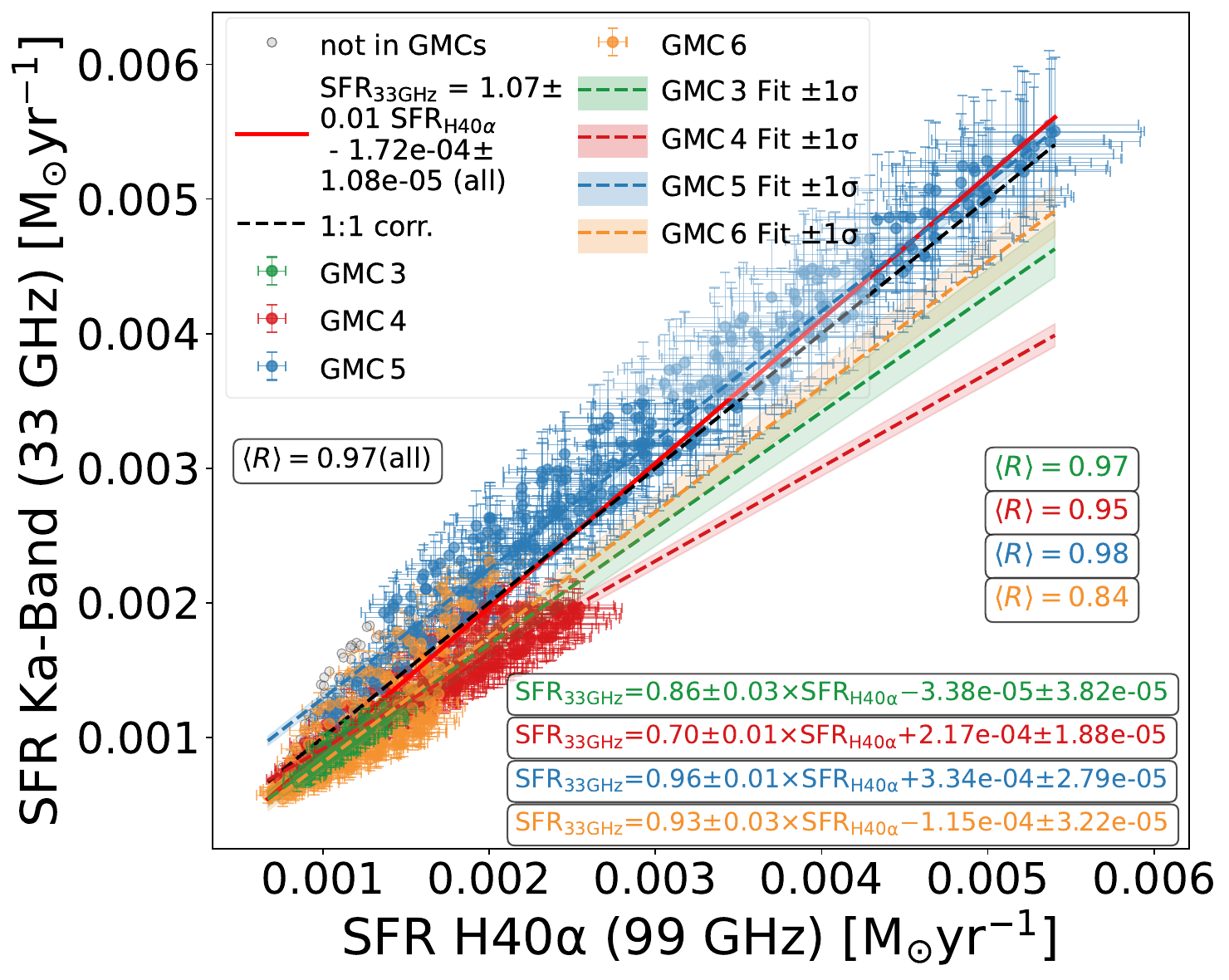}
\caption{Star formation rate (SFR) of the inner regions of NGC\,253's CMZ. In the x-axis, the SFR is obtained from the H$40\alpha$ RRL (see Subsect.~\ref{subsubsec.RRL}). In the y-axis, the SFR was obtained from the Ka EVLA Band applying the best-fitting relation from our linear regression, presented in Eq.~\ref{eq.SFR_from_Ka_band}. The resulting fit considering all data points (all) and per GMC (color-coded) are labeled along with the Pearson correlation coefficient, $\langle R \rangle$. All points include 10\% uncertainties, with regression handling errors in both axes via Markov Chain Monte Carlo, as described in our Subsect.~\ref{subsubsec.lin_regrees}.} 
\label{fig.comparing_SED_with_H40a}
\end{figure}

\begin{figure*}[!ht]
\centering
\includegraphics[width=\textwidth, trim={0 0 0 0}, clip]{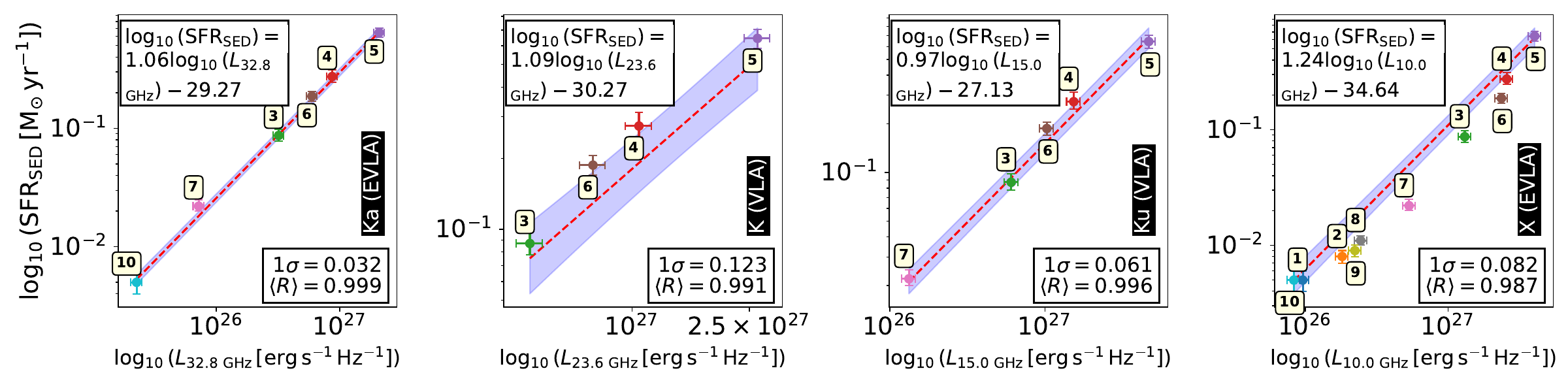}
\caption{Star formation rate (SFR) from our SED fitting using GalaPy versus fluxes from our retrieved (E)VLA continuum images, using apertures of 3$\arcsec$ in available GMCs (see Table~\ref{tab.phot_points}). Best-fit linear regressions in the logarithmic scale are indicated in the top leftmost corners of each subplot, this best-fit is indicated with a dashed red line. The 1$\sigma$ dispersion from the best-fit to the different measurements is indicated in the bottom right corners of each subplot and also as shaded blue areas around the linear regression. (E)VLA Band names are labeled with black background legends. GMCs are numbered and also color-coded following legend in Fig.~\ref{fig:comparison_grid}. We assume a conservative 10\% flux uncertainty for (E)VLA observations.}
\label{Fig.SFR_GalaPy_vs_radio}
\end{figure*}

\begin{figure*}[!ht]
\centering
\includegraphics[width=\textwidth, trim={0 0 0 0}, clip]{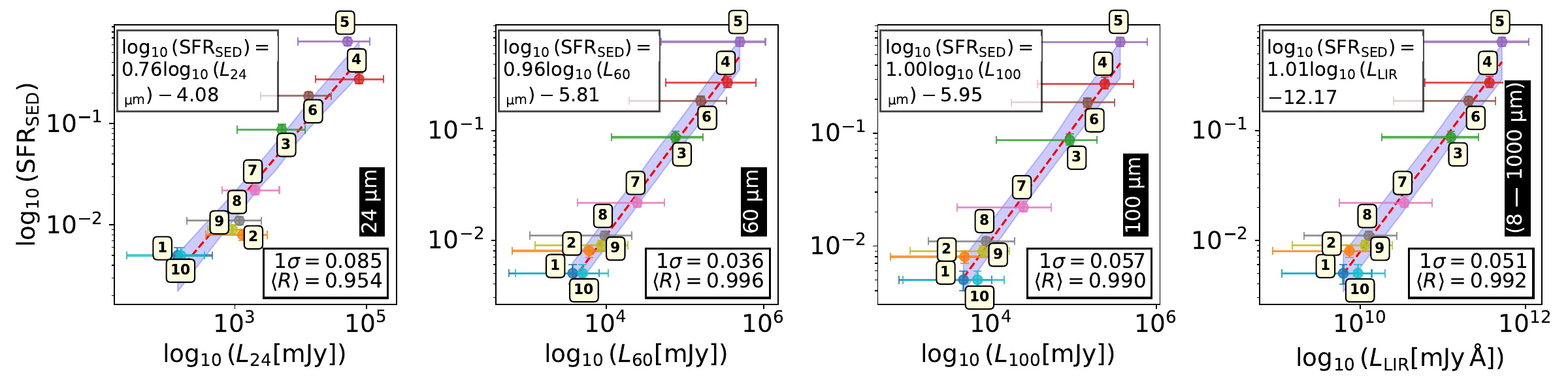}
\caption{Same as in Fig.~\ref{Fig.SFR_GalaPy_vs_radio} but this time using IR SED fitting values, both from monochromatic fluxes (at 24, 60, and 100$\mu$m) and from integrating the total infrared luminosity range (LIR; 8--1000$\mu$m). GMCs are numbered and also color-coded following the legend in Fig.~\ref{fig:comparison_grid}. 
}
\label{Fig.SFR_GalaPy_vs_IR}
\end{figure*}

\subsubsection{Radio recombination lines}
\label{subsubsec.RRL}

There are several methods to estimate the SFR of an astronomical source. One of the most accurate techniques, unaffected by dust extinction, relies on radio recombination lines (RRLs). Following the approach of \citet{Bendo2015}, who used the H$40\alpha$ line at 99.02\,GHz, we derived the SFR for the nuclear GMCs (specifically GMCs~3 to 6). Figure~\ref{Fig.SFR_H40a} displays the SFR density in units of M$_{\odot}$\,yr$^{-1}$\,spaxel$^{-1}$, after selecting the free-free emission only, namely, after applying a correction of 70\% to H40$\alpha$'s total emission, as described in \citet{Bendo2015}.

To contextualize our results, we compared our SED-based estimates (GalaPy; Table~\ref{Tab.star_formation_results_GalaPy}) with those derived from the H40$\alpha$ line in \citet{Bendo2015} with observations at a similar resolution. While our aperture is larger than theirs (see their Fig.~1, leftmost panel), the regions are comparable: their West, Central, and East regions correspond to our GMCs~4, 5, and 6, respectively. The SFRs derived from the H40$\alpha$ line in \citet{Bendo2015} (their Table~1) are on average between 6\% (total H40$\alpha)$ and 25\% (free-free corrected) lower than the ones derived by GalaPy, and this can be due to our larger apertures.

Eliminating the effect of different aperture sizes by taking the total emission in NGC~253's CMZ, we conciliate the results: \citet{Bendo2015} reported a total SFR of $1.73\pm 0.12~\text{M}_{\odot}\,\text{yr}^{-1}$ for NGC~253's CMZ within a $20 \times 10$\,arcsec region. A separate $9\arcsec$-radius SED (GalaPy) analysis (including Herschel PACS data) encompassing GMCs~4--6 (Fig.~\ref{fig.FIR_SEDs_attempts}, bottom inset in the left panel) yields a SFR of $1.620^{+0.201}_{-0.156}~\text{M}_{\odot}\,\text{yr}^{-1}$. Including the contribution from GMCs~3 ($0.087^{+0.011}_{-0.009}~\text{M}_{\odot}\,\text{yr}^{-1}$) and 7 ($0.022^{+0.003}_{-0.002}~\text{M}_{\odot}\,\text{yr}^{-1}$), which also fall within the mentioned region, the total SFR raises to $1.729^{+0.201}_{-0.156}~\text{M}_{\odot}\,\text{yr}^{-1}$, consistent with \citet{Bendo2015}'s (free-free) value within uncertainties.

Since the H40$\alpha$ line is only detectable in the inner regions of the studied CMZ, at SFRs larger than about 0.870~$\text{M}_{\odot}~\text{yr}^{-1}$ (GalaPy) or 0.262~$\text{M}_{\odot}~\text{yr}^{-1}$ (CIGALE), considering GMC~3's SFR as the lower limit, we wanted to probe its derived SFR at spaxel levels. With that aim, we decided to consider the EVLA Ka-Band, at 32.8~GHz, which is the Band that better correlates with our SED analysis, as we will see in the next Subsection~\ref{subsubsection.SFR_from_radio_cm_and_FIR}. To derive the SFR from that band, we applied the best-fit from the linear regression labeled in Fig.~\ref{Fig.SFR_GalaPy_vs_radio} upper panels, which for the case of the mentioned Band is:
\begin{equation}
\label{eq.SFR_from_Ka_band}
   \mathrm{SFR_{SED}} [M_{\odot}~\mathrm{yr}^{-1}] = L_{32.8~\mathrm{GHz}}^{1.06}~[\mathrm{erg s^{-1}~Hz^{-1}}] \cdot 10^{-29.27},
\end{equation}
\noindent
in linear scale. The resulting SFR value is what we used in the y-axis of Fig.~\ref{fig.comparing_SED_with_H40a} while, on the x-axis, we plot the same values as shown in Fig.~\ref{Fig.SFR_H40a}.

Fig.~\ref{fig.comparing_SED_with_H40a} shows a variety of fits, the best fit found considering all spaxels is shown as a solid red line while the best-fitting values per GMCs are color-coded. In general, the correlations are very strong, with the largest Pearson correlation coefficient ($\langle R \rangle$) and the closest slope to unity found for GMC~5. Taking the dashed black line, which denotes a 1:1 correspondence (``corr.'' in the Fig.), as a reference, we find that the H40$\alpha$ derived SFR tends to be slightly overestimated in GMC~5 and slightly underestimated in the other GMCs. In general, the correlation is strong ($\langle R \rangle =$0.97) and the values tend to be slightly lower than the derived ones from the 33~GHz band, with a general slope of 1.07$\pm$0.01.

We note that GalaPy measures the instantaneous SFR, whereas RRLs trace star formation over the last $\sim$5--10\,Myr \citep{Scoville2013,Emig2020}. The tight correlation between the Ka-band and H$40\alpha$ SFRs (Figure~\ref{fig.comparing_SED_with_H40a}) suggests that both tracers reliably probe recent star formation activity, and that the comprised timescales should be similar.

\subsubsection{Radio cm continuum and FIR}
\label{subsubsection.SFR_from_radio_cm_and_FIR}

As summarized in \citet{Murphy11}, there are several SFR tracers in the wavelength range considered in this work. According to these authors, the 33~GHz continuum emission is a reliable indicator of current star formation, as it primarily arises from thermal free-free emission. Although this frequency can sometimes be affected by rotating dust, this is not the case for NGC~253 at the studied scales, as shown in the SEDs. We indeed find a strong correlation between the emission in the EVLA Ka Band, as originally published in \citet{Kepley2011}, and the total SFR computed by GalaPy. Furthermore, accounting for scaling offsets, we also observe generally good correlations—close to a 1:1 correspondence—between cm tracers and the GalaPy results, as seen in Fig.~\ref{Fig.SFR_GalaPy_vs_radio}. The latter was obtained using the same approach as for the 33~GHz data, namely, by performing a linear regression in logarithmic space with all retrieved (E)VLA continuum images that detect at least four GMCs (see Table~\ref{Tab.observations}), aiming to obtain the best-fit assembly valid for the GMCs in the starburst galaxy NGC~253. A caveat should be noted, as different galaxy types (e.g., MW-like, dwarf galaxies, AGN hosts) may present different relationships. In this regard, an immediate prospect for the current work is to evaluate the different relations between SFR tracers.

Accounting for IR points as SFR tracers, as can be seen in Fig.~\ref{Fig.SFR_GalaPy_vs_IR}, we generally find a comparable dispersion and quality of fit (in terms of mean $R$) to that obtained with radio bands. However, it is important to note that, since these IR data points are directly derived from the GalaPy SED fitting employed to estimate the SFR, they are not entirely independent of the latter. On the other hand, while the flux uncertainties (along the x-axis) associated with the IR data are considerably larger than those of the radio bands, this is compensated by the availability of IR points for all ten GMCs. Notwithstanding this potential interdependence, the general conclusion drawn from the IR tracers is that the best ones correspond to the integrated IR emission between 8 and 1000~$\mu$m, commonly referred to as total IR emission, and the IR emission at 60~$\mu$m.

For the radio bands, unlike optical lines, we do not assume a relation factor a priori, we only convert our mJy measurements to erg s$^{-1}$ Hz$^{-1}$ for an easy comparison between this study and the literature. For the mJy to erg s$^{-1}$ Hz$^{-1}$ conversion we have assumed a distance of 3.5~Mpc to NGC~253 \citep{Rekola2005}. For the IR bands, we retained the original units of mJy or mJy~$\AA$ to allow for straightforward comparison with other studies.

\section{Discussion}
\label{sec.discussion}

\subsection{Reliability of our SED model along the FIR window}
\label{Sec.reliability_of_SED_fit_in_the_FIR}

One of the critical, if not the most critical, challenges that astronomers have to face when attempting to fit a panchromatic SED at high angular resolution is the impossibility to acquire far infrared (FIR) observations at resolutions below 6$\arcsec$. Current facilities in the regime, such as the James Webb (JWST) and Euclid telescopes, cover mainly the near-IR and mid-IR parts of the electromagnetic spectrum, with the largest wavelength window ranging from 0.5 (Euclid) to 28~$\mu$m (JWST). Indeed, only the JWST can keep angular resolutions below $\sim$0.8$\arcsec$ at 28~$\mu$m. However, the FIR bump primarily occurs at wavelengths larger than those, mainly from about 50 to 300~$\mu$m.

While we made use of our well covered sub-mm regime, with ALCHEMI data, to fine-tune the Rayleigh-Jeans tail, at the emergence of the FIR bump on the radio side. We also used VLT/NACO data at 11 and 18.7~$\mu$m on the MIR side for GMCs 3--7, we know the caveat that this is likely not enough to properly ensure the fit on the IR bump itself, relying on our SED model extrapolations. This is why we further retrieved archival Herschel PACS observations selecting level 2.5 PACS images at 70, 100, and 160~$\mu$m. While these FIR observations are not able to resolve the targeted GMCs, they are sufficient to fit emission from the inner part of the CMZ, covering at most GMCs 3 to 6 in the worst-case scenario (at 11.3$\arcsec$), and partially covering GMCs~4 to 6 at 6 and 9$\arcsec$ (see Appendix~\ref{apen.testing_at_9_arcsec} for more information on the latter).

In Fig.~\ref{fig.FIR_SEDs_attempts} it can be noted that the far infrared peak agrees between the 3$\arcsec$ aperture SED, devoid of FIR photometric points and the SED at 9$\arcsec$, whose diffuse dust temperature, $T_{\rm{DD}}$, of 68.8$_{-13.9}^{+7.8}$~K is in agreement to the ones of the covered GMCs (GMCs 4--6), as noted in Table~\ref{Tab.star_formation_results_GalaPy}. Specifically, we get $T_{\rm{DD}}$ estimations of 84.5$_{-12.6}^{+6.8}$, 67.9$_{-3.9}^{+4.0}$, and 63.3$_{-2.8}^{+3.2}$~K for GMCs 4, 5, and 6, respectively. All in all, this arguments in favor of our 3$\arcsec$ apertures as the FIR bump and the SED shape in general peak at a very similar point. It is worth noting that the SED shape is responsible for specific star formation rate estimates and, thus, for the stellar mass and SFR estimates as well \citep{Conroy2013}. 

Given the lack of FIR photometric points in the FIR regime at 3$\arcsec$, our SED models experienced considerable freedom in this specific wavelength range. We found that GalaPy produced SED shapes that were closer, compared to CIGALE's models, to those observed at 9$\arcsec$, where Herschel PACS observations are available. Nevertheless, it is worth emphasizing that CIGALE performs well in fitting the SED of GMC~5 at 9$\arcsec$ (which also includes GMCs~4 and 6), both with and without an AGN component, as discussed in Appendix~\ref{apen.CIGALE_AGN}.

\subsection{Comparison to high-z dusty starbursts}
\label{Sect.comparison_to_high_z}
The question of whether local starbursts can serve as analogues of primeval dusty galaxies at high redshifts ($z>4-8$) remains a topic of active debate. A useful parameter in this context is the dust-to-stellar mass ratio, which provides insight into the balance between dust production and destructive processes such as supernova-driven shocks and astration (\citealt{nanni20}, \citealt{donevski20}, \citealt{kokorev21}, \citealt{donevski23}, \citealt{witstok23}, \citealt{palla24}, \citealt{sawant25}). 

To place the ISM conditions of NGC~253 into perspective against those of distant dusty galaxies, we compare the dust-to-stellar mass ratios ($M_{\rm dust}/M_{\star}$) derived in this work to those inferred from ALMA observations of high-$z$ starbursts (\citealt{donevski20}, \citealt{salak24}, \citealt{sawant25}). Regardless of the SED fitting code used, all GMCs in NGC~253 exhibit relatively high dust-to-stellar mass ratios, with a median value of $\log(M_{\rm dust}/M_{\star}) = -2.32 \pm 0.29$ (values from CIGALE), or $\log(M_{\rm dust}/M_{\star}) = -2.89 \pm 0.45$, if values from GalaPy are considered. These ratios are close to those observed in young dusty starburst galaxies in the distant Universe. Specifically, the average dust-to-stellar mass ratio is $\log(M_{\rm dust}/M_{\star}) = -2.13 \pm 0.3$ at $z \sim 2-5$ (\citealt{donevski20}) and $\log(M_{\rm dust}/M_{\star}) = -2.58 \pm 0.4$ at $z \sim 4-6$ (\citealt{sawant25}).
A caveat in comparing the dust-to-stellar mass ratio found here with literature is in the method used to derive galaxy properties. Both in \cite{donevski20} and \cite{sawant25}, the SED fitting is performed using CIGALE, which results in average lower values with respect to GalaPy, as also reported above.

Moreover, inferred attenuation values in the GMCs of NGC~253 are comparable to those observed in high-$z$ dusty starbursts but also significantly higher than those typically found in JWST-detected galaxies at $z > 6-8$ (e.g., \citealt{nanni20}, \citealt{alvarez24}, \citealt{markov25}). It is important to note that, despite their comparable "dustiness," GMCs in NGC~253 have $M_{\star}$, SFR and sSFR several orders of magnitude lower than their high-$z$ counterparts. In such low stellar mass regimes, dust production is generally expected to be dominated solely by stellar sources, such as Type II supernovae (\citealt{galliano19}, \citealt{sommovigo20}). 

However, recent models suggest that dust production in young starbursts may be significantly diminished due to ISM removal in the presence of strong stellar outflows, similar to those observed in NGC~253. To reconcile the high dust-to-stellar mass ratios, these models propose efficient growth of large dust grains through ISM accretion in rapidly cooling gas within stellar outflows (\citealt{hirashita23}, \citealt{donevski23}, \citealt{romano24}, \citealt{palla24}). While a detailed exploration of this intriguing possibility lies beyond the scope of this work, we note that the high dust-to-stellar mass ratios and attenuations observed in the GMCs may signal that NGC~253 is undergoing a shift in the dominant dust production mechanisms, as proposed by recent simulations (see e.g., \citealt{narayanan25}, also \citealt{schneider24} for a review).

\section{Conclusions and prospects}
\label{sec.conclusions}

This study represents a significant step forward in understanding the CMZ of NGC~253, utilizing the first panchromatic SED analysis at a 51-pc resolution towards extragalactic regions. The integration of diverse modeling approaches, from GalaPy to CIGALE and \textsc{starlight} on the stellar continuum features, enabled a thorough characterization of physical properties such as star formation rates, dust and stellar masses, and attenuation effects across ten GMCs using both parametric and non-parametric models.

The integration of multi-wavelength SED modeling tools, such as GalaPy and CIGALE plus the integration of mono-wavelength tracers in the optical using \textsc{starlight} and the sub-mm using ALMA observations, provided detailed insights into star formation histories, attenuation, and stellar population properties at different timescales. High extinction values (e.g., A\(_V\) exceeding 5 magnitudes in internal GMCs) were robustly derived, aligning well with other methods and showcasing the importance of panchromatic approaches.

We summarize our main findings below.

\begin{enumerate}
    \item Dynamical and chemical diversity:
        The Central Molecular Zone (CMZ) of NGC~253 exhibits significant dynamical and chemical distinctions between internal (nuclear) and external GMCs.
        Internal GMCs (e.g., GMCs 3--6) are characterized by higher star formation rates (SFR), stellar masses, and dust masses, consistent with active star formation, high-excitation regions, and higher metallicities.
        External GMCs (e.g., GMCs 1, 2, 7--10) display lower density, slower and weaker shocks, and chemical signatures indicative of less energetic processes as compared to the internal GMCs.

    \item Comparison of internal vs. external GMCs:
        Internal GMCs exhibit stellar masses in the range of \(3.7^{+1.1}_{-0.7}\)--\(7.1^{+5.5}_{-1.4} \times 10^8 M_\odot\) and dust masses in the range of \(1.6^{+0.7}_{-0.3}\)--\(5.8^{+3.5}_{-3.0} \times 10^5 M_\odot\), both greatly surpassing (doubling) maximum values observed in external GMCs.
        Specific star formation rates (sSFR) further underscore these differences, exhibiting a trend of enhanced values in the nuclear regions.
    
    \item Insights from age-metallicity relations:
        Differences in stellar ages and metallicities highlight the varied evolutionary states of the studied GMCs. Depending on the adopted software, these values can vary by up to two orders of magnitude in age (\textsc{starlight}; Tab~\ref{Tab.STARLGIGHT_results}) and one order of magnitude in metallicity (GalaPy; Tab~\ref{Tab.star_formation_results_GalaPy}).

    \item Emission line diagnostics:
        The BPT and WHAN diagrams indicate shock signatures in the nuclear region of NGC~253, placing it in the composite zone. However, these shocks likely result from stellar winds and supernovae in the starburst environment, not from AGN activity.

    \item Caveats on H$\alpha$'s derived SFR:
        While H$\alpha$ remains a popular SFR tracer, our analysis reveals its significant limitations in NGC 253's dust-obscured CMZ, where attenuation causes large discrepancies compared to more robust radio (e.g., the band at 33~GHz) and panchromatic SED tracers, suggesting H$\alpha$ should be supplemented with dust-insensitive indicators in similar environments.
    
    \item Star formation, panchromatic vs. monochromatic results:
        We have found a strong 1:1 correlation between cm tracers and SED derived SFR values, being the Ka-band (at 33~GHz) the best among the available dataset. This is in line with previous works based on entire galaxy properties \citep[e.g.,][]{Murphy11}, and is now confirmed at GMC scales. Additionally, we validate the effectiveness of RRLs (H40$\alpha$) as a SF tracer, although limited to the brightest sources (GMCs~3--6).
    
\end{enumerate}

The current study provides a robust framework for leveraging spatially-resolved multi-wavelength datasets to investigate the interplay between star formation, gas dynamics, and attenuation in starburst galaxies. In addition, this work not only advances our knowledge of starburst environments but also serves as a benchmark for studying high-redshift galaxies, where similar extreme conditions prevail. 

In the future, we plan to focus on several key directions to broaden the applicability of these findings. These include utilizing JWST instead of Spitzer observations, achieving aperture diameters as fine as 1\arcsec\ in the mid-IR, enabling more detailed analyses of star-forming regions and/or augmenting the redshift regime for our potential targets. We plan to apply the same approach to other well-known starburst like M~82. Alternatively, we want to improve the robustness of our fit by restricting ourselves to very nearby galaxies where FIR data from Herschel can still resolve GMC-scale structures, that is, if we assume a GMC size of 50~pc, and a Herschel aperture of 9$\arcsec$, we can utilize Herschel/PACS FIR observations of Milky Way's and Andromeda's satellites, yielding approximately 20 objects. 

Additionally, this framework has the potential to be extended to diverse systems, including Milky Way like galaxies, LINERs, ULIRGs, and AGN, to explore differences in star formation processes across various galactic environments. Furthermore, focused studies of GMCs in our Galaxy will offer a critical baseline for contrasting extragalactic GMC properties, contributing to a thorough understanding of star formation under varying conditions. As a final remark, we also aim at providing future GalaPy users with a full set of priors well suited for GMC scales in extreme, intensively star-forming and highly obscured environments, which will also be helpful for Galactic analogues such as Sgr~B2 mini-starbursting system \citep{Belloche2013,Schwoerer2019} and the distant (high-$z$) universe.

\begin{acknowledgements}
We thank the anonymous referee for their helpful comments, questions, and suggestions on revising the manuscript. P.K.H. gratefully acknowledges the Fundação de Amparo à Pesquisa do Estado de São Paulo (FAPESP) for the support grant 2023/14272-4. P.K.H. acknowledges L. Barcos for discussions on the project's initial scope, E. Caux for insights into the FIR cross-validation methodology, G. Ortiz-León for support with the EVLA X-band data reduction, and D. Hazarika for guidance on the proper implementation of the MCMC algorithm for the linear regression analysis. We thank A. Kepley for providing us with the 33\,GHz (Ka-band) EVLA image.
S.D has been supported by the Polish National Science Center project UMO-2023/51/D/ST9/00147
TR is supported by the Italian Research Center on High Performance Computing Big Data and Quantum Computing (ICSC), project funded by European Union - NextGenerationEU - and National Recovery and Resilience Plan (NRRP) - Mission 4 Component 2 within the activities of Spoke 3 (Astrophysics and Cosmos Observations). D.D. acknowledges support from the National
Science Center (NCN) grant SONATA (UMO-2020/39/D/ST9/00720). JAFO acknowledges financial support by the Spanish Ministry of Science and Innovation (MCIN/AEI/10.13039/501100011033), by ``ERDF A way of making Europe'' and by ``European Union NextGenerationEU/PRTR'' through the grants PID2021-124918NB-C44 and CNS2023-145339; MCIN and the European Union -- NextGenerationEU through the Recovery and Resilience Facility project ICTS-MRR-2021-03-CEFCA. M.H. acknowledges the support by the National Science Centre, Poland (UMO-2022/45/N/ST9/01336).
V.M.R. and L.C. acknowledge support from the grant PID2022-136814NB-I00 by the Spanish Ministry of Science, Innovation and Universities/State Agency of Research MICIU/AEI/10.13039/501100011033 and by ERDF, UE.  V.M.R also acknowledges support from the grant RYC2020-029387-I funded by MICIU/AEI/10.13039/501100011033 and by "ESF, Investing in your future", and from the Consejo Superior de Investigaciones Cient{\'i}ficas (CSIC) and the Centro de Astrobiolog{\'i}a (CAB) through the project 20225AT015 (Proyectos intramurales especiales del CSIC); and from the grant CNS2023-144464 funded by MICIU/AEI/10.13039/501100011033 and by “European Union NextGenerationEU/PRTR”. A.R.L. acknowledges financial support from Consejo Nacional de Investigaciones Científicas y Técnicas (CONICET), Agencia I+D+i (PICT 2019–03299) and Universidad Nacional de La Plata (Argentina). 
A.C.K thanks Fundação de Amparo à Pesquisa do Estado de São Paulo (FAPESP) for the support grant 2024/05467-9. SP is supported by the international Gemini Observatory, a program of NSF NOIRLab, which is managed by the Association of Universities for Research in Astronomy (AURA) under a cooperative agreement with the U.S. National Science Foundation, on behalf of the Gemini partnership of Argentina, Brazil, Canada, Chile, the Republic of Korea, and the United States of America.
R.D. gratefully acknowledges support by the ANID BASAL project FB210003

For the near-UV, optical, and near-IR regimes, this work utilizes data from the S-PLUS collaboration \citep{MendesdeOliveira2019}, as well as VLT and HST observations reported by \citet{FernandezOntiveros2009}.

Additionally, this study is partially based on archival data obtained with the Spitzer Space Telescope, which was operated by the Jet Propulsion Laboratory, California Institute of Technology, under a contract with NASA.

For radio observations, this work includes the following ALMA data: : ADS/JAO.ALMA\#2017.1.00161.L and ADS/JAO.ALMA\#2018.1.00162.S. ALMA is a partnership of ESO (representing its member states), NSF (USA) and NINS (Japan), together with NRC (Canada), MOST and ASIAA (Taiwan), and KASI (Republic of Korea), in cooperation with the Republic of Chile. The Joint ALMA Observatory is operated by ESO, AUI/NRAO and NAOJ. This study also made use of (extended) Karl G. Jansky Very Large Array archival observations provided by the National Radio Astronomy Observatory (NRAO), a facility of the National Science Foundation operated under cooperative agreement by Associated Universities, Inc.

In constructing the SED fitting at 9$\arcsec$ resolution, this work incorporates observations from the ESA Herschel Space Observatory \citep{Pilbratt2010}, specifically using the PACS instrument \citep{Poglitsch2010}.
\end{acknowledgements}
    
\bibliographystyle{aa}
\bibliography{aanda} 


\begin{appendix}
    
\section{CIGALE SED modeling with AGN component}
\label{apen.CIGALE_AGN}

\begin{figure*}[htp!]
\setlength{\tabcolsep}{-9pt} 
\renewcommand{\arraystretch}{0} 
    \centering
    \begin{tabular}{ccc} 
        \includegraphics[width = 0.38\textwidth]{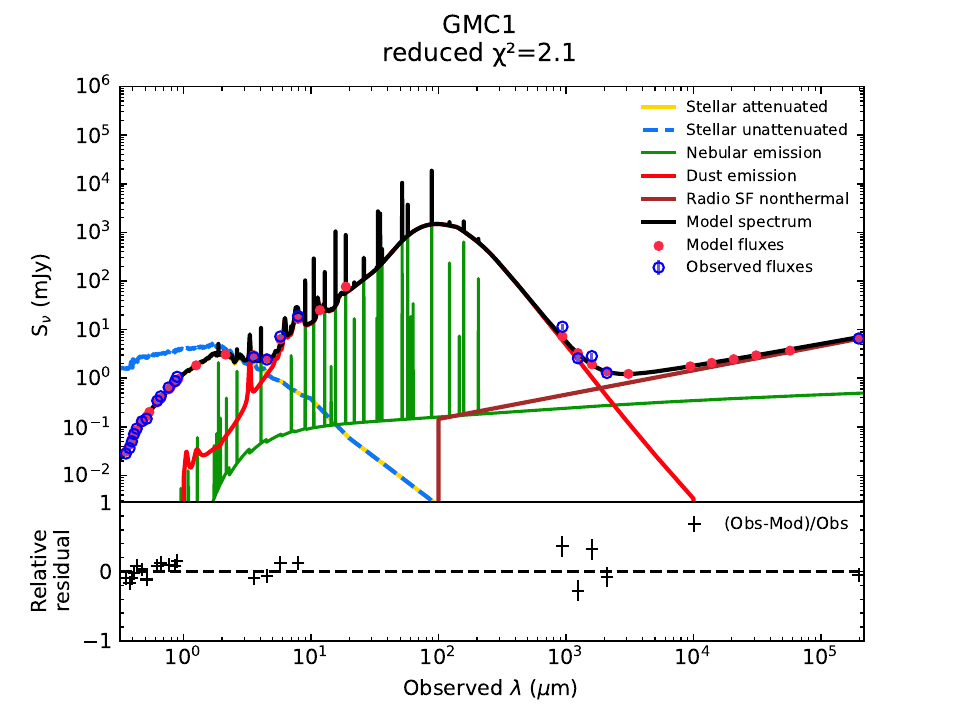} &
        \includegraphics[width = 0.38\textwidth]{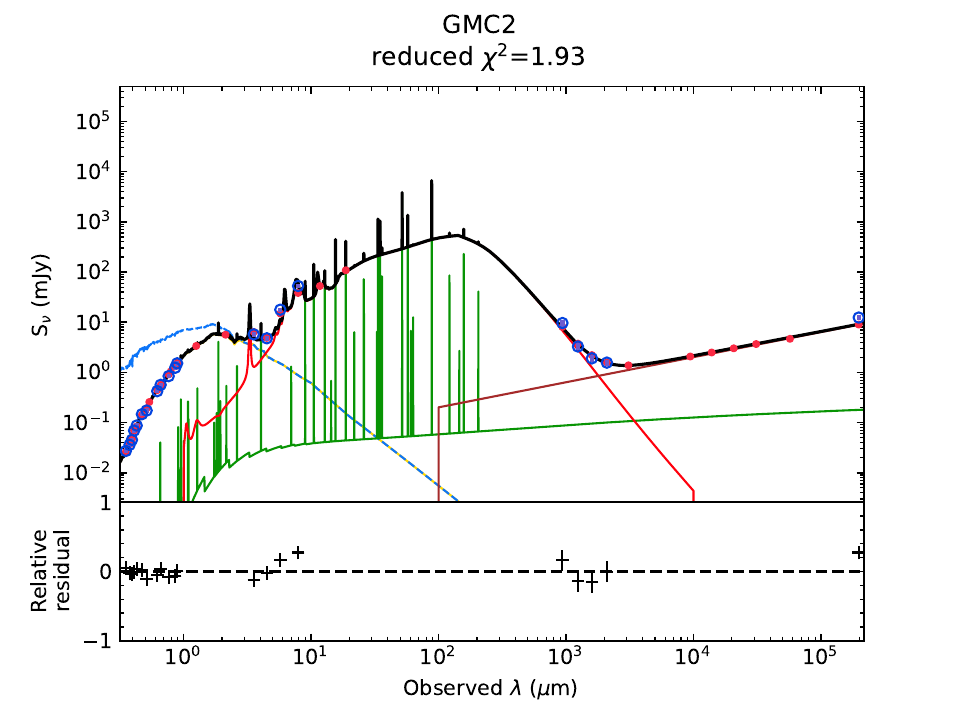} &
        \includegraphics[width = 0.38\textwidth]{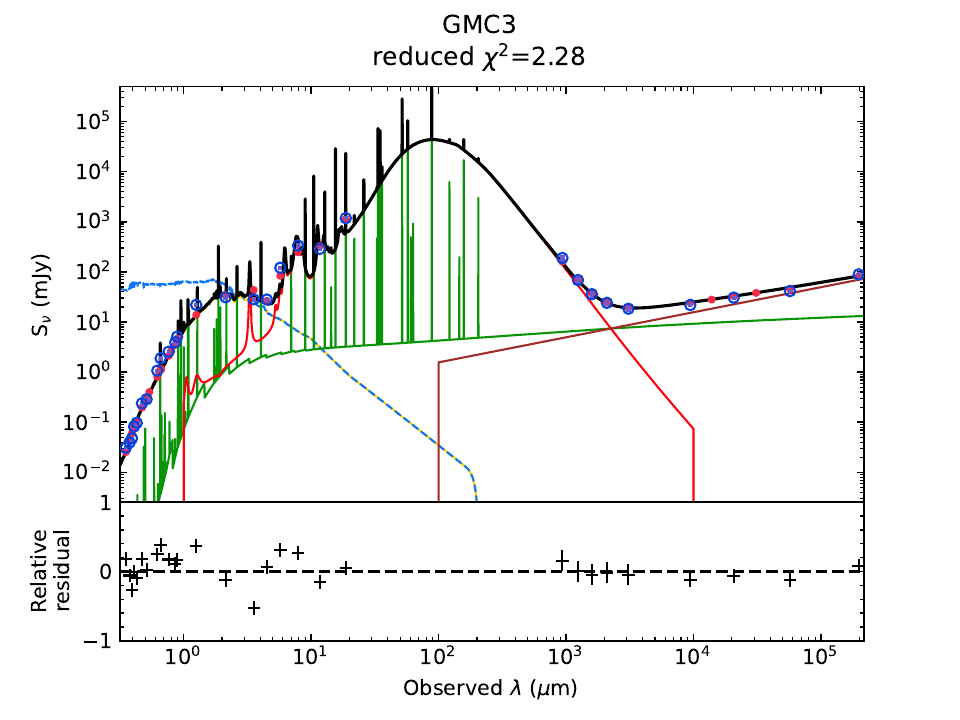} \\
        \includegraphics[width = 0.38\textwidth]{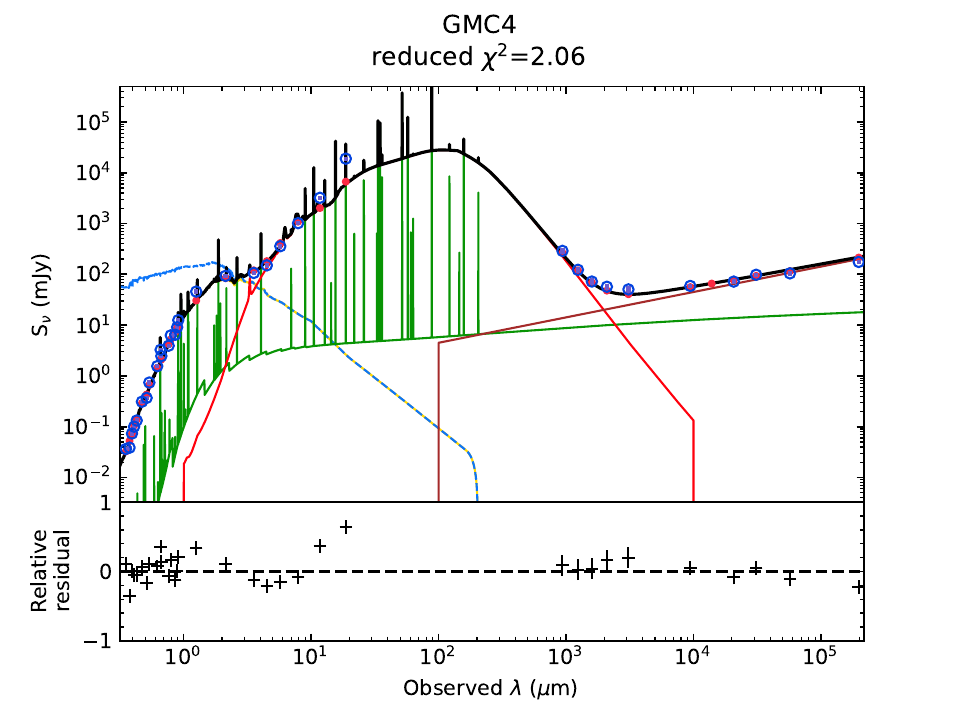} &
        \includegraphics[width = 0.38\textwidth]{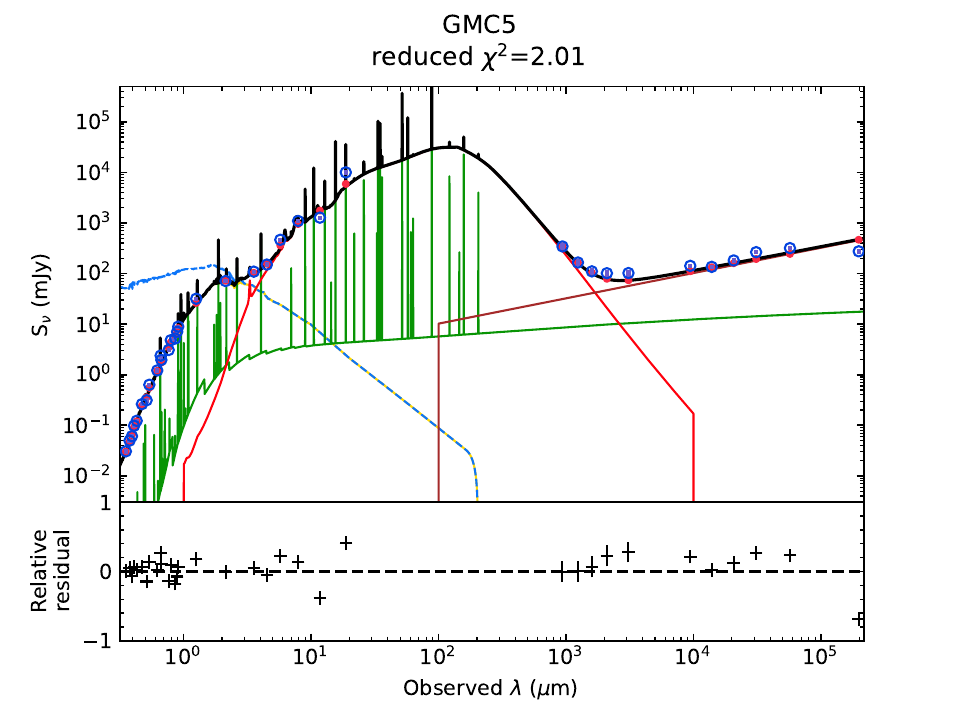} &
        \includegraphics[width = 0.38\textwidth]{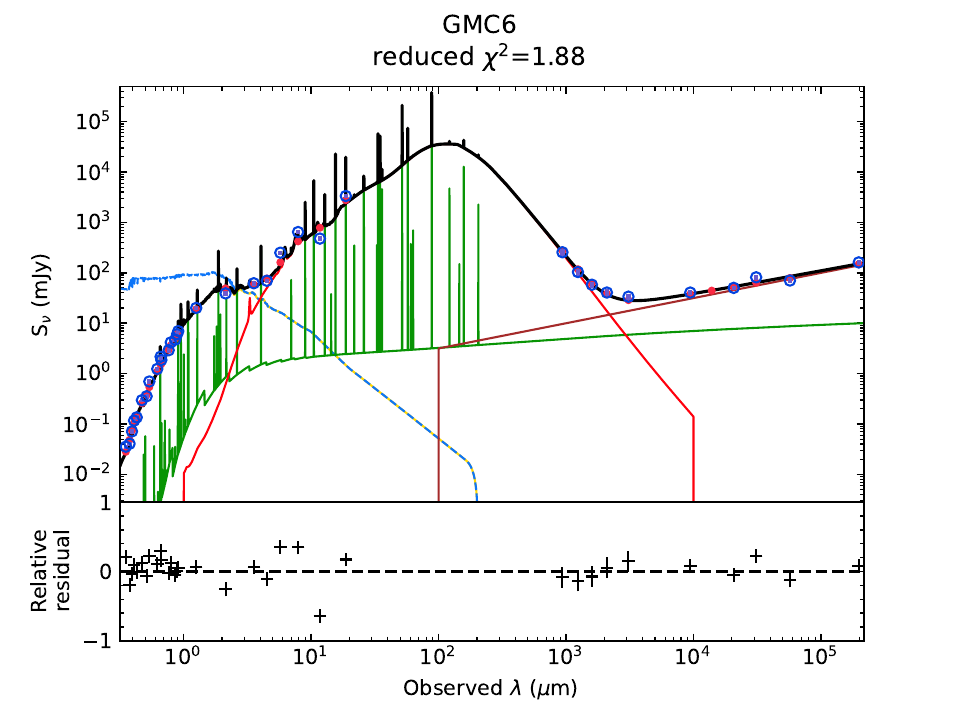} \\
        \includegraphics[width = 0.38\textwidth]{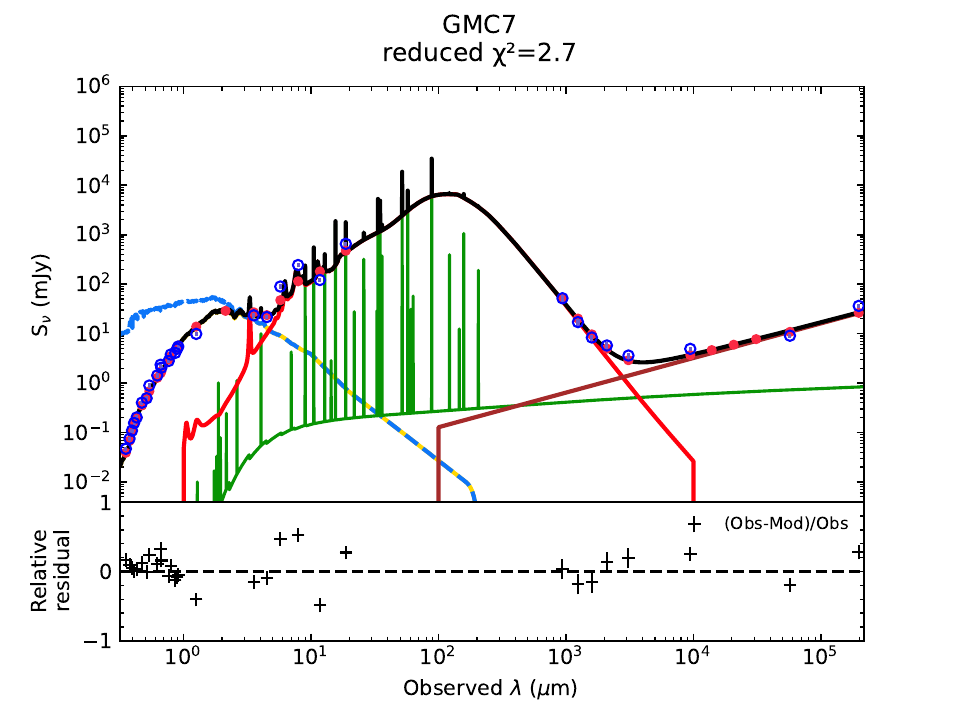} &
        \includegraphics[width = 0.38\textwidth]{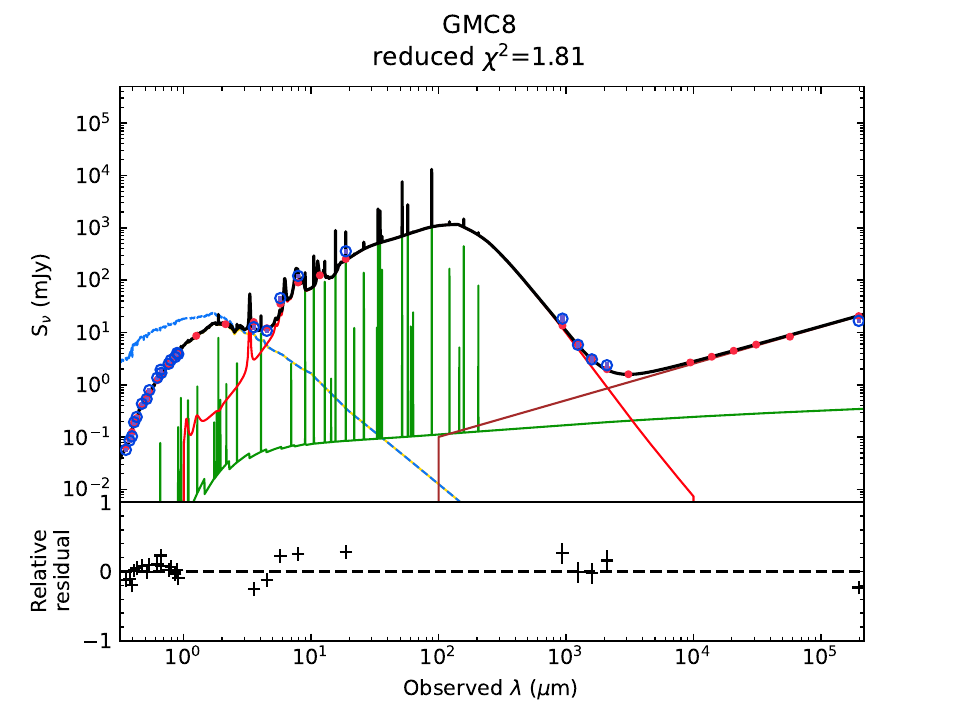} &
        \includegraphics[width = 0.38\textwidth]{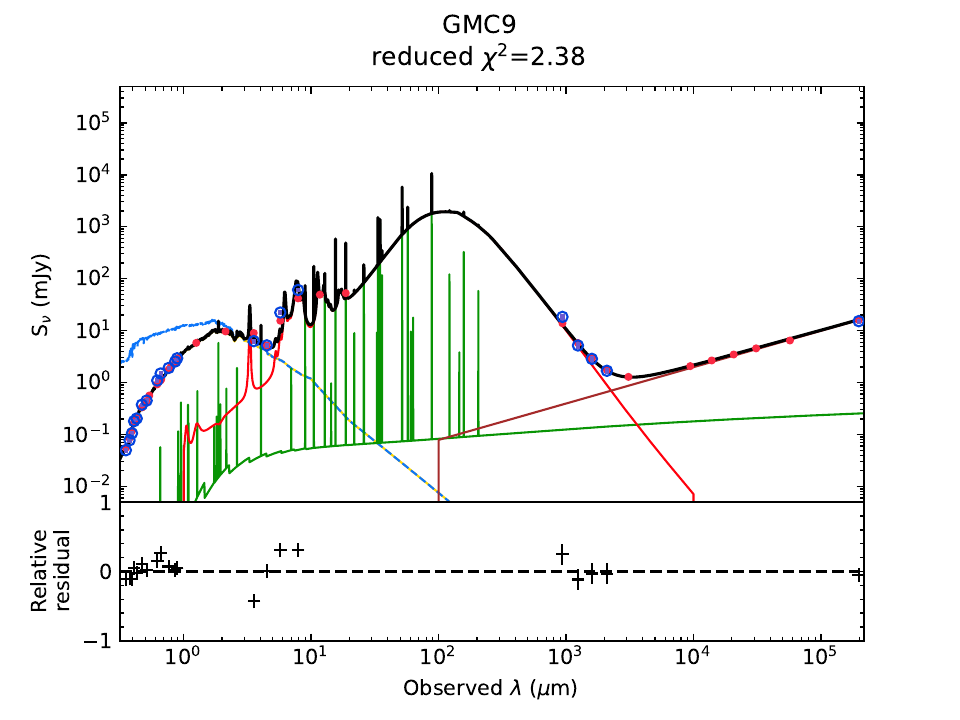} \\
        \includegraphics[width = 0.38\textwidth]{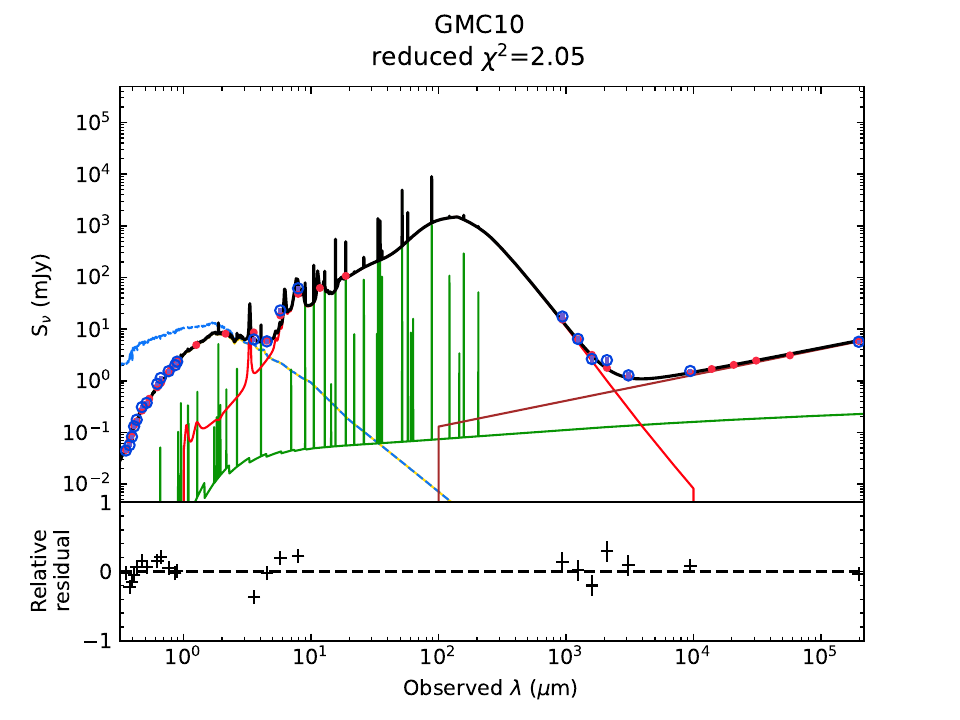} &
        \includegraphics[width = 0.38\textwidth]{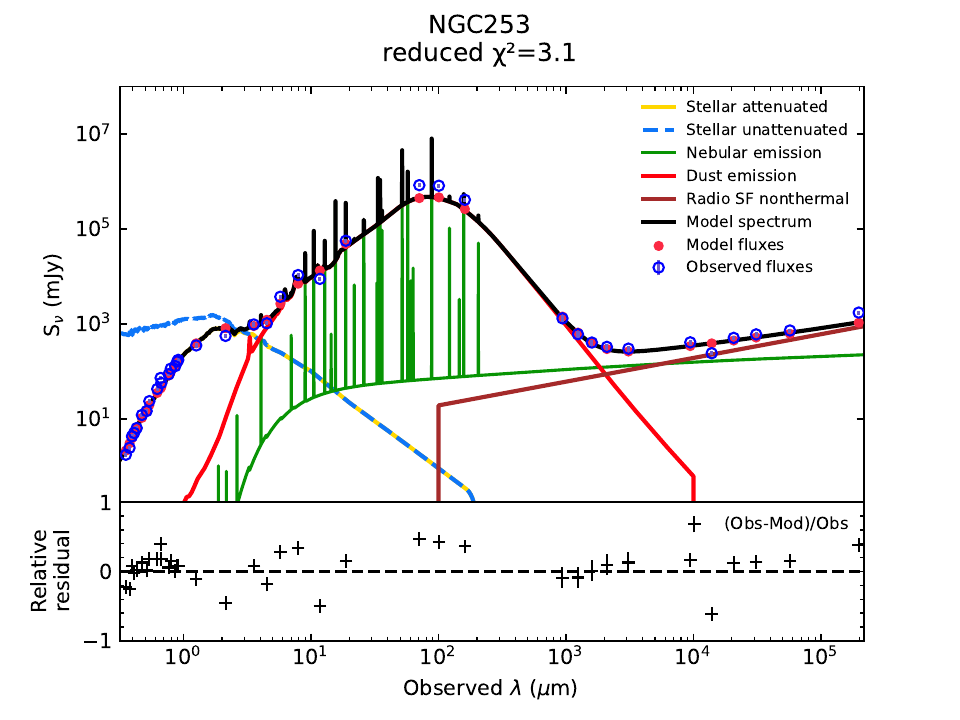} &        
        
    \end{tabular}
\caption{Best-fits CIGALE SEDs without an AGN component of the 10 GMCs studied here plus a 9\arcsec\ aperture SED for the nuclear region, also devoid of an AGN component, in the bottom middle panel. Open blue circles are the observed fluxes while filled red points indicate modeled fluxes whose observed values may or may not be fitted, depending on their availability (see Table~\ref{tab.phot_points}). The goodness of fit is estimated by the reduced $\chi^2$ shown at the top of each panel, along with the name and redshift of the GMC. In almost all cases, SEDs are well modeled, giving reasonable estimates of astrophysical properties. In Fig.~\ref{apen.fig:cigalefitting_AGN} we show our best-fits CIGALE models considering an AGN component.}\label{apen.fig:cigalefitting}
\end{figure*}

\begin{figure*}[htp!]
\setlength{\tabcolsep}{-9pt} 
\renewcommand{\arraystretch}{0} 
    \centering
    \begin{tabular}{ccc} 
        \includegraphics[width = 0.38\textwidth]{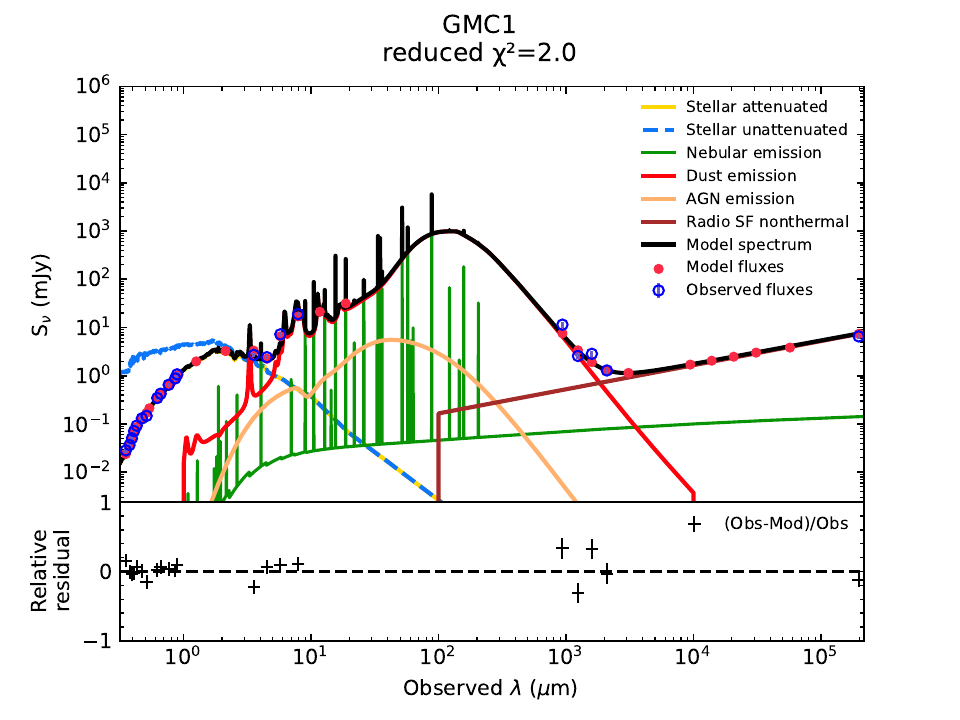} &
        \includegraphics[width = 0.38\textwidth]{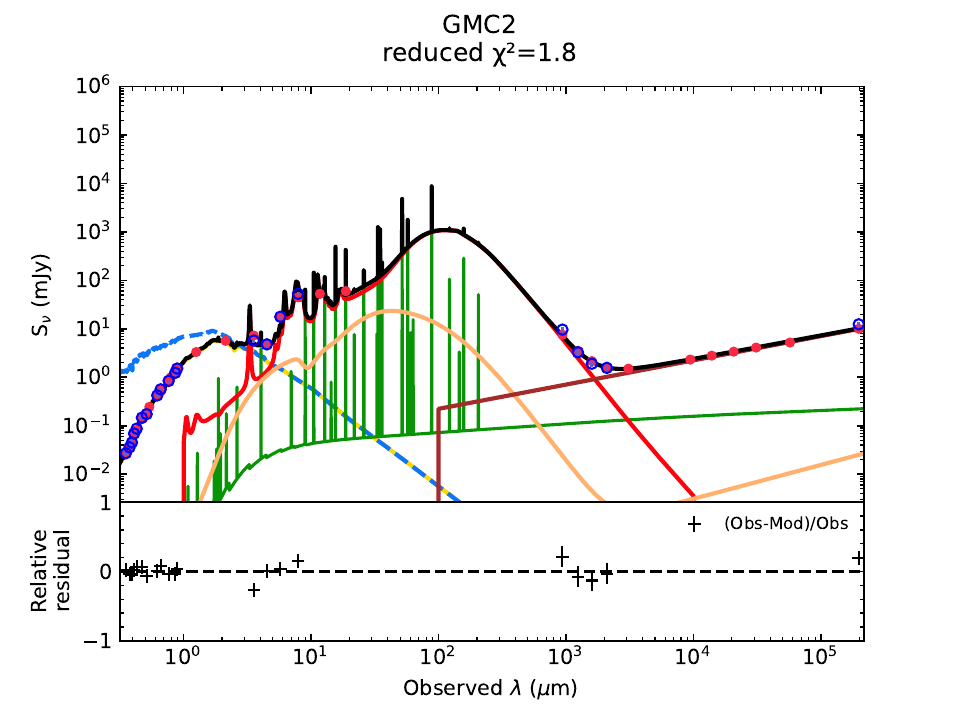} &
        \includegraphics[width = 0.38\textwidth]{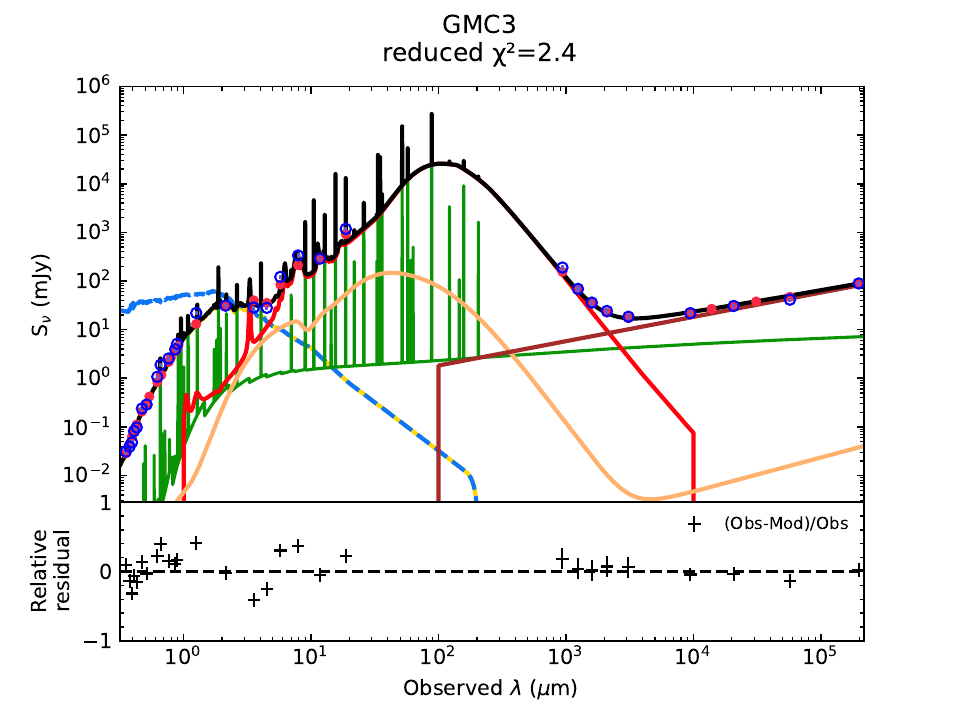} \\
        \includegraphics[width = 0.38\textwidth]{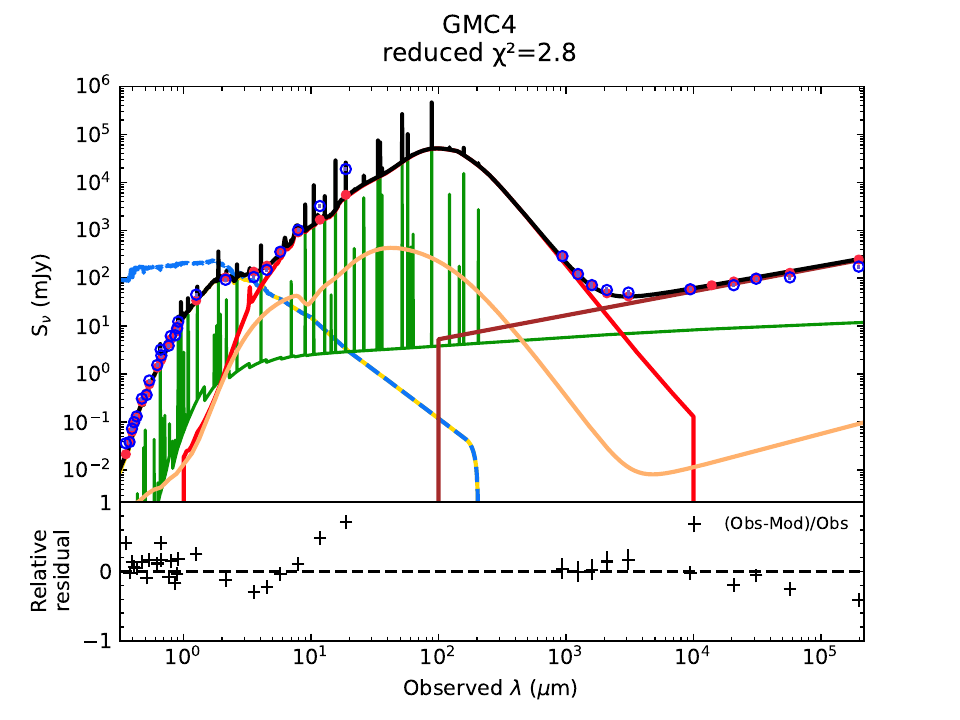} &
        \includegraphics[width = 0.38\textwidth]{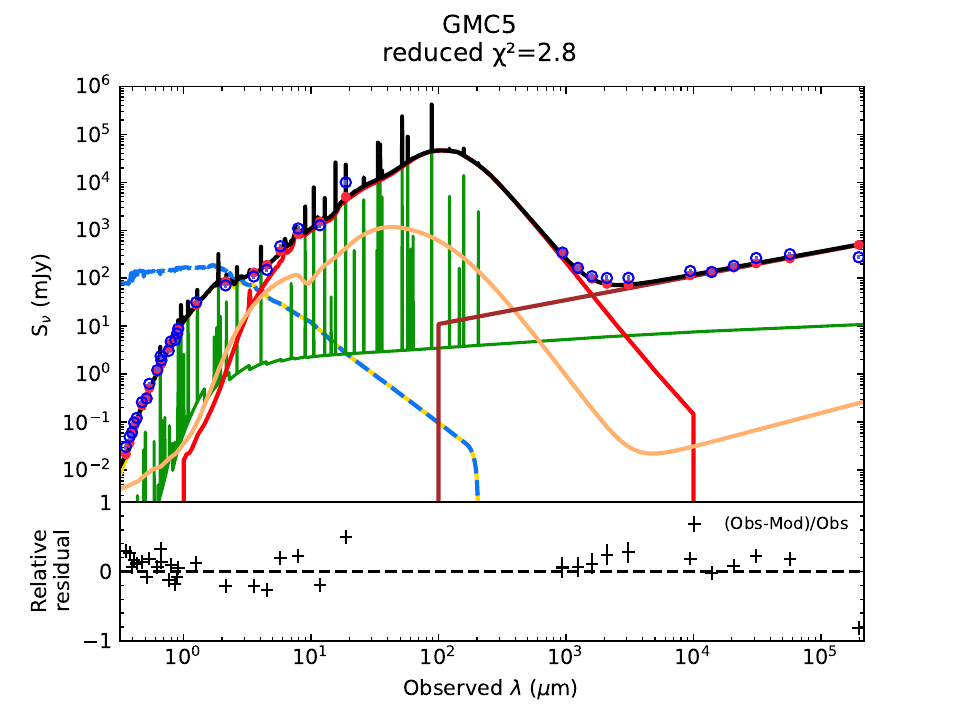} &
        \includegraphics[width = 0.38\textwidth]{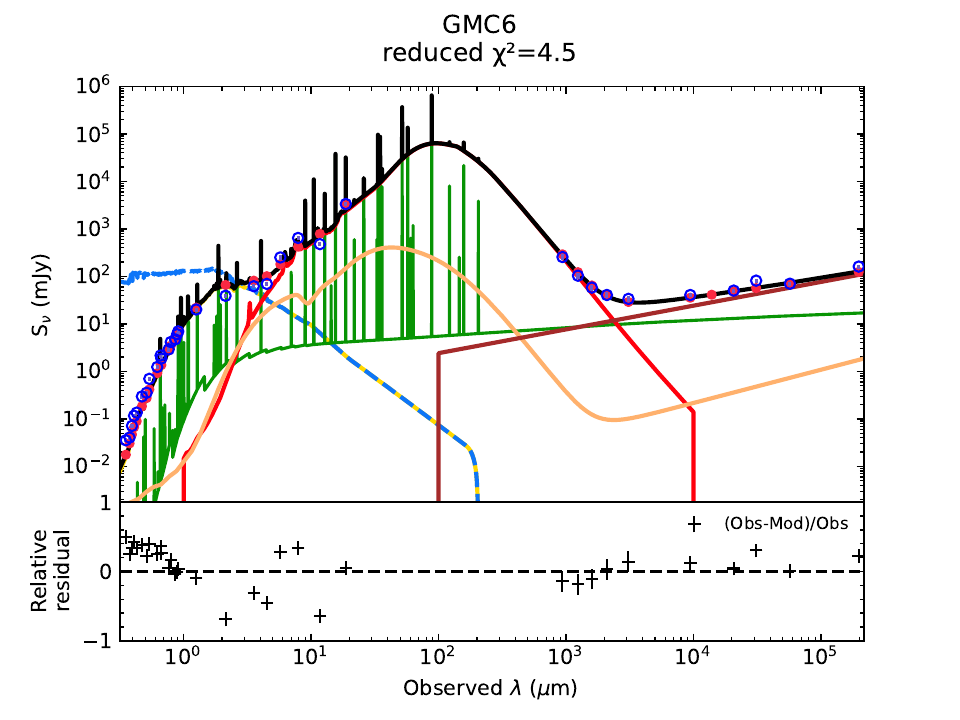} \\
        \includegraphics[width = 0.38\textwidth]{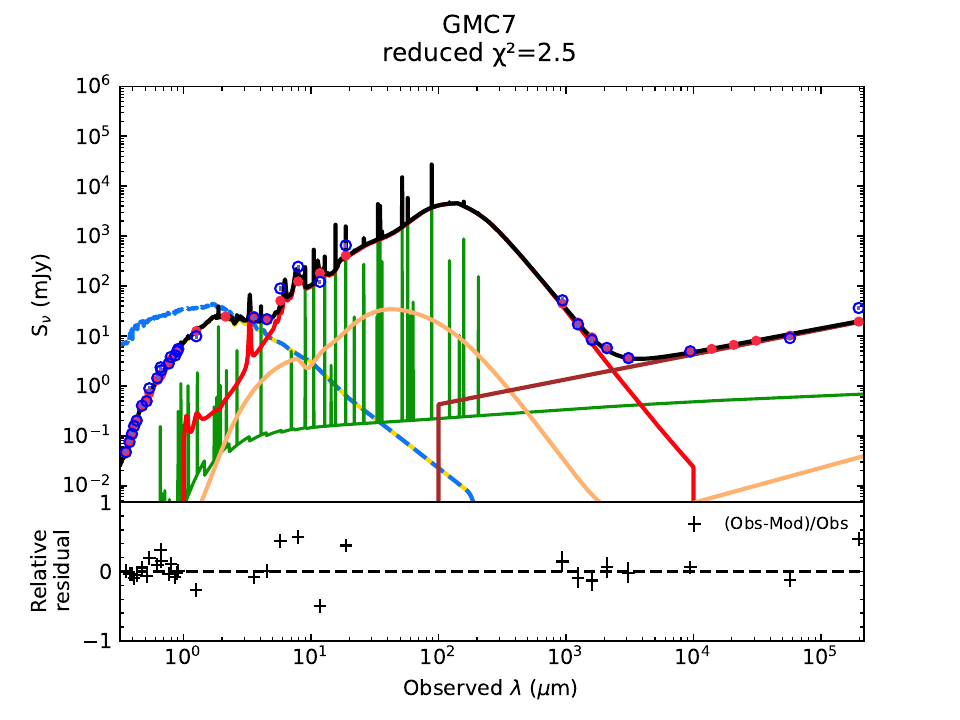} &
        \includegraphics[width = 0.38\textwidth]{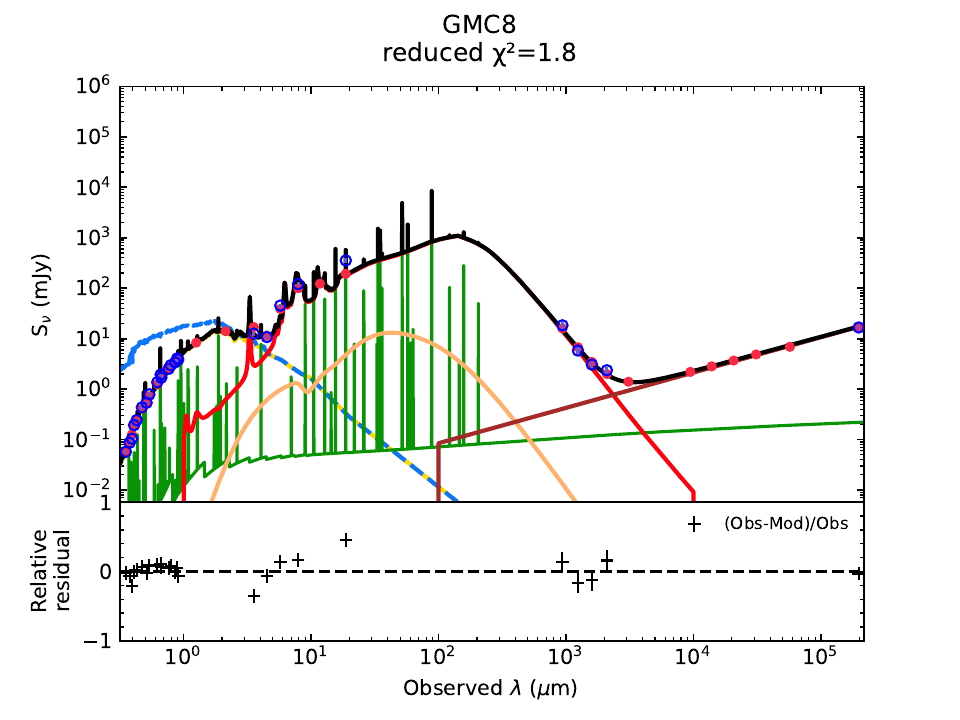} &
        \includegraphics[width = 0.38\textwidth]{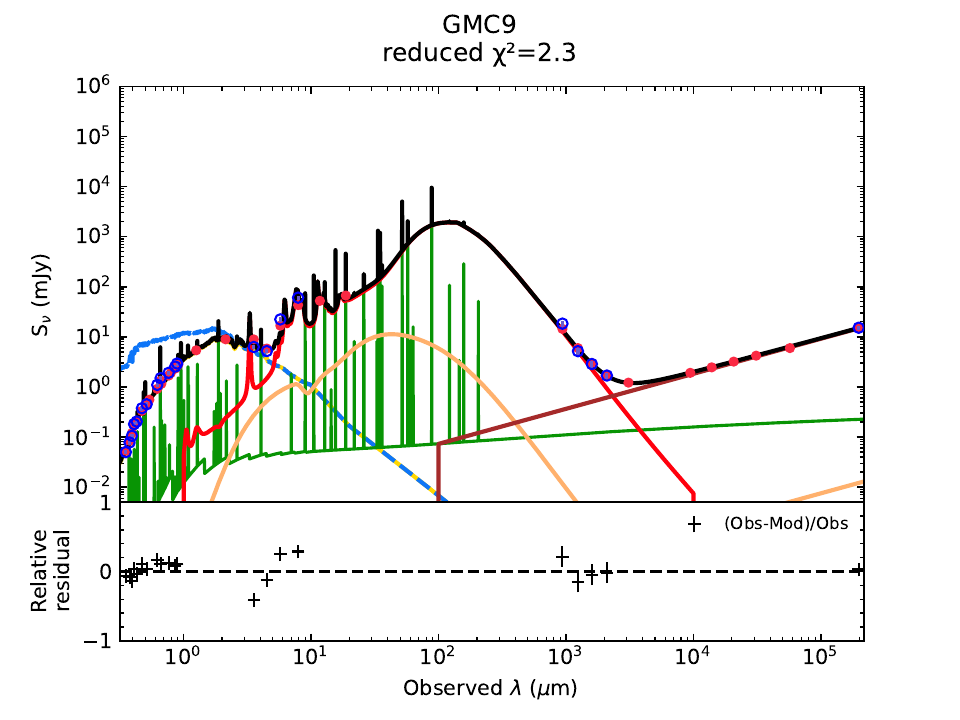} \\
        \includegraphics[width = 0.38\textwidth]{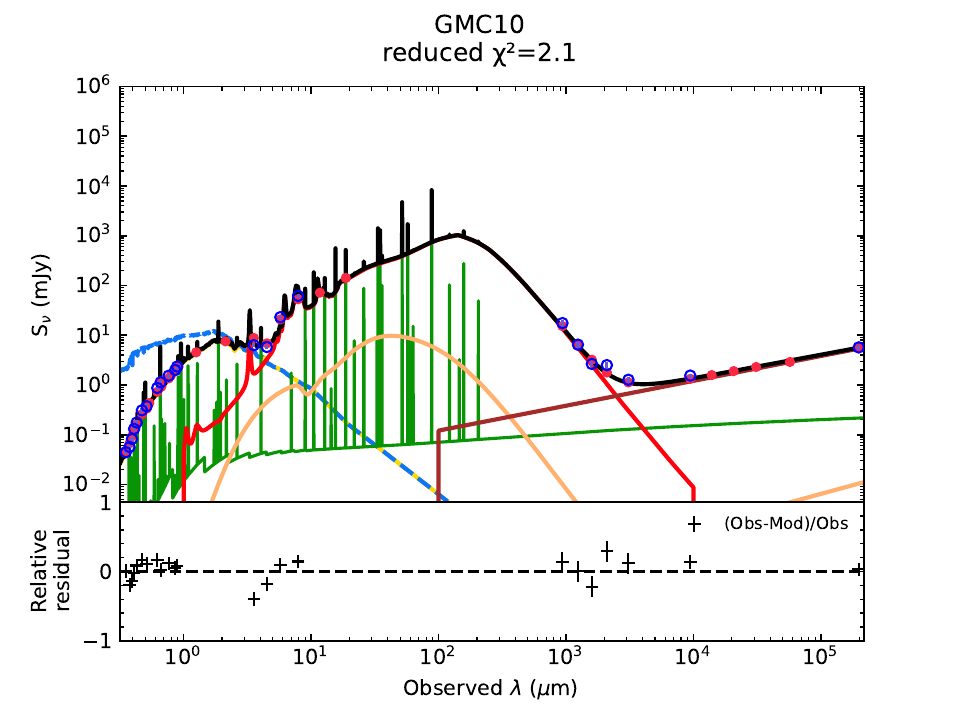} &
        \includegraphics[width = 0.38\textwidth]
        {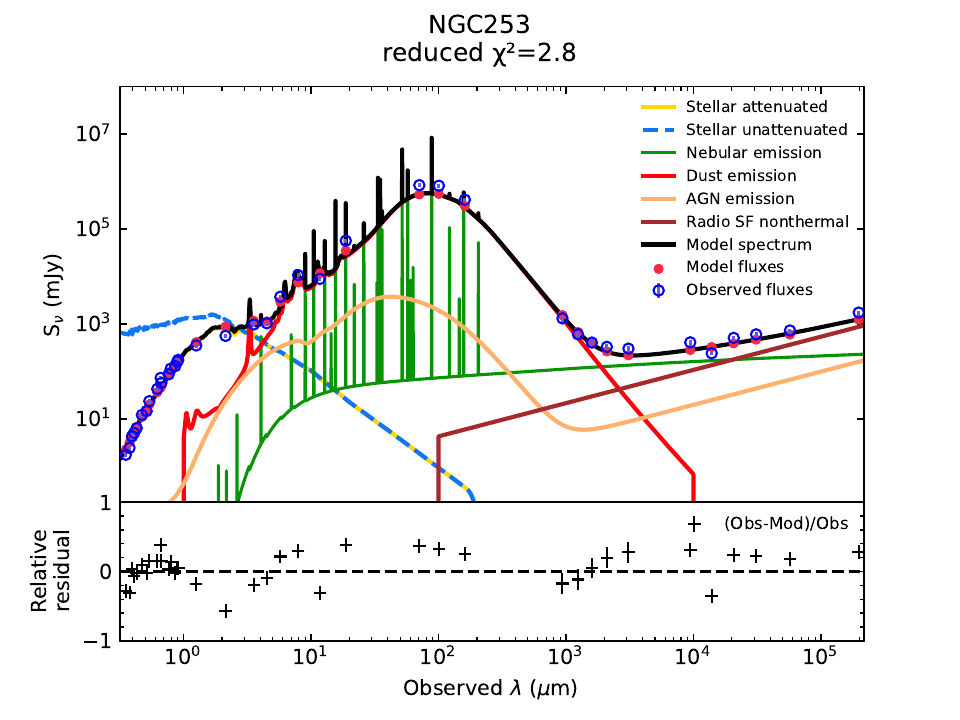}
    \end{tabular}
\caption{Best-fit CIGALE$+$AGN SEDs of the 10 GMCs studied here plus a 9\arcsec\ aperture SED for the nuclear region, also considering an AGN component, in the bottom middle panel. Open blue circles are the observed fluxes while filled red points indicate modeled fluxes whose observed values may or may not be fitted, depending on their availability (see Table~\ref{tab.phot_points}). The goodness of fit is estimated by the reduced $\chi^2$ shown at the top of each panel, along with the name and redshift of the GMC. In almost all cases, SEDs are well modeled, giving reasonable estimates of astrophysical properties.}\label{apen.fig:cigalefitting_AGN}
\end{figure*}

To assess potential AGN contributions, we incorporated the SKIRTOR module \citep{Stalevski2012, Stalevski2016}. SKIRTOR is based on the 3D radiative transfer code SKIRT \citep{Baes2011}. It includes obscuration by dusty torus and obscuration by dust settled along with the polar directions.
The 2022.1 version of CIGALE offers significant advancements over its predecessors, particularly through its enhanced radio module, which includes both thermal (free-free emission from nebular gas) and nonthermal (synchrotron emission from star formation and AGN activity) contributions. The updated module also incorporates a new AGN component, enabling us to estimate the radio loudness, $R_{\rm{AGN}}$, defined as the ratio of AGN luminosities measured at 5 GHz and 2500\AA. Additionally, it provides the slope of the AGN power-law (PL) radiation, which is assumed to be isotropic \citep[][]{Yang22}.\\
CIGALE 2022.1 models the radio luminosities from star formation and AGN components using the FIR/radio correlation coefficient (q$_{\rm IR}$) and radio loudness (R$_{\rm AGN}$) parameters in scaling relations. At 1.4 GHz, the radio luminosity from star formation is predominantly nonthermal synchrotron emission, normalized using the q$_{\rm IR}$ parameter. The q$_{\rm IR}$ determines the normalization at this frequency, while a single power-law describes the synchrotron emission from star formation. Variations in the q$_{\rm IR}$ parameter contribute to uncertainties in modeling the radio component with CIGALE. The nonthermal synchrotron emission from the AGN is also described using a single PL form governed by the R$_{\rm AGN}$ parameter. However, the galaxy-averaged radio SEDs of starburst galaxies are rarely represented by single-PL models, showing low-frequency turnover due to free-free absorption \citep[e.g.,][]{Galvin18, Dey2022,Dey2024}. It would be interesting to compare the physical conditions responsible for radio emission on the galaxy-wide scale to GMC scales, and that will be the aim of future studies using spatially-resolved radio SEDs of NGC\,253. The list of input parameters to CIGALE analysis without and with AGN component is given in Table~\ref{tab:sedpar}. For the CIGALE modeling including the AGN component, we only added the AGN module to the existing input parameters. 
These CIGALE inputs have yielded a mean reduced $\chi^2$ of 2.2 for the ten GMCs studied in this work. The resulting SEDs are shown in Figs. \ref{apen.fig:cigalefitting} (without an AGN component) and \ref{apen.fig:cigalefitting_AGN} (considering an AGN component). These Figs. also include, in the bottom middle panel each, the SED fitting centered at GMC~5 at 9$\arcsec$ aperture diameter, which allows to include Herschel PACS observations (see Appendix~\ref{apen.testing_at_9_arcsec}). The results on the main physical properties for each GMC are shown in Table \ref{Tab.CIGALE_main_results}, those regarding attenuation effects are also in Table~\ref{Apen.Tab.Extinction_Factors_CIGALE} while additional results can be found in Tables~\ref{Apen.Tab.Modified_Parameters_CIGALE}, and \ref{Apen.Tab.CIGALE_AGN}.

\label{apen.CIGALE_results}

\begin{table*}
\caption{List of input parameters for CIGALE modeling}\label{tab:sedpar}

  \begin{center}
 
    \begin{tabular}{l c c}
     \hline\hline
    \textbf{\textit{Parameters}} & & \textbf{\textit{Values}}\\
     \hline\hline
     \multicolumn{3}{c}{\textbf{delayed star formation history + additional burst} \citep{ciesla2015}}   \\
     \hline
     e-folding time of the main stellar population model [Myr] &$\tau_{\rm{main}}$ & 300-5000 by a bin of 300\\
    e-folding time of the late starburst population model [Myr] &$\tau_{\rm{burst}}$   & 50-500 by a bin of 50\\
     mass fraction of the late burst population & $f_{\rm{burst}}$ &  0.01, 0.03, 0.05, 0.1, 0.3, 0.6, 0.9\\
   Age of the main stellar population in the galaxy [Myr]   & age   & 1000, 2000, 3000, 4500, 5000, 6500, 10000, 12000\\
    Age of the late burst [Myr] & age$_{\rm{burst}}$  & 10-350 by a bin of 50\\
    \hline\hline
    \multicolumn{3}{c}{\textbf{stellar synthesis population} \citep{Bruzual2003}}   \\
    \hline
    Initial mass function & IMF & \citep{Salpeter1995}\\
    Metallicity & $Z$ &  0.02 \\
    Separation age &    &  1\,Myr\\
    \hline\hline
    \multicolumn{3}{c}{\textbf{dust attenuation laws} \citep{Calzetti2000}}   \\
    \hline
    \hline
    Colour excess of young stars & E(B-V) & 0.1-2 by a bin of 0.2 \\
    Reduction factor$^{(iii)}$ & $f_{\rm{att}}$ & 0.3, 0.44, 0.6, 0.7 \\
    \hline
    \hline
    \multicolumn{3}{c}{\textbf{dust emission} \citep{Draine2014}}   \\
    \hline
    Mass fraction of PAH  & $q_{\rm{PAH}}$ &  0.47, 1.77, 2.50, 3.19, 4.58, 5.26, 6.63, 7.32 \\ 
    Minimum radiation field & $U_{\rm{min}}$ & 0.4, 0.8, 1.2, 1.6, 2.0, 2.5, 4, 8, 15 \\ 
    power law index of the radiation & $\alpha$ & 2 \\
    fraction illuminated from $U_\mathrm{min}$ to $U_\mathrm{max}$ & $\gamma$ & 0.02, 0.06, 0.1, 0.15, 0.2 \\
    \hline\hline
    \multicolumn{3}{c}{\textbf{active nucleus model; Skirtor} \citep{Stalevski2012,Stalevski2016}} \\
    \hline
    optical depth at 9.7$~\mu$m & $\tau_{9.7}$ & 3.0, 7.0 \\
    torus density radial parameter  & pl & 1.0 \\
    torus density angular parameter & q & 1.0 \\
    angle between the equatorial plan and edge of the torus [deg] & oa & 40 \\
    ratio of outer to inner radius & R & 20 \\
    fraction of total dust mass inside clumps [\%] & Mcl & 97 \\
    inclination (viewing angle) [deg] & i & 30 (type 1), 70 (type 2) \\
    AGN fraction & & 0.0-0.4 by a bin of 0.05\\
    extinction law of polar dust && SMC \\
    E(B-V) of polar dust && 0.01-0.7 by a bin of 0.5\\
    Temperature of the polar dust & K & 100 \\
    Emissivity index of the polar dust && 1.6 \\
   \hline \hline
   \multicolumn{3}{c}{\textbf{Radio}}   \\
    \hline
    SF FIR/radio parameter$^*$ & $q_{\rm{IR}}$ &  1.3 - 2.9 by a bin of 0.5 \\
    SF Power-law slope (Flux $\propto$ Frequency$^{\alpha_{synch}}$) & $\alpha_{\rm{SF}}$ & $-$2.0 to $-$0.2 by a bin of 0.5 \\
    Radio-Loudness parameter$^{**}$ & $R_{\rm{AGN}}$ & 0.01, 0.05, 0.1, 0.5... 1000, 5000\\
    AGN Power-law slope & $\alpha_{\rm{AGN}}$ & $-${2.0} to $-$0.2 by a bin of 0.5 \\
    \hline
    \end{tabular}
    \end{center}
    $^*$ computed as $\log_{10}L_{\rm{IR}(8-1000\mu\rm m)}$ -$\log_{10}L_{1.4\,\rm{GHz}}$, where $L_{1.4\,\rm{GHz}}$ is the radio luminosity at 1.4\,GHz.\\
    $^{**}$ defined as a ratio of AGN luminosities measured at 5 GHz and 2500\AA.
    \end{table*}

\begin{table*}[htbp]
\centering
\small
\caption{The extinction-related results obtained from SED modeling with CIGALE, including the AGN component.}
\begin{adjustbox}{max width=\textheight}    
\setlength{\tabcolsep}{0.1cm}
\begin{tabular}{c c c c c c}
\hline \hline
GMC & V\_B90 & B\_B90 & E\_BV\_factor & E\_BV\_lines & E\_BVs \\ 
      & [mag]  & [mag] & [mag]         & [mag]        & [mag]  \\ \hline
1  & $2.87 \pm 0.04$ & $3.58 \pm 0.05$ & $0.10 \pm 0.00$ & $7.00 \pm 0.09$ & $0.70 \pm 0.01$ \\ 
2  & $3.00 \pm 0.17$ & $3.70 \pm 0.21$ & $0.33 \pm 0.33$ & $5.00 \pm 2.80$ & $0.73 \pm 0.05$ \\ 
3  & $6.14 \pm 0.04$ & $6.20 \pm 0.02$ & $0.40 \pm 0.00$ & $3.00 \pm 0.00$ & $1.20 \pm 0.00$ \\ 
4  & $6.10 \pm 0.01$ & $7.60 \pm 0.01$ & $0.45 \pm 0.09$ & $3.50 \pm 0.90$ & $1.50 \pm 0.00$ \\ 
5  & $6.10 \pm 0.01$ & $7.60 \pm 0.01$ & $0.45 \pm 0.09$ & $3.50 \pm 0.90$ & $1.50 \pm 0.00$ \\ 
6  & $4.90 \pm 0.01$ & $6.10 \pm 0.01$ & $0.40 \pm 0.00$ & $3.00 \pm 0.00$ & $1.20 \pm 0.00$ \\ 
7  & $3.20 \pm 0.01$ & $4.00 \pm 0.01$ & $0.80 \pm 0.01$ & $1.00 \pm 0.05$ & $0.80 \pm 0.01$ \\ 
8  & $3.00 \pm 0.17$ & $3.70 \pm 0.21$ & $0.29 \pm 0.31$ & $5.40 \pm 2.70$ & $0.73 \pm 0.04$ \\ 
9  & $2.80 \pm 0.11$ & $3.50 \pm 0.14$ & $0.10 \pm 0.02$ & $7.00 \pm 0.34$ & $0.70 \pm 0.03$ \\ 
10 & $2.90 \pm 0.03$ & $3.60 \pm 0.04$ & $0.10 \pm 0.05$ & $7.00 \pm 0.41$ & $0.70 \pm 0.01$ \\ 
\hline
\end{tabular}
\end{adjustbox}
\tablefoot{The columns represent the GMC number, $A_{\rm{V}}^{\rm{Balmer}}$ as presented in Table~\ref{tab:attenuation_comparison}, the slope of the attenuation curve, the scaling factor between E(B-V) for lines and continuum, the color excess for nebular emission lines, and the color excess for the stellar continuum.}
\label{Apen.Tab.Extinction_Factors_CIGALE}
\end{table*}

\begin{table*}[htbp]
\centering
\small
\caption{Dust and SFH related results obtained from SED modeling with CIGALE, including AGN component.}
\begin{adjustbox}{max width=\textheight}    
\setlength{\tabcolsep}{0.1cm}
\begin{tabular}{c c c c c c c c c c}
\hline \hline
GMC & $M_{\rm dust}$ & U$_{\rm min}$ & $\gamma$ & $q_{\rm PAH}$ & age$_{\rm burst}$ & $f_{\rm burst}$ & $\tau_{\rm burst}$ & $\tau_{\rm main}$ \\ 
     & [log$_{10}$ ($M_{\odot}$)] &       &       &      & [Myr]    &          & [Myr]    & [Myr]     \\ \hline
1    & $4.47 \pm 0.09$ & $4.39 \pm 1.51$ & $0.025 \pm 0.033$ & $2.236 \pm 0.590$ & $7.87 \pm 99.45$ & $0.03 \pm 0.02$ & $142.23 \pm 85.13$ & $1170.23 \pm 645.23$ \\ 
2    & $4.49 \pm 0.09$ & $3.06 \pm 1.84$ & $0.057 \pm 0.067$ & $4.821 \pm 1.218$ & $8.10 \pm 32.48$ & $0.02 \pm 0.01$ & $210.26 \pm 75.10$ & $290.93 \pm 141.61$ \\ 
3    & $5.89 \pm 0.07$ & $3.25 \pm 0.72$ & $0.011 \pm 0.010$ & $1.118 \pm 0.036$ & $7.74 \pm 98.75$ & $0.03 \pm 0.02$ & $126.62 \pm 81.91$ & $864.58 \pm 527.05$ \\ 
4    & $6.19 \pm 0.04$ & $0.68 \pm 0.16$ & $0.089 \pm 0.038$ & $0.470 \pm 0.001$ & $8.16 \pm 8.01$  & $0.01 \pm 0.00$ & $246.94 \pm 60.90$ & $203.13 \pm 30.50$ \\ 
5    & $6.23 \pm 0.09$ & $0.90 \pm 0.39$ & $0.078 \pm 0.025$ & $0.471 \pm 0.026$ & $7.92 \pm 0.01$  & $0.01 \pm 0.02$ & $180.44 \pm 84.63$ & $213.33 \pm 61.83$ \\ 
6    & $6.08 \pm 0.06$ & $0.58 \pm 0.34$ & $0.189 \pm 0.033$ & $1.036 \pm 0.312$ & $7.93 \pm 101.79$ & $0.03 \pm 0.02$ & $170.89 \pm 86.57$ & $1520.01 \pm 683.09$ \\ 
7    & $5.13 \pm 0.11$ & $9.03 \pm 1.47$ & $0.192 \pm 0.029$ & $3.116 \pm 0.871$ & $7.68 \pm 93.06$ & $0.03 \pm 0.02$ & $134.24 \pm 83.65$ & $844.18 \pm 544.03$ \\ 
8    & $4.79 \pm 0.21$ & $2.65 \pm 1.02$ & $0.160 \pm 0.053$ & $3.384 \pm 1.132$ & $7.75 \pm 88.29$ & $0.02 \pm 0.01$ & $176.04 \pm 86.66$ & $258.82 \pm 125.10$ \\ 
9    & $4.73 \pm 0.23$ & $5.37 \pm 1.72$ & $0.012 \pm 0.014$ & $3.105 \pm 0.838$ & $7.71 \pm 93.91$ & $0.03 \pm 0.02$ & $137.08 \pm 83.70$ & $592.04 \pm 305.75$ \\ 
10   & $4.83 \pm 0.19$ & $1.87 \pm 0.71$ & $0.051 \pm 0.042$ & $3.412 \pm 0.918$ & $7.82 \pm 103.06$ & $0.02 \pm 0.02$ & $170.82 \pm 86.23$ & $589.30 \pm 352.26$ \\ 
\hline
\end{tabular}
\end{adjustbox}
\tablefoot{From left to right columns are: the GMC number, the dust mass ($M_{\rm dust}$), the minimum radiation field (U$_{\rm min}$), the fraction illuminated from U$_{\rm min}$ to the maximum radiation field ($\gamma$), the mass fraction of PAH ($q_{\rm PAH}$), the age of the late burst (age$_{\rm burst}$), the mass fraction of late burst population ($f_{\rm burst}$), the e-folding time of the late starburst population model ($\tau_{\rm burst}$) and that of the main stellar population model ($\tau_{\rm main}$).}
\label{Apen.Tab.Modified_Parameters_CIGALE}
\end{table*}

\begin{table*}[htbp]
\centering
\small
\caption{Results of AGN and radio SED modeling using CIGALE.}
\begin{adjustbox}{max width=\textheight}    
\setlength{\tabcolsep}{0.1cm}
\begin{tabular}{c c c c c c c c c c}
\hline \hline
GMC & $f_{\rm{AGN}}$ & accretion power & $L_{\rm disk}$ & polar dust luminosity & $R_{\rm AGN}$ & $\alpha_{\rm AGN}$ & $\alpha_{\rm SF}$ & $q_{\rm IR}$ \\ 
    & [\%]    & [log$_{10}$ erg~s$^{-1}$] & [log$_{10}$ erg~s$^{-1}$] & [log$_{10}$ erg~s$^{-1}$] &       &            &           &          \\ \hline
1   & $2.52 \pm 1.37$ & $32.70 \pm 0.24$ & $30.40 \pm 0.24$ & $31.97 \pm 0.24$ & $0.04 \pm 0.04$ & $0.70 \pm 0.00$ & $0.56 \pm 0.11$ & $2.20 \pm 0.03$ \\ 
2   & $3.86 \pm 2.56$ & $33.01 \pm 0.30$ & $30.70 \pm 0.31$ & $32.28 \pm 0.30$ & $0.04 \pm 0.04$ & $0.70 \pm 0.00$ & $0.59 \pm 0.12$ & $2.20 \pm 0.00$ \\ 
3   & $2.00 \pm 0.01$ & $34.01 \pm 0.02$ & $31.56 \pm 0.10$ & $33.28 \pm 0.02$ & $0.04 \pm 0.04$ & $0.70 \pm 0.00$ & $0.51 \pm 0.04$ & $2.21 \pm 0.05$ \\ 
4   & $2.05 \pm 0.46$ & $34.48 \pm 0.10$ & $32.01 \pm 0.10$ & $33.75 \pm 0.10$ & $0.01 \pm 0.01$ & $0.70 \pm 0.00$ & $0.50 \pm 0.00$ & $2.20 \pm 0.00$ \\ 
5   & $5.34 \pm 1.48$ & $34.87 \pm 0.12$ & $32.40 \pm 0.12$ & $34.13 \pm 0.12$ & $0.01 \pm 0.01$ & $0.70 \pm 0.00$ & $0.50 \pm 0.00$ & $2.20 \pm 0.00$ \\ 
6   & $4.66 \pm 2.68$ & $34.31 \pm 0.26$ & $31.92 \pm 0.24$ & $33.57 \pm 0.26$ & $0.04 \pm 0.04$ & $0.70 \pm 0.00$ & $0.50 \pm 0.00$ & $2.20 \pm 0.01$ \\ 
7   & $7.54 \pm 2.85$ & $33.79 \pm 0.18$ & $31.47 \pm 0.25$ & $33.06 \pm 0.18$ & $0.04 \pm 0.04$ & $0.70 \pm 0.00$ & $0.69 \pm 0.11$ & $2.64 \pm 0.08$ \\ 
8   & $2.56 \pm 1.54$ & $33.18 \pm 0.24$ & $30.85 \pm 0.25$ & $32.45 \pm 0.24$ & $0.04 \pm 0.04$ & $0.70 \pm 0.00$ & $0.73 \pm 0.12$ & $2.22 \pm 0.09$ \\ 
9   & $2.17 \pm 0.93$ & $32.95 \pm 0.14$ & $30.62 \pm 0.22$ & $32.22 \pm 0.14$ & $0.04 \pm 0.04$ & $0.70 \pm 0.00$ & $0.79 \pm 0.11$ & $2.20 \pm 0.01$ \\ 
10  & $2.51 \pm 1.42$ & $32.97 \pm 0.26$ & $30.64 \pm 0.25$ & $32.23 \pm 0.26$ & $0.04 \pm 0.04$ & $0.70 \pm 0.00$ & $0.50 \pm 0.00$ & $2.58 \pm 0.01$ \\ 
\hline
\end{tabular}
\end{adjustbox}
\tablefoot{From left to right, the columns represents the GMC number, AGN fraction ($f_{\rm{AGN}}$), accretion power, disk luminosity ($L_{\rm disk}$), polar dust luminosity of AGN, radio loudness ($R_{\rm AGN}$), AGN power-law slope ($\alpha_{\rm AGN}$), SF power-law slope ($\alpha_{\rm SF}$), and radio-IR correlation parameter ($q_{\rm IR}$).}
\label{Apen.Tab.CIGALE_AGN}
\end{table*}

\section{Individual GMC properties}
\label{apen.individual_GMCs}

GMC~1: Located at the boundaries of the CMZ, where x$_{1}$/x$_{2}$ interactions are present, GMC~1 traces weak shocks at low-frequency transitions \citep{Humire2022,Harada2022,Huang2023} and strong shocks at high-$J$ frequency transitions. Indeed, the strongest SiO(5--4)/HNCO(10$_{0,10}-9_{0,9}$) ratios are observed at this position \citep{Humire2022}. The whole picture may also be that weak and strong shocks are actually mixed over the entire CMZ \citep{Bao2024}. The shocked environment results in some of the more prominent Class~I methanol masers (MMCIs) in the $J_{-1}\rightarrow(J-$ 1)$_{0}-E$ and $J_{0}\rightarrow(J-$ 1)$_{1}-A$ line series \citep{Ellingsen2017,Gorski2019,Humire2022}. Devoid of a strong continuum, but showing the strongest maser over LTE emission \citep[][their Fig.~13]{Humire2022}, GMC~1 appears to be an extragalactic example of what is seen in the Milky Way object GMC G$+$0.693 \citep{Zeng2020}. In line with the latter, this cloud possesses the lowest amount of dust and stellar masses in the sample, according to GalaPy (see Table~\ref{Tab.SED_outputs_GalaPy}).

Contrary to the others GMCs in our sample, GMC~1 is the only one whose SED fitting could also deliver ages of the order of 10$^{5}$ years. However, since we observe CO emission, a molecule that needs a few million years to be produced \citep[e.g.,][]{Walmsley1989}, a permitted range of 10$^{6-11}$ or 10$^{6-12}$ years was imposed to all GMCs. This restriction yielded SED shapes that are closer to the observed ones compared to those with an age of $\sim$10$^{5}$~years, and significantly improved the results at millimeter and centimeter wavelengths by better reproducing the observed flux densities.

\begin{table*}[htbp]
\centering
\caption{Gas, star-formation (SF) related parameters and dust-related parameters for all SED results derived from GalaPy.}
\label{Tab.SED_outputs_GalaPy}
\begin{tabular}{llllllll}
\hline \hline
GMC & $T_{\rm{MC}}$ & $M_{\rm{gas}}$ & ism.R\_MC & $Z_{\rm{gas}}$ & sfh.psi\_max & ism.f\_MC & ism.tau\_esc \\
\hline
   & [K] & [log$_{10}$ ($M_{\odot}$)] & [log$_{10}$ (pc)] & [$\times 10^{-3}$] & [log$_{10}$ ($M_{\odot}$ yr$^{-1}$)] &  & [log$_{10}$ (yrs)] \\
\hline
1  & 50.365$_{-6.532}^{+11.259}$ & 8.573$_{-0.173}^{+0.095}$ & 0.90$_{-0.48}^{+0.60}$ & 0.53$_{-0.12}^{+0.27}$ & -1.69$_{-0.24}^{+0.23}$ & 0.01$_{-0.01}^{+0.01}$ & 7.75$_{-0.43}^{+0.34}$ \\
2  & 67.295$_{-4.749}^{+6.536}$  & 7.126$_{-0.119}^{+0.129}$ & 1.02$_{-0.13}^{+0.13}$ & 4.11$_{-0.16}^{+0.20}$ & -1.67$_{-0.06}^{+0.07}$ & 0.02$_{-0.01}^{+0.01}$ & 9.19$_{-0.26}^{+0.17}$ \\
3  & 66.739$_{-6.464}^{+7.394}$  & 8.010$_{-0.439}^{+0.260}$ & 0.79$_{-0.18}^{+0.20}$ & 8.77$_{-0.94}^{+2.50}$ & -0.54$_{-0.17}^{+0.38}$ & 0.04$_{-0.02}^{+0.09}$ & 8.83$_{-1.96}^{+0.29}$ \\
4  & 61.899$_{-10.657}^{+22.586}$ & 6.925$_{-0.124}^{+0.135}$ & 1.27$_{-0.40}^{+0.41}$ & 25.38$_{-2.09}^{+3.44}$ & 1.15$_{-0.14}^{+0.22}$ & 0.43$_{-0.23}^{+0.13}$ & 7.98$_{-0.45}^{+0.25}$ \\
5  & 54.833$_{-5.765}^{+6.791}$   & 7.386$_{-0.134}^{+0.133}$ & 1.60$_{-0.35}^{+0.28}$ & 21.14$_{-1.45}^{+1.57}$ & 0.84$_{-0.12}^{+0.12}$ & 0.20$_{-0.10}^{+0.17}$ & 6.98$_{-0.68}^{+0.71}$ \\
6  & 35.814$_{-5.174}^{+13.118}$  & 7.441$_{-0.121}^{+0.095}$ & 1.53$_{-0.60}^{+0.34}$ & 14.70$_{-1.08}^{+1.09}$ & 0.26$_{-0.12}^{+0.11}$ & 0.35$_{-0.12}^{+0.13}$ & 6.18$_{-0.13}^{+0.40}$ \\
7  & 40.327$_{-7.569}^{+10.562}$  & 6.833$_{-0.164}^{+0.288}$ & 1.41$_{-0.51}^{+0.37}$ & 8.77$_{-1.32}^{+1.00}$ & -0.54$_{-0.25}^{+0.17}$ & 0.38$_{-0.19}^{+0.16}$ & 7.64$_{-1.39}^{+1.10}$ \\
8  & 59.476$_{-2.716}^{+3.305}$   & 7.320$_{-0.163}^{+0.147}$ & 1.05$_{-0.11}^{+0.13}$ & 3.04$_{-0.32}^{+0.41}$ & -0.51$_{-0.09}^{+0.10}$ & 0.02$_{-0.01}^{+0.01}$ & 8.12$_{-0.28}^{+0.25}$ \\
9  & 64.423$_{-8.143}^{+16.924}$  & 7.079$_{-0.175}^{+0.234}$ & 0.54$_{-0.33}^{+0.38}$ & 4.77$_{-0.54}^{+0.65}$ & -1.45$_{-0.18}^{+0.19}$ & 0.04$_{-0.02}^{+0.03}$ & 9.22$_{-0.15}^{+0.12}$ \\
10 & 41.421$_{-2.957}^{+3.811}$   & 7.250$_{-0.247}^{+0.213}$ & 1.53$_{-0.65}^{+0.31}$ & 3.48$_{-0.31}^{+0.37}$ & -1.92$_{-0.14}^{+0.15}$ & 0.11$_{-0.05}^{+0.09}$ & 9.02$_{-0.56}^{+0.40}$ \\
\hline
\end{tabular}
\tablefoot{From left to right columns are: selected GMC, the molecular cloud (MC) temperature ($T_{\rm{MC}}$) obtained from a gray body fit, the MC radius (ism.R\_MC), the gas metallicity ($Z_{\rm{gas}}$), the maximum SFR, which happens at certain time ism.tau\_esc, known as the characteristic timescale (last column). The previous to last column corresponds to MC's fraction into the ISM (ism.f\_MC). For main SF related parameters, see Table~\ref{Tab.star_formation_results_GalaPy}.}

\bigskip

\begin{tabular}{llllllllll}
\hline \hline
GMC & $T_{\rm{DD}}$ & $M_{\rm{dust}}$ & ism.Rdust & ism.f\_PAH & noise.f\_cal & $\chi^{2}_{red}$ \\
\hline
   & [K] & [log$_{10}$ ($M_{\odot}$)] & [log$_{10}$ (pc)] & & & \\
\hline
1  & 34.643$_{-3.512}^{+2.957}$ & 4.410$_{-0.195}^{+0.210}$ & 1.08$_{-0.09}^{+0.10}$ & 0.17$_{-0.06}^{+0.10}$ & -0.64$_{-0.11}^{+0.11}$ & 2.27 \\
2  & 33.528$_{-2.875}^{+4.437}$ & 4.435$_{-0.105}^{+0.115}$ & 1.08$_{-0.07}^{+0.07}$ & 0.32$_{-0.07}^{+0.06}$ & -2.25$_{-0.14}^{+0.14}$ & 1.80 \\
3  & 34.110$_{-3.781}^{+14.707}$ & 5.766$_{-0.315}^{+0.204}$ & 1.64$_{-0.20}^{+0.11}$ & 0.12$_{-0.07}^{+0.04}$ & -0.61$_{-0.06}^{+0.07}$ & 1.47 \\
4  & 84.532$_{-12.551}^{+6.791}$ & 5.195$_{-0.108}^{+0.153}$ & 1.17$_{-0.08}^{+0.15}$ & 0.04$_{-0.02}^{+0.02}$ & -0.55$_{-0.07}^{+0.07}$ & 1.32 \\
5  & 67.929$_{-3.896}^{+4.023}$ & 5.566$_{-0.107}^{+0.133}$ & 1.43$_{-0.09}^{+0.08}$ & 0.03$_{-0.01}^{+0.02}$ & -0.62$_{-0.08}^{+0.08}$ & 1.14 \\
6  & 63.318$_{-2.800}^{+3.233}$ & 5.453$_{-0.102}^{+0.079}$ & 1.35$_{-0.08}^{+0.06}$ & 0.04$_{-0.01}^{+0.01}$ & -0.73$_{-0.08}^{+0.10}$ & 1.35 \\
7  & 62.911$_{-10.806}^{+4.631}$ & 4.474$_{-0.135}^{+0.145}$ & 1.08$_{-0.10}^{+0.12}$ & 0.06$_{-0.03}^{+0.02}$ & -0.94$_{-0.09}^{+0.10}$ & 1.38 \\
8  & 41.419$_{-2.629}^{+4.510}$ & 4.492$_{-0.127}^{+0.147}$ & 1.21$_{-0.09}^{+0.08}$ & 0.17$_{-0.05}^{+0.07}$ & -1.38$_{-0.10}^{+0.11}$ & 1.35 \\
9  & 43.383$_{-4.396}^{+5.060}$ & 4.679$_{-0.124}^{+0.133}$ & 1.17$_{-0.09}^{+0.08}$ & 0.15$_{-0.04}^{+0.07}$ & -1.11$_{-0.08}^{+0.09}$ & 1.29 \\
10 & 41.532$_{-3.377}^{+5.290}$ & 4.574$_{-0.117}^{+0.128}$ & 1.20$_{-0.08}^{+0.08}$ & 0.19$_{-0.05}^{+0.06}$ & -1.25$_{-0.09}^{+0.10}$ & 1.31 \\
\hline
\end{tabular}
\tablefoot{From left to right, the columns represent the GMC number, the temperature diffuse dust (DD) temperature ($T_{\rm{DD}}$), the dust mass ($M_{\rm{dust}}$), the radius of the DD region (ism.Rdust), the fraction of the total emission from DD contributed by PAH (ism.f\_PAH), the calibration error (noise.f\_cal) which can introduce up to 30\% uncertainties in the photometric points, accounting for observation uncertainties while effectively relaxing the constraints on the parameter space and allowing for a broader range of possible solutions. The last column is the the reduced $\chi^{2}$ computed at the best-fitting values of parameters. For main SF related parameters, see Table~\ref{Tab.star_formation_results_GalaPy}.}
\end{table*}

GMC~2: This molecular cloud also overlaps with x$_{1}$/x$_{2}$ interactions and presents physical and chemical conditions that are similar to GMC~1, although GMC~2 is more massive in terms of stellar and dust mass (Table~\ref{Tab.SED_outputs_GalaPy}; see also \citealt{Humire2022} and \citealt{Bouvier2024}), older (Table~\ref{Tab.SED_outputs_GalaPy}), and shows weaker MMCIs exclusively in the $J_{-1}\rightarrow(J-$ 1)$_{0}-E$ family \citep{Humire2022}. With slightly lower values than GMC~1, GMc~2 presents the lowest dusty radius, as calculated by GalaPy, and the lowest stellar mass, according to \textsc{starlight}, as indicated in Tables~\ref{Tab.SED_outputs_GalaPy} and \ref{Tab.STARLGIGHT_results}, respectively.

GMC~3 appears to be the source of the stellar outflow best viewed by high-density tracers such as CN, HCN \citep{Walter2017}, or an increment in the CO/$^{13}$CO $J=$1--0 ratios owed to a more transparent environment \citep{Bao2024}. Although dense molecular gas possess a velocity dispersion of $\sim$40~km~s$^{-1}$ in the streamer \citep{Walter2017}, the stellar velocity dispersion is the lowest among the sample (see Table~\ref{Tab.STARLGIGHT_results}) at the GMC position. This corresponds to the depiction of cold/low-dispersion gas actively producing stars at the convergence of the ring and the bar, as observed in M~100 \citep{Allard2005}. A more evolved and therefore, more easily observed stellar stream is in line with this position being the oldest among the nuclear regions, as deduced from our SED fitting in Table~\ref{Tab.SED_outputs_GalaPy}. In the same table, and also in Figure\ref{Fig.Mstar_vs_Mdust}, it can be seen than GMC~3 presents the largest dust reservoir and extension.

GMC~4 exhibits the highest SiO(2--1)/HNCO(4$_{0,4}-3_{0,3}$) ratio in the sample \citep{Meier2015,Humire2022}, indicative of strong shocks. It is also the second youngest molecular cloud in the sample according to GalaPy and the youngest according to \textsc{starlight} fits, respectively (see Table~\ref{Tab.star_formation_results_GalaPy} and Figs.~\ref{Fig.SFHs} and \ref{fig.stellar_ages_GalaPy_STARLIGHT_CIGALE}). From CIGALE and GalaPy, it also possess the second largest SFR. This intense stellar formation, despite the absence of a distinct molecular stellar streamer, may be a consequence of its youthful nature or simply that radio emission is not strong enough to trace it (although the stellar streamers originating from the central GMCs is clearly seen in H$\alpha$; see second to last upper panel of Fig.~\ref{Fig.summary_plot}). Additionally, and from GalaPy's results, GMC~4 has the largest $M_{\rm{dust}}/M_{\rm{gas}}$ ratio in the sample ($\log_{10} \left( \frac{M_{\rm{dust}}}{M_{\rm{gas}}} \right) = -1.730$; see Table~\ref{Tab.SED_outputs_GalaPy}), which could suggest that material is being rapidly funneled toward the nucleus via the bar. Similarly, GMCs~5 and 6 are also among the youngest in the sample, with the second and third largest dust-to-gas mass ratios ($\log_{10} \left( \frac{M_{\rm{dust}}}{M_{\rm{gas}}} \right) = -1.820$ and -1.988, respectively). A possible explanation is that material is being concentrated in these GMCs due to inner-inner Lindblad resonances caused by the nuclear bar \citep{Anantharamaiah1996,Cohen2020}. Indeed, looking at the left panel of Fig.~11 in \citet{Humire2022}, and after considering our 3$\arcsec$ photometric aperture and modified GMC positions, we can clearly see a peak in the SiO(2--1)/HNCO(4$_{0,4}-3_{0,3}$) ratios around GMC~5, this latter showing a gap due to SiO(2--1) self-absorption.

Our mid-IR data from NACO and VISIR at the VLT show the highest fluxes in the sample at this position (see Fig.~\ref{fig.SEDs}). GMCs~2 and 4 are the only giant molecular clouds where 10--36\% of the stellar mass was produced in the last $\sim$2 million years. In contrast, the bulk of stellar mass in the remaining GMCs was produced approximately 1 to 10 billion years ago (see Fig.~\ref{Fig.SFHs}).

GMC~5: Among our sample, GMC~5 has the best signal across the different wavelengths, being the most clearly observed at centimeter wavelengths, with its emission covering almost the entire aperture. It is closest to the strongest radio continuum source, TH2 \citep{Turner_and_Ho1985}, which also produces some (self-)absorption features in methanol \citep{Humire2022}, in contrast to the outermost GMCs, where methanol absorption is primarily attributed to anti-inversion against the cosmic microwave background (CMB; \citealt{Bulatek2023}). The strong continuum emission also leads to self-absorption in other molecules, such as SiO $J$=2--1, as inferred from \citet[][their Fig.~11, left panel]{Humire2022}, although this has not been investigated in detail. According to GalaPy and CIGALE results, GMC~5 exhibits by far the largest instantaneous and time-averaged (over 10 and 100~Myr) SFRs in our sample ($\sim$0.7 and 0.2$M_{\odot}$~yr$^{-1}$, according to GalaPy and CIGALE, respectively; see Tables~\ref{Tab.star_formation_results_GalaPy} and \ref{Tab.CIGALE_main_results}), creating about three times more stars than GMC~4, which shows the closest characteristics among the ten GMCs studied.

GMC~6: Most species peak at this location, including CO isotopologues \citep{Harada2024}. Clear exceptions are the sub-mm continuum emissions at Bands 3-7, RRLs, CN, HCN, and CO ($J=$2--1, and $J=$3--2), whose emission is strongest in GMC~5. It shows the highest column densities in both methanol symmetric types and presents the second highest SiO(2--1)/HNCO(4$_{0,4}-3_{0,3}$) ratios after GMC~4, indicative of a strongly perturbed environment due to the presence of strong shocks \citep{Meier2015,Humire2022,Huang2023}.

GMC~7 hosts the strongest methanol emission at 84.53 and 132.89~GHz, which is enhanced by maser emission. It also presents the strongest difference between E- and A-CH$_{3}$OH from a single LTE component analysis, with a E/A ratio of 3.16$^{+0.17}_{-0.18}$, which can only be related to shocks since theoretical values are not expected to surpass 1.0, unless methanol is not formed from CO on cold grain surfaces in NGC~253 \citep{Wirstrom2010}. We note that E/A ratios of up to 6.0 have been seen in massive star-forming regions in our galaxy \citep[see, e.g.,][their Fig.~10(a)]{Zhao2023}.

GMC~8: Using a two-component LTE analysis, this GMC shows one of the largest E/A-CH$_{3}$OH ratios in the cold layer (3.14$\pm$0.58). This may be due to low temperatures, where E-CH$_{3}$OH is more populated as their upper level energies ($E_{\rm{up}}$) are in general lower than its A-CH$_{3}$OH counterparts \citep{Humire2022,Humire2024}. The above is in line with cold dust emission increments, as can be inferred from enhanced $~$100~GHz continuum emission, in contrast to its spurious signal in GMCs~1 and 2, as accounted for in our Fig.~\ref{Fig.Cont_maps_ALCHEMI}. A similar conclusion can also be obtained from the distribution of sulfurated molecules seen in ALMA Band~3 (84--116~GHz), whose emission is also enhanced toward the northeast segment of NGC~253's CMZ \citep{Meier2015}.

GMC~9 presents a mild stellar velocity dispersion ($\sigma_{\bigstar}$; see Table~\ref{Tab.STARLGIGHT_results}) that may account for a relative low formation of stars \citep{Allard2006} and be a result of its position in between the spiral arms and the inner bar, where (x$_{1}$/x$_{2}$) interactions and inner Lindblad resonances take place (see dashed gray ellipses in Fig.~\ref{Fig.Cont_maps_ALCHEMI}). A similar origin for the velocity dispersion can be inferred for GMCs~1, 2, and 8--9, in contrast to the larger $\sigma_{\bigstar}$ measured at GMCs~3, 5, 6, and 7, that are possibly due to a vigorous stellar formation instead \citep{Bouvier2024,Bao2024}.

GMC~10: We do not see much high-excitation methanol lines in this region, that shows the lowest excitation temperatures and number of methanol lines, in addition to the lowest column density and line-widths (FWHM $\sim$23~km~s$^{-1}$) of both methanol symmetric types among the ten GMCs studied in this work. Using high density sulfured molecules, \citet{Bouvier2024} also determined the lowest line-widths (FWHM $\sim$20~km~s$^{-1}$) and rest velocities (100--150~km~s$^{-1}$) among the GMCs in the sample.

\section{Sulfur ratios and Age-metallicity relation in GMC~5}
\label{sec.AMR}
While many simple abundant molecules are prone to undergo strong opacity effects, impeding a confident derivation of line ratios, isotopologues of less abundant elements are normally optically thin. Moreover, line ratios can be used to derive metallicities such as [Fe/H], defined concerning the Solar metallicity by the following relation:
\begin{equation}
    [\rm{Fe/H}] = log_{10}([N_{\rm{Fe}/N_{\rm{H}}}])_{\rm{star}} - log_{10}([N_{Fe}/N_{\rm{H}}])_{\rm{Sun}}.
\end{equation}

One way to obtain [Fe/H] through optically thin molecules is using the relation in Table~3 of \citet{Kobayashi2011}, well defined across the -0.5$\leq$[Fe/H]$\leq$0.0 range:

\begin{equation}
    \frac{\it{N}\rm{(^{32}S})}{\it{N}\rm{(^{34}S)}} = -19.8\times [\rm{Fe/H}]+23.2.
\label{eq.Kabayashi11}
\end{equation}

\begin{figure*}[!ht]
    \centering
    \begin{subfigure}[b]{\textwidth}
        \centering
        \includegraphics[width=\textwidth, trim={0 0 0 0}, clip]{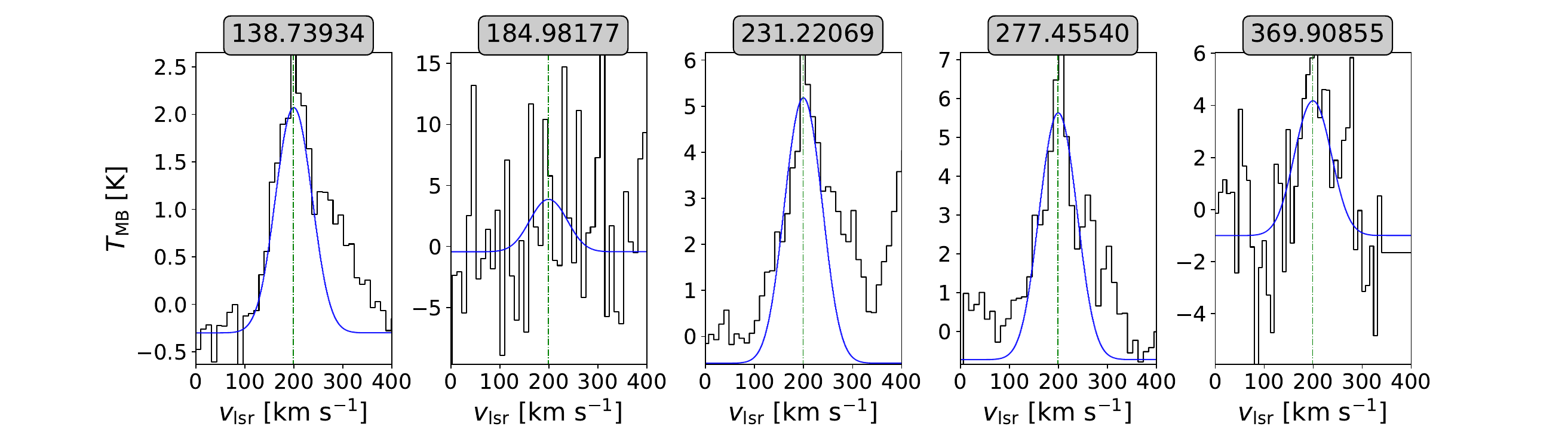}
        \label{Fig.13CS}
    \end{subfigure}
    \vspace{0cm} 
    \begin{subfigure}[b]{\textwidth}
        \centering
        \includegraphics[width=\textwidth, trim={0 0.3cm 0 0.3cm}, clip]{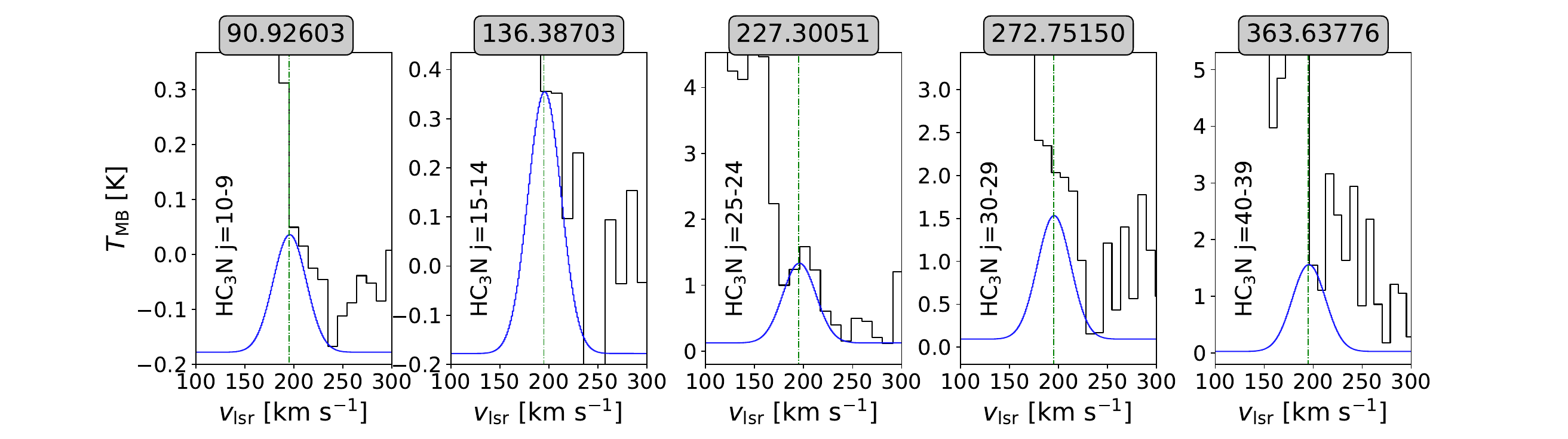}
        \label{Fig.13C34S}
    \end{subfigure}
    \caption{Top panel: $^{13}$CS. Bottom panel: $^{13}$C$^{34}$S. Frequencies, in GHz, are labeled at the top of each sub-panel. Best-fit $V_{\rm{LSR}}$ values (see Table~\ref{tab.best_fit_sulfur_isotopologues}) are indicated by green-dashed vertical lines. The HC$_{3}$N blending feature is indicated to the left of each $^{13}$CS transition.}
    \label{Fig.13CS_and_13C34S_LTE_modeling}
\end{figure*}

This equation relates the column densities, $N$, of sulfur isotopologues with the metallicities. With the purpose of obtaining the required column density ratio, we employed the $^{13}$CS and $^{13}$C$^{34}$S isotopologues, of which $^{13}$C$^{34}$S is only detectable in GMC~5 from our sub-mm data. Within the ALCHEMI frequency coverage (84.4--373.3~GHz), $^{13}$C$^{34}$S is always close but not fully blended with HC$_{3}$N (see Fig.~\ref{Fig.13CS_and_13C34S_LTE_modeling}, bottom panels). Its transitions at $J_{u}=$2 to 8, when present, fall $\sim$160~km~s$^{-1}$ away (redshifed) from HC$_{3}$N at $J_{u}=$2$x$ to 8$x$, with $x$ being equal to 5 and $J_{u}$ being always the unity above the lower $J$; for example $J_{u}=4$ being equal to the $J=$4--3 transition.

Assuming LTE conditions, we fitted the observed spectra with a single component to account for the very dense regions that CS isotopologues trace. To this end, we employed the Centre d'Analyse Scientifique de Spectres Instrumentaux et Synthétiques (CASSIS), a software developed by CESR/IRAP\footnote{\url{http://cassis.irap.omp.eu/?page=cassis}} \citep[see, e.g.,][]{Vastel2015} and designed for the analysis and synthesis of molecular spectra in astrophysical research. It enables spectral modeling, line identification, and data interpretation for astronomical observations, with the capability of performing both local thermodynamic equilibrium (LTE) and non-LTE modeling. When performing non-LTE modeling, CASSIS functions as a wrapper for the RADEX code \citep{vanderTak2007}, facilitating its utilization for users. CASSIS is not restricted to the radio regime \citep[e.g.,][]{HernandezGomez2019} as it has also been used to distinguish between LTE and non-LTE emission, enabling the discovery of new maser transitions \citep{Humire2024}. While originally designed to process single spectra, the code can be adapted to handle data cubes \citep{Harada2022}

To produce the synthetic spectrum that better reproduces the CS-related lines, we have utilized the Cologne Database for Molecular Spectroscopy \citep[CDMS;][]{Mueller2005} for spectroscopic data and restricted to LTE conditions, since these molecules are expected to trace dense gas. Our best-results are shown in Fig.~\ref{Fig.13CS_and_13C34S_LTE_modeling}.

\begin{figure*}[!ht]
\centering
\includegraphics[width=\textwidth, trim={0 0 0 0}, clip]{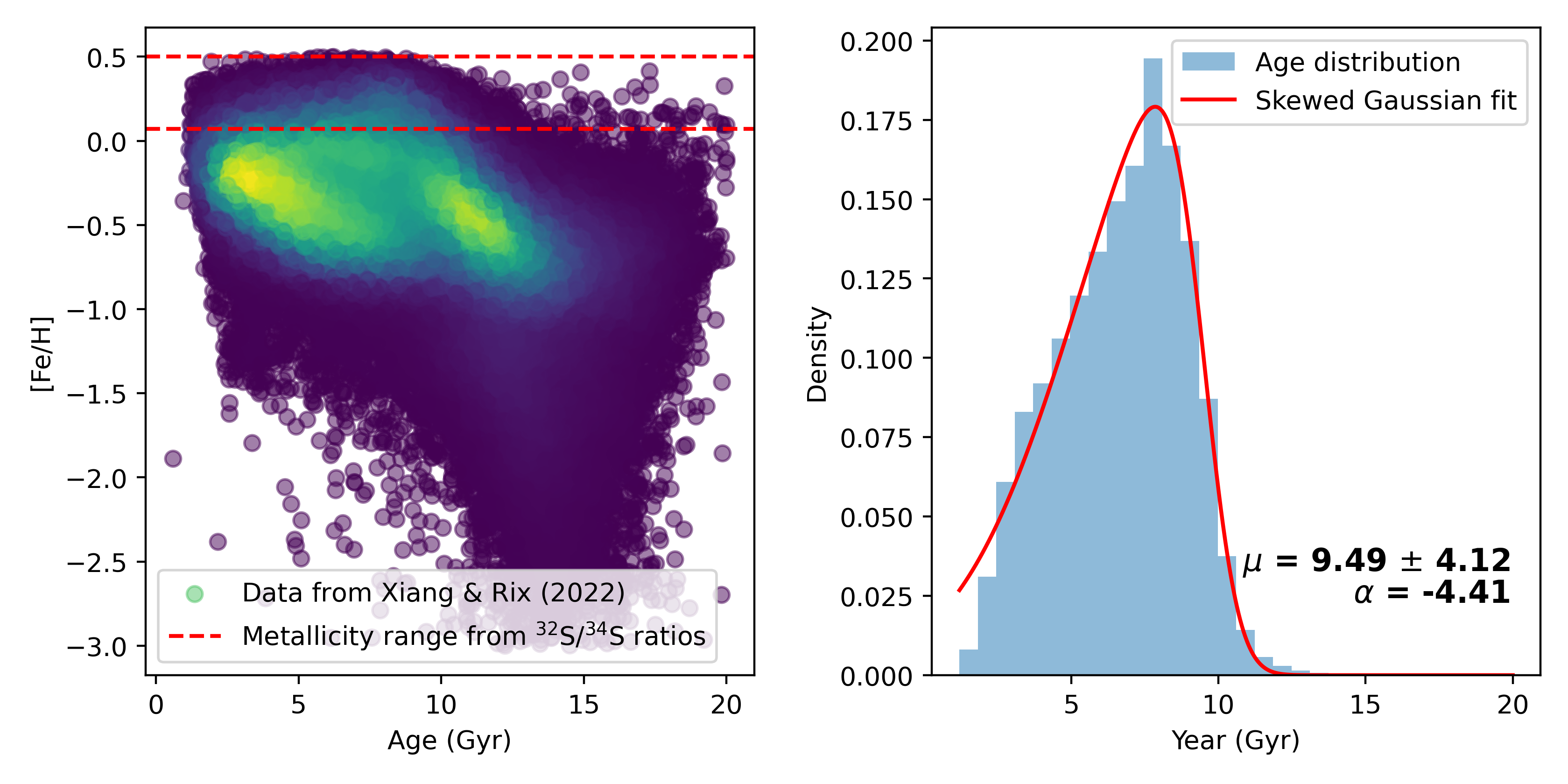}
\caption{Metallicities and ages of the sample presented in \citet{Xiang2022}. Left panel: Metallicity-age relation color-coded by density with our derived metallicity range from sulfur isotopologue ratios. Right: Age distribution for the considered metallicity range indicating its mean value and 1$\sigma$ uncertainty from a single Gaussian fit.}
\label{Fig.MAR}
\end{figure*}

This way, we obtained different solutions ranging from $^{32}$S/$^{34}$S rates much lower than expected in the Milky Way ($\sim$5) to more reasonable values ($\sim$17). We also find that the line intensities can be reproduced under a large range of excitation temperatures. Due to this degeneracy, we decided to limit the $^{32}$S/$^{34}$S range to values fulfilling the criterion: [Fe/H]$\leq$0.5, motivated by the metallicity values obtained from a sample of $\sim$247,000 sub-giant stars in the Milky Way (see \citet{Xiang2022} and Fig.~\ref{Fig.MAR}). This way, we expect $^{32}$S/$^{34}$S isotopologue ratios no lower than 13.3, namely, within the range of values measured in most of the galaxy \citep[see][their Fig.~5]{Humire2020b}. Considering that Sgr~B2(N) shows properties comparable to a mini-starburst system \citep{Belloche2013,Schwoerer2019}, and that the best $^{32}$S/$^{34}$S column density ratio measured there is 16.3$^{+2.1}_{-1.7}$ \citep{Humire2020b}, our assumption of $^{32}$S/$^{34}$S$\geq$13.3 is well justified.

Our best-fit LTE modeling parameters from the spectra of $^{13}$CS and $^{13}$C$^{34}$S are presented in Table~\ref{tab.best_fit_sulfur_isotopologues}. The resulting $^{32}$S/$^{34}$S column density ratio is 14.26$^{7.53}_{-4.72}$, and we set it to be 14.26$^{7.53}_{-0.96}$ considering our lowest threshold. Using Eq.~\ref{eq.Kabayashi11}, the resulting [Fe/H] metallicity values are 0.45$^{+0.05}_{-0.38}$, namely, between 0.07 and 0.5. Comparing this latter result with the age-metallicity relation presented in Fig.~2a of \citet{Xiang2022}, reproduced in the left panel of Fig.~\ref{Fig.MAR}, we can estimate an age range. To this aim, we utilized a skewed Gaussian distribution model due to the uni-modal and asymmetric distribution of ages within our estimated metallicity range, as can be seen in the right panel of Fig.~\ref{Fig.MAR}. The fitting process involved optimizing the parameters of the skewed Gaussian function, including amplitude \(A\), mean \( \mu \), standard deviation \( \sigma \), and skewness \( \alpha \), using a least-squares approach. The skewed Gaussian distribution function is defined as:
\begin{equation}
f(x) = \frac{A}{\sigma \sqrt{\pi}} e^{-\frac{(x - \mu)^2}{2\sigma^2}} \left(1 + \text{erf}\left(\alpha \frac{x - \mu}{\sigma \sqrt{2}}\right)\right),
\end{equation}
\noindent
where the error function, \(\text{erf}(z)\), is defined as:
\begin{equation}
\text{erf}(z) = \frac{2}{\sqrt{\pi}} \int_0^z e^{-t^2} \, dt.
\end{equation}
\noindent
Our analysis revealed a negative skewness parameter \( \alpha \), indicating a non-symmetric age distribution with a higher density toward young ($\sim$2.5--7.5~Gyr) sub-giant stars, withing the age-metallicity relation distribution used here (Fig.~\ref{Fig.MAR}). The resulting estimated age, namely, the distribution peak and its FWHM, is 7.85$^{+0.18}_{-5.39}$~Gyrs, reflecting a stellar population 31 times older than the one traced by our SED fitting in GMC~5, summarized in Table~\ref{Tab.SED_outputs_GalaPy}, whose value is of 10$^{8.40}$ or $\sim$251~Myr. 

\begin{table}[!t]
\caption{Best fit parameters from our LTE plus non-LTE modeling for CARMA-7.}
\label{tab.best_fit_sulfur_isotopologues}
\setlength{\tabcolsep}{0.03cm}
\begin{center}
\begin{tabular}{llllll}
\hline \hline
Species & $N$(Sp)                                    & $T_{\rm ex}$            & FWHM                               & $V_{\rm LSR}$                       & Size\\
&   \multicolumn{1}{c}{[$\times$10$^{14}$cm$^{-2}$]}   & \multicolumn{1}{l}{[K]} & \multicolumn{1}{l}{[km\,s$^{-1}$]} & \multicolumn{1}{l}{[km\,s$^{-1}$]}  & \multicolumn{1}{l}{[$\arcsec$]}\\ 
\hline \\
$^{13}$CS         & 29.67$^{+4.12}_{-9.63}$ & 36.17$^{+46.95}_{-10.14}$         &83.87$^{+86.64}_{-22.81}$      &198.89$^{+39.20}_{-12.99}$  & 1.08$^{+1.91}_{-0.25}$\\
$^{13}$C$^{34}$S  & 2.08$^{+0.02}_{-0.53}$  & 60.97$^{+101.06}_{-23.88}$        &39.83$^{+0.16}_{-14.66}$       &195.00$^{+1.00}_{-1.00}$    & 2.93$^{+0.07}_{-1.96}$ \\
\hline \\
\end{tabular}
\tablefoot{Best-fitting parameters obtained through our LTE modeling of sulfur isotopologues. From left to right columns correspond to: Name of the species, $N$(Sp) stands for column density of species, $T_{\rm{ex}}$ is the excitation temperature, the line-width measured from its FWHM, the local standard of rest velocity, $V_{\rm{LSR}}$, and the best-fit source size in arcseconds. Uncertainties correspond to a 3$\sigma$ level.}
\end{center}
\end{table}

Although magnetohydrodynamic models for sulfur may not yet be well constrained for the bulge of galaxies, the most likely explanation for sulfur ratios in the very center of our Milky Way is that its large-scale bar is funneling material from its edges toward the galaxy's nucleus or bulge \citep{Humire2020b}. Following this reasoning, the $^{32}$S/$^{34}$S abundances can still be modeled according to \citet{Kobayashi2011} (C. Kobayashi, priv. comm.), by assuming that the origin of $^{32}$S/$^{34}$S abundances occurs at a certain level in the disk of galaxies.

In a more recent paper, \citet{Kobayashi2020} presented an updated version of their age-metallicity relation (their Fig. 2b), which, as the SFH models used in GalaPy and CIGALE, also considers a single burst in its SFH model. This updated model integrates a bulge outflow component that is more suited for the NGC~253 central molecular zone (CMZ), where stars drive material outside the bar through significant stellar outflows, likely originating from supernovae. Applying the 0.0–0.5 iron metallicity range to the bulge outflow model yields ages starting from one Gyr, while the bulge component alone provides a more restricted range of one to five Gyr, peaking at 3~Gyr (or 10$^{9.48}$ years). Both results are in good agreement with the AMR-derived estimations.

The discrepancy arises because metallicities measured by molecular emitters provide a broader perspective, reflecting the average contributions of all stellar populations that have expelled material to the interstellar medium (ISM). In contrast, values derived from SED fitting are predominantly influenced by recent star formation, which emits most of its radiation in the mid-infrared (MIR; 1.5--50$\mu$m). Consequently, we anticipate deriving an average age estimate from the sub-millimeter regime and a younger age estimate from SED fitting. Additionally, employing molecular tracers at higher vibrational modes ($v=1$ or 2) and angular resolutions allows for tracing the youngest stellar population nicely tracing the last burst in the light-weighted SFHs of Fig.~\ref{Fig.SFHs}, representing up to 43\% of the total stellar light emitted in the optical regime (see dashed blue vertical lines in Fig.~\ref{Fig.SFHs}) \citep{RicoVillas2020, Butterworth2024}.

However, it might be that the MIR emission we account for in the SED fitting mainly originates from low-mass/old stellar populations. If that is the case, we do not have a clear answer for this discrepancy.

\section{Comparison among different methodologies}

\begin{figure*}[!ht]
\centering
\includegraphics[width=\textwidth, trim={0 0 0 0}, clip]{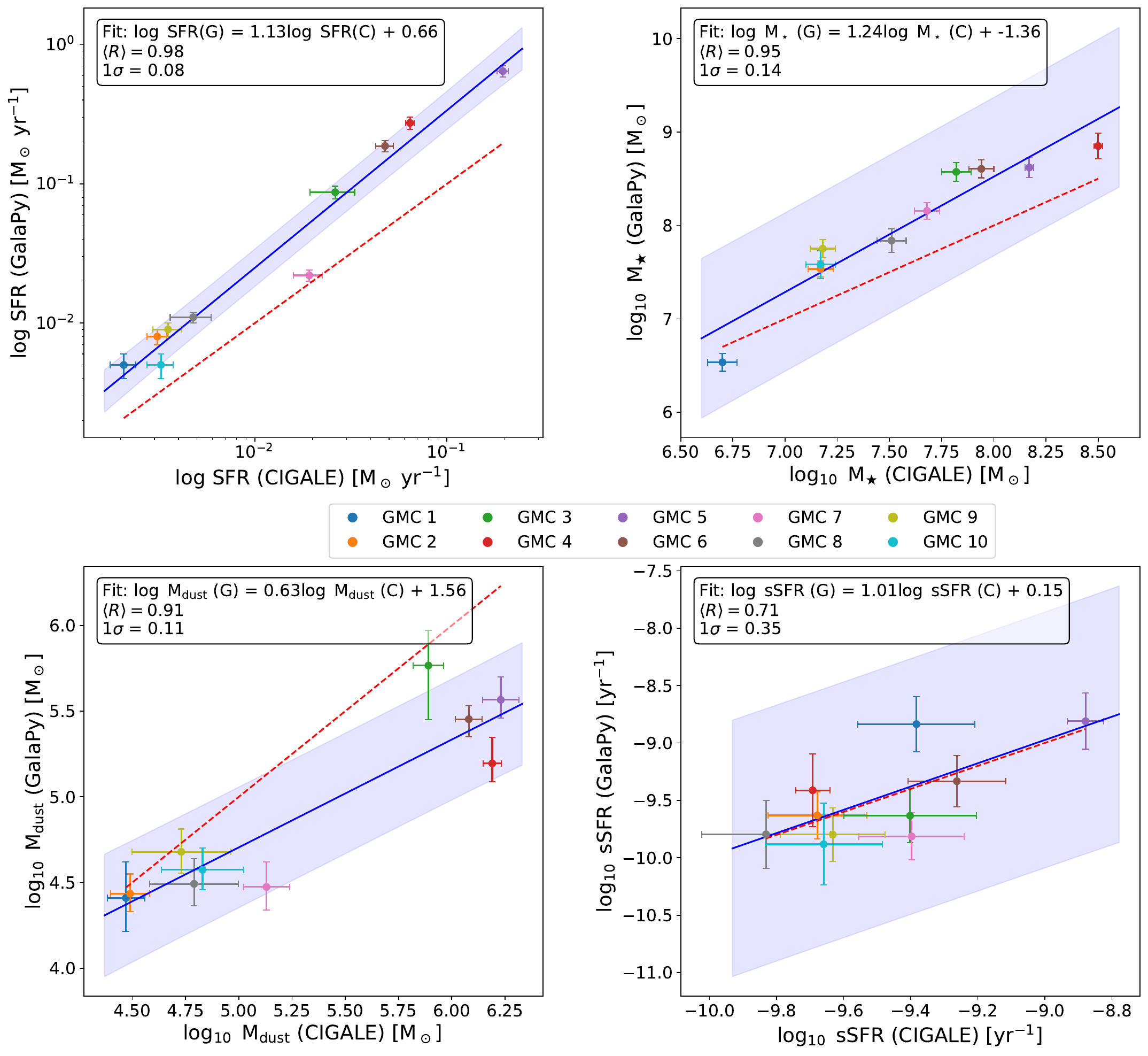}
\caption{Comparison between the instantaneous star formation rate (SFR), the stellar mass ($M_{\bigstar}$), the dust mass ($M_{\rm{dust}}$), and the specific star formation rate (sSFR) derived from GalaPy (G; y-axes) and CIGALE (\textbf{C;} x-axes). GMCs are color-coded following the legend at the middle panel. All fits were performed on log-transformed data, except for the SFR, which was fitted using the linear scale. Linear regression was initially performed using the {\tt linregress} function from the {\tt SciPy} package \citep{Gommers2022}, followed by Bayesian inference through Markov chain Monte Carlo (MCMC) sampling implemented in the {\tt emcee} package \citep{Foreman-Mackey2013}.}\label{fig:comparison_grid}
\end{figure*}

Given the different SED modeling approaches, that differ in terms of star formation histories and several other assumptions, here we provide a few comparison plots. Depending directly on the SED shape, we will obtain a sSFR \citep{Conroy13}. Then, each model assumes a certain mass-to-light ratio to derive the stellar mass and a specific SFH to derive the stellar ages, combining both results, mass and time, to obtain the SFR. Given the variation between GalaPy and CIGALE modeled SED shapes, as seen in Fig.~\ref{fig.SEDs}, we expect dispersion in their resulting sSFRs, but a relative tendency to be a one-to-one relation. Such a tendency is seen in the bottom right panel of our Fig.~\ref{fig:comparison_grid}, whose values are listed in Table~\ref{tab:sSFR_comparison}, meaning that, at this level of comparison, both codes share similarities. From that same Figure it can be noted that GalaPy tends to produce larger SFRs and stellar masses, as also previously noted by \citet{Ronconi2024}. The dust mass is mostly consistent between the GalaPy and CIGALE models within 1$\sigma$ uncertainties for the less massive GMCs but systematically shifts toward the lower values of CIGALE compared to the higher values of GalaPy, as evidenced by the 1:1 red line in the bottom left panel of Fig.~\ref{fig:comparison_grid}. This trend becomes more pronounced in the more nuclear and massive regions, specifically GMCs~4–6 and also GMC~7.

\begin{table}
    \caption{Specific star formation rates (sSFRs).}
    \label{tab:sSFR_comparison}
    \centering
    \begin{tabular}{lcc}
    \hline \hline
    \textbf{GMC} & \textbf{sSFR (GalaPy)} & \textbf{sSFR (CIGALE) } \\
         & [log$_{10}$~\text{yr}$^{-1}$] & [log$_{10}$~\text{yr}$^{-1}$] \\
    \hline
    1 & $-8.84 \pm 0.09$ & $-9.38 \pm 0.07$ \\
    2 & $-9.63 \pm 0.05$ & $-9.68 \pm 0.05$ \\
    3 & $-9.63 \pm 0.05$ & $-9.40 \pm 0.11$ \\
    4 & $-9.41 \pm 0.04$ & $-9.69 \pm 0.02$ \\
    5 & $-8.81 \pm 0.04$ & $-8.88 \pm 0.03$ \\
    6 & $-9.33 \pm 0.04$ & $-9.26 \pm 0.05$ \\
    7 & $-9.81 \pm 0.04$ & $-9.40 \pm 0.07$ \\
    8 & $-9.80 \pm 0.04$ & $-9.83 \pm 0.11$ \\
    9 & $-9.80 \pm 0.05$ & $-9.63 \pm 0.07$ \\
    10 & $-9.88 \pm 0.09$ & $-9.66 \pm 0.07$ \\
    \hline
    \end{tabular}
    \tablefoot{Comparison of specific star formation rates (sSFRs) in log$_{10}$ scale derived from GalaPy and CIGALE in our ten selected GMCs.}
\end{table}

\begin{figure*}
\centering
\includegraphics[width=\textwidth, trim={0 0 0 0}, clip]{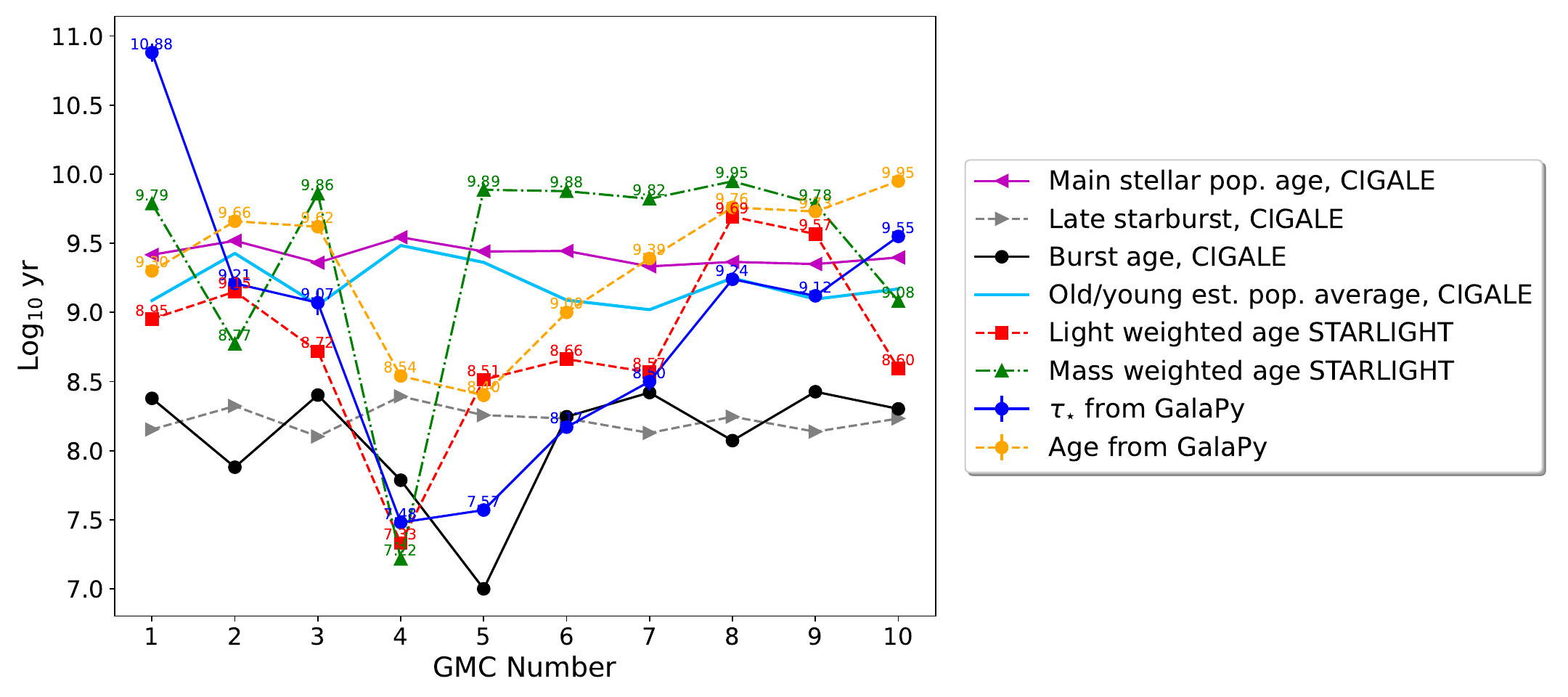}
\caption{GMC stellar ages and burst estimations as traced by GalaPy, {\textsc{starlight}, and CIGALE. SF peaks (SF characteristic timescale, $\tau_{\star}$, from GalaPy in blue and last burst age from CIGALE in black) are also given for comparison.}}
\label{Fig.GMC_ages}\label{fig.stellar_ages_GalaPy_STARLIGHT_CIGALE}
\end{figure*}

In Fig.~\ref{fig.stellar_ages_GalaPy_STARLIGHT_CIGALE}, we compare the stellar population ages, burst ages, SF characteristic timescales, and ages weighted by light and mass, as provided by the different models utilized in this work. Specifically, CIGALE provides the main stellar population age, the late starburst age, the burst age, and the average age of the young and old stellar populations. From \textsc{starlight}, we include the averaged stellar ages weighted by light and by mass, which primarily correspond to the young and old stellar populations, respectively. Additionally, we present the stellar ages derived by GalaPy, along with the characteristic star-formation timescale, $\tau_{\star}$, obtained from the same code.

\section{Testing the FIR bump accuracy}
\label{apen.testing_at_9_arcsec}

We have performed an SED fitting with both GalaPy and CIGALE considering a 9\arcsec\ aperture diameter to include Herschel PACS data in the far infrared. This approach ensures that the infrared bump more accentuated in GalaPy than in CIGALE models (see Fig.~\ref{fig.SEDs} for a clear comparison) at around 100$\mu$m is properly fit without extrapolations. Additionally, it allows us to test whether our extrapolation was consistent with real observations, which cannot be verified at 3$\arcsec$ due to instrumental limitations. We found a similar dust temperature, which is a result driven by the FIR SED peak, in both cases, at 3 and 9$\arcsec$, as already pointed out in Sect.~\ref{Sec.reliability_of_SED_fit_in_the_FIR}. Overall, the SED model used in GalaPy for GMC~5 correlates well with the one derived for the larger 9\arcsec\ aperture centered at the same GMC.

In contrast, while CIGALE models underpredict the FIR bump in several GMCs compared to GalaPy, incorporating Herschel photometric points at 70, 100, and 160$\mu$m enables the code to fit all SED points accurately, regardless of whether an AGN component is assumed. The 9$\arcsec$ SED fitting using CIGALE is shown in the bottom middle panel of Figs.~\ref{apen.fig:cigalefitting} and \ref{apen.fig:cigalefitting_AGN} for the cases without and with an AGN component, respectively. Interestingly, the SFRs differ by less than 5\% between these models, reaffirming that the presence of an AGN is irrelevant from the SED perspective, as previously discussed in Sect.~\ref{subsec.WHAN_BPT}. 

\begin{figure*}[!ht]
\centering
\begin{subfigure}{0.48\textwidth}
    \centering
    \includegraphics[width=\textwidth, trim={0 0 0 0}, clip]{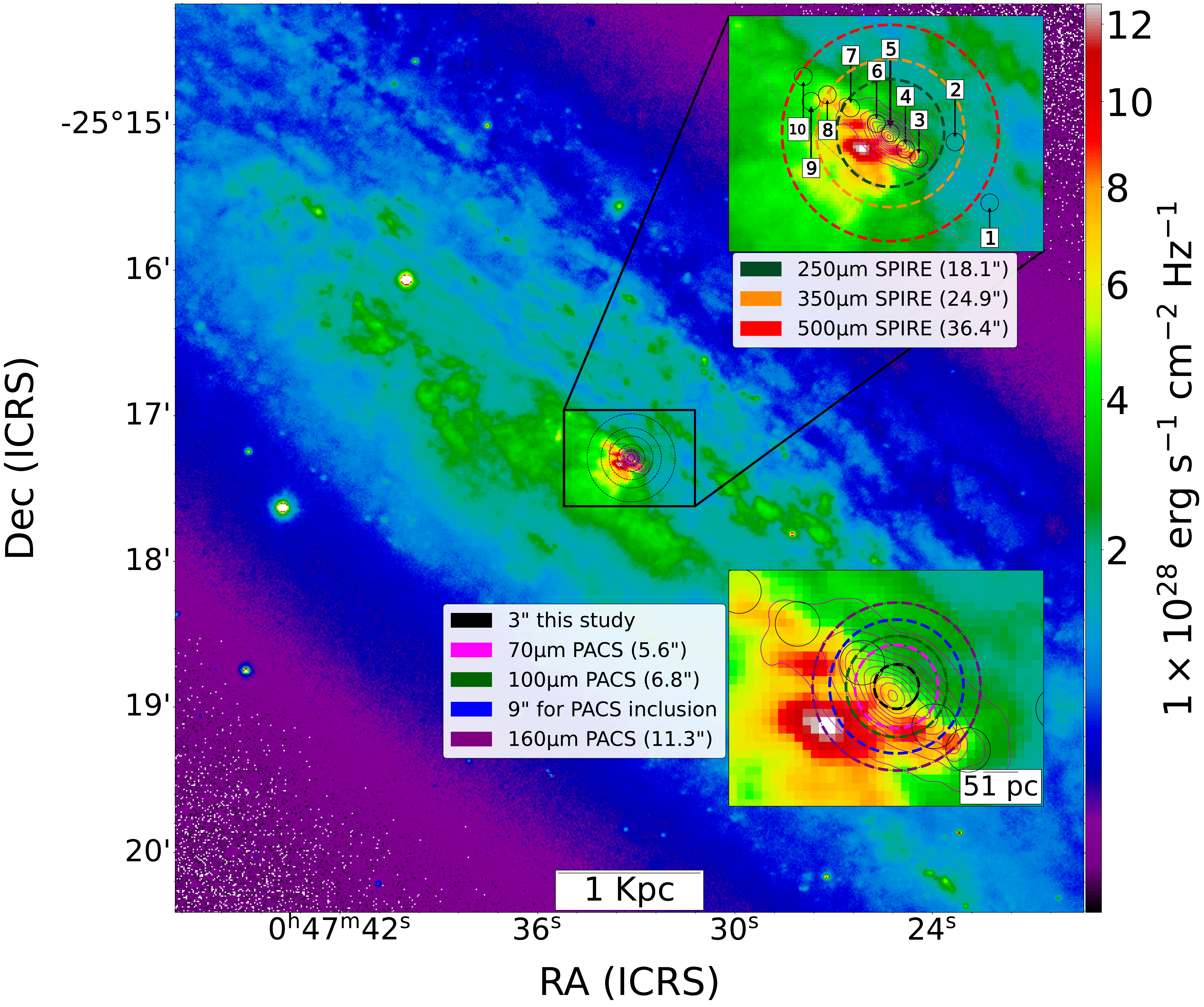}
    \label{subfig:apertures}
\end{subfigure}
\hfill
\begin{subfigure}{0.515\textwidth}
    \centering
    \includegraphics[width=\textwidth, trim={0 0 0 1.8cm}, clip]{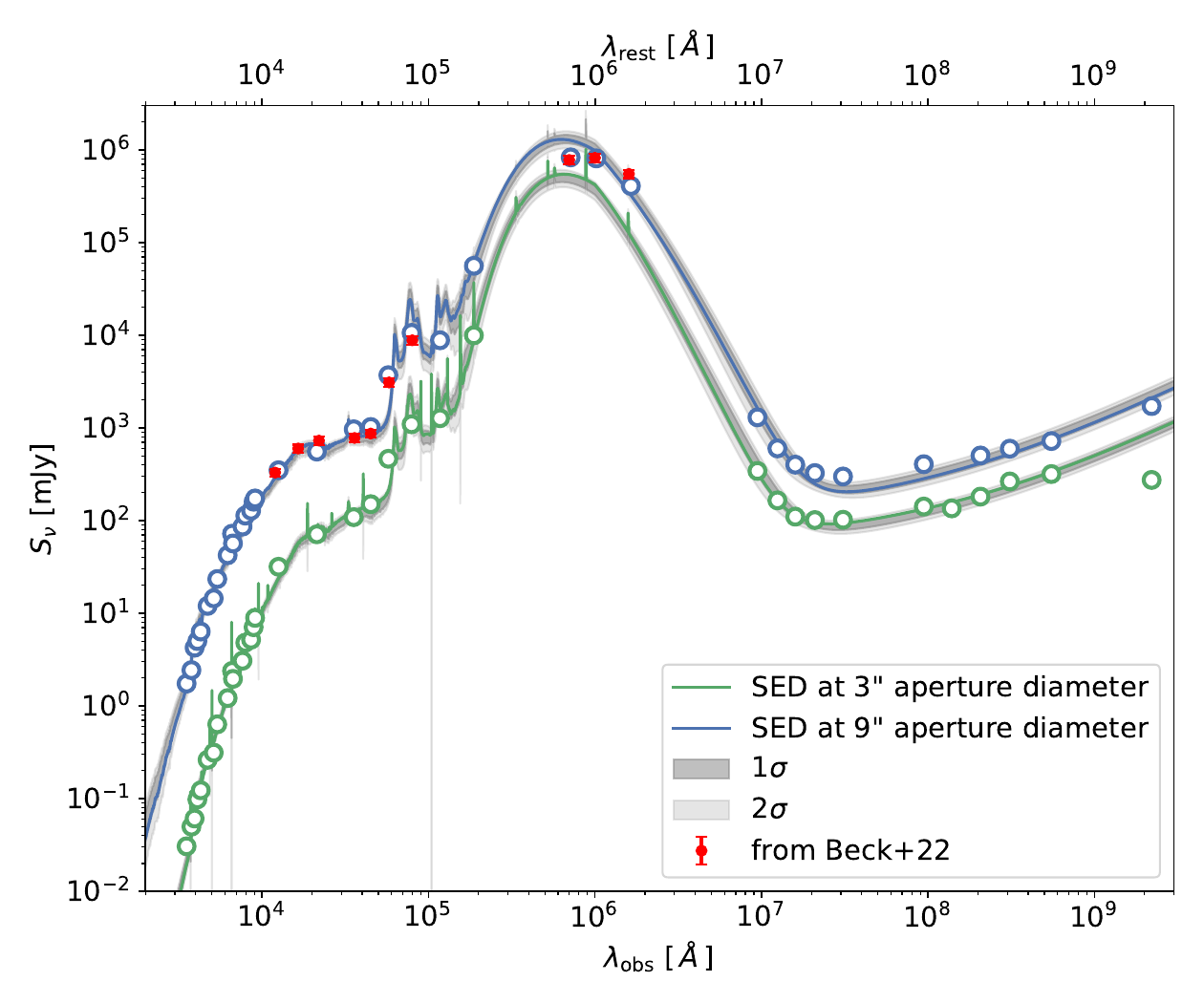}
    \label{subfig:sed_results}
\end{subfigure}
\caption{Testing the far infrared bump with Herschel PACS instruments. Left: Flux density of the S-PLUS R-filter band following the levels indicated in the colorbar. The two insets provide a closer look into the inner regions, where the CMZ is located. The upper inset shows the ten GMCs studied in this work, along with the ellipses where Lindblad resonances were detected by \citet{Iodice2014}. The labeled apertures show how coarse (factor of 6--12 compared to the 3$\arcsec$ aperture used in this work) the angular resolution of Herschel SPIRE is from tracing our observations. The bottom inset performs a further 40\% zooming on the nuclear regions, with Herschel PACS instrument apertures overlaid on the regions, and showing how GMCs~3 to 6 are partially or fully covered at 70, 100, and 160~$\mu$m observations beams. The contours show the L-Band VLA observations (1.4~GHz) to indicate the maximum extent at which the rightmost point in our SEDs, namely, the largest wavelength, is available. Right: Visualization of Herschel PACS apertures and SED combined results. 3 and 9$\arcsec$ aperture SEDs obtained by GalaPy and centered at the GMC~5 position are in green and blue, respectively. We note that 6$\arcsec$ and 9$\arcsec$-aperture-extracted SED do not vary significantly (6$\arcsec$ SED is not shown) indicating that HST, VLT, and ALMA observations correspond to the central emission that does not increment with larger apertures. We consider only the 9$\arcsec$-aperture-extracted SED for comparison with the 3$\arcsec$ one as it better covers Herschel PACS observations at 70 and 100~$\mu$m bands, whose respective angular resolutions are of 5.6$\arcsec$ and 6.8$\arcsec$. Red points with an assumed 10\% uncertainty correspond to the dataset used in \citet{Beck2022}, their Table~4, to produce the SED of their Fig.~8, excluding GALEX observations as we have not used UV information for this work.}
\label{fig.FIR_SEDs_attempts}
\end{figure*}

Finally, the 3$\arcsec$ and 9$\arcsec$ SED models from GalaPy are shown in the right panel of Fig.~\ref{fig.FIR_SEDs_attempts}. The 3$\arcsec$ aperture SED corresponds to what was already presented in Fig.~\ref{fig.SEDs} for GMC~5.

\end{appendix}
\end{document}